\documentclass[sigconf,screen]{acmart}
\AtBeginDocument{%
  }

\usepackage{xcolor}
\usepackage{booktabs}
\usepackage{listings}
\usepackage{xspace}
\usepackage{array}

\usepackage{algorithm}
\usepackage{algpseudocode}
\usepackage{balance}
\usepackage{wrapfig}
\usepackage[table]{xcolor}
\usepackage{hyperref}

\usepackage{enumitem}
\usepackage{listings}
\usepackage{tabularx}
\usepackage{multirow}
\usepackage{booktabs}
\usepackage{graphicx}
\usepackage{caption}
\usepackage{subcaption}

\usepackage{enumitem}
\usepackage{xcolor}    % for \textcolor
\usepackage{pifont}    % alternative checkmark \ding{51}
\usepackage[most]{tcolorbox}
\usepackage{svg}
\usepackage{makecell}
\usepackage{tabularx}
\usepackage{dsfont}    % Type-1 double-stroke; bbm is bitmap-only (Type-3)

\tcbuselibrary{listings,breakable,skins}

\definecolor{jupyterinbg}{RGB}{248,248,248} % Light grey for input background
\definecolor{jupyteroutbg}{RGB}{255,255,255} % White for output background
\definecolor{inpromptcolor}{RGB}{0,100,0}  % Dark green for In[] prompt
\definecolor{outpromptcolor}{RGB}{150,0,0} % Dark red for Out[] prompt
\definecolor{bordercolor}{RGB}{200,200,200} % Light grey for border

\copyrightyear{2026}
\acmYear{2026}
\setcopyright{cc}
\setcctype{by-nc-nd}
\acmConference[ASE '26]{Proceedings of the 41st IEEE/ACM International Conference on Automated Software Engineering}{October 12--16, 2026}{Munich, Germany}
\acmBooktitle{Proceedings of the 41st IEEE/ACM International Conference on Automated Software Engineering (ASE '26), October 12--16, 2026, Munich, Germany}
\acmDOI{10.1145/3832783.3834390}
\acmISBN{979-8-4007-2882-2/2026/10}

\begin{document}

\definecolor{WowColor}{rgb}{.75,0,.75}
\definecolor{SubtleColor}{rgb}{0,0,.50}
\newcommand{\cmark}{\ding{51}\xspace}%
\newcommand{\xmark}{\ding{55}\xspace}%

% general
%\renewcommand{\comment}[1]{}
\ifdefined\Comment
        \renewcommand{\Comment}[1]{}
\else
        \newcommand{\Comment}[1]{}
\fi
% inline
\newcommand{\NA}[1]{\textcolor{SubtleColor}{ {\tiny \bf ($\star$)} #1}}
\newcommand{\Fix}[1]{\textcolor{red}{[#1]}}
\newcommand{\sd}[1]{\textcolor{red}{[#1]}}
\newcommand{\MN}[1]{\textcolor{blue}{#1}}
\newcommand{\LATER}[1]{\textcolor{SubtleColor}{#1}}
\newcommand{\TBD}[1]{\textcolor{SubtleColor}{ {\tiny \bf (!)} #1}}
\newcommand{\PROBLEM}[1]{\textcolor{WowColor}{ {\bf (!!)} {\bf #1}}}

% as margin notes
\newcounter{margincounter}
\newcommand{\displaycounter}{{\arabic{margincounter}}}
\newcommand{\incdisplaycounter}{{\stepcounter{margincounter}\arabic{margincounter}}}

\newcommand{\fTBD}[1]{\textcolor{SubtleColor}{$\,^{(\incdisplaycounter)}$}\marginpar{\tiny\textcolor{SubtleColor}{ {\tiny $(\displaycounter)$} #1}}}

\newcommand{\fPROBLEM}[1]{\textcolor{WowColor}{$\,^{((\incdisplaycounter))}$}\marginpar{\tiny\textcolor{WowColor}{ {\bf $\mathbf{((\displaycounter))}$} {\bf #1}}}}

\newcommand{\fLATER}[1]{\textcolor{SubtleColor}{$\,^{(\incdisplaycounter\dagger)}$}\marginpar{\tiny\textcolor{SubtleColor}{ {\tiny $(\displaycounter\dagger)$} #1}}}

\newcommand{\mypara}[1]{\vspace{.03in}\noindent \textbf{#1.}}
\newcommand{\empara}[1]{\vspace{.03in}\noindent \emph{#1.}}

\lstset{
    language=Python,
    basicstyle=\ttfamily\footnotesize, % Font style and size
    numbers=left, % Line numbers on the left
    numberstyle=\tiny\color{gray}, % Style of line numbers
    stepnumber=1, % Line number interval
    numbersep=8pt, % Distance between line numbers and code
    backgroundcolor=\color{white}, % Background color
    showspaces=false,
    showstringspaces=false,
    showtabs=false,
    frame=single, % Frame around the code
    rulecolor=\color{black},
    tabsize=4, % Tab size
    captionpos=b, % Position of the caption (b: bottom)
    breaklines=true, % Automatic line breaking
    breakatwhitespace=false,
    keywordstyle=\color{blue}, % Keyword color
    commentstyle=\color{green!60!black}, % Comment color
    stringstyle=\color{orange}, % String color
    morekeywords={as}, % Additional keywords
    escapechar=|
}

\definecolor{ghgreen}{rgb}{0.90,1,0.93}
\definecolor{ghred}{rgb}{1,0.88,0.94}

\definecolor{codegreen}{rgb}{0,0.6,0}
\definecolor{codegray}{rgb}{0.5,0.5,0.5}
\definecolor{codepurple}{rgb}{0.58,0,0.82}
\definecolor{backcolour}{rgb}{0.95,0.95,0.92}

\lstset{  
    language=Java,  
    %backgroundcolor=\color{backcolour},
    commentstyle=\color{codegreen},
    keywordstyle=\color{codepurple},
    numberstyle=\tiny\color{codegray},
    stringstyle=\color{blue},
    basicstyle=\footnotesize\ttfamily,
    breakatwhitespace=false,
    breaklines=true,
    captionpos=b,
    keepspaces=true,
    numbers=left,
    numbersep=5pt,
    tabsize=4,
    columns=fullflexible
}

\definecolor{codegreen}{rgb}{0,0.6,0}
\definecolor{codegray}{rgb}{0.5,0.5,0.5}
\definecolor{codepurple}{rgb}{0.58,0,0.82}
\definecolor{backcolour}{rgb}{0.95,0.95,0.92}

\newcommand{\circledsup}[1]{%
  \tikz[baseline=(char.base)]{%
    \node[shape=circle,draw,inner sep=0.5pt] (char) {#1};}%
}

\newcolumntype{H}{>{\setbox0=\hbox\bgroup}c<{\egroup}@{}}

\newcommand{\tool}{NBTest\xspace}
\newcommand{\toolgen}{NBTestGen\xspace}
\newcommand{\framework}{NBTest\xspace}
\newcommand{\toolpylib}{\tool{}-lib\xspace}
\newcommand{\tooljupyterplugin}{\tool{}-lab-extension\xspace}

\newcommand{\assertgenplugin}{\MakeLowercase{\tool{}}-gen}

\newcommand{\onlyifflag}[2]{\ifcsname#1\endcsname#2\fi}
\newcommand{\hideifflag}[2]{\ifcsname#1\endcsname\else#2\fi}
\newcommand{\ifflagelse}[3]{\ifcsname#1\endcsname#2\else#3\fi}

% revision
%\newcommand{\added}[1]{\textcolor{blue}{#1}}
\newcommand{\added}[1]{#1}
\newcommand{\deleted}[1]{\textcolor{red}{}}

\newcommand{\inlinecode}[1]{\lstinline[basicstyle=\ttfamily\color{blue}]|#1|}
\renewcommand{\inlinecode}[1]{\lstinline[basicstyle=\ttfamily\color{blue}]|#1|}

\newcommand{\tablesize}{\footnotesize}
%components
\newcommand{\propfinder}{PropertyFinder\xspace}
\newcommand{\nbrunner}{NotebookRunner\xspace}
\newcommand{\entity}{property\xspace}
\newcommand{\entities}{properties\xspace}
\newcommand{\assertgen}{Assertion Generator\xspace}

\newcommand{\cellscoped}{cell-scoped\xspace}
\newcommand{\Cellscoped}{Cell-scoped\xspace}

\newcommand{\elaine}[1]{\textcolor{blue}{#1}}
\newcommand{\todelete}[1]{\textcolor{orange}{#1}}

\newcommand{\userstudy}{17\xspace}

\newcommand{\nbval}{nbval\xspace}
\newcommand{\NBval}{Nbval\xspace}

% assertion types

\newcommand{\datasetAssert}{Dataset\xspace}

\newcommand{\modelArchAssert}{Model Arch.\xspace}
\newcommand{\modelPerfAssert}{Model Perf.\xspace}

% mutations
\newcommand{\outliers}{Add Outliers\xspace}
\newcommand{\repetition}{Repeat Data\xspace}
\newcommand{\addednull}{Add Nulls\xspace}
\newcommand{\labelerrors}{Modify Labels\xspace}
\newcommand{\datashift}{Data Shift\xspace}

\newcommand{\preprocdataleakage}{Add Data Leakage\xspace}
\newcommand{\removetorcheval}{Remove Torch Eval\xspace}
\newcommand{\removezerograd}{Remove Zero Grad\xspace}
\newcommand{\modifyhyperparams}{Modify Hyper-parameters\xspace}
\newcommand{\removehyperparams}{Remove Hyper-parameters\xspace}
\newcommand{\metricswap}{Swap APIs\xspace}
\newcommand{\layerinsert}{Add Layers\xspace}
\newcommand{\layerremove}{Remove Layers\xspace}
\newcommand{\replacelayer}{Replace Layer\xspace}
\newcommand{\zerolayer}{Zero Layer\xspace}

% --- Define Jupyter Input Code Block Style ---
\newtcblisting{jupytercode}[1]{%
    % #1 will be the cell number (e.g., 1, 2, ...)
    listing only,
    breakable, % Allows the box to break across pages if content is long
    colback=jupyterinbg, % Background color for the code
    colframe=bordercolor, % Frame color
    frame style={line width=0.5pt}, % Thickness of the frame
    arc=0mm, % Sharp corners
    boxsep=0pt, % No extra spacing between box and content
    left=3mm, right=3mm, top=2mm, bottom=2mm, % Padding inside the box
    listing options={
        language=Python, % Set the language for syntax highlighting
        basicstyle=\ttfamily\small, % Font style for code
        breaklines=true, % Allow long lines to break
        showstringspaces=false, % Don't show spaces as visible characters
        tabsize=4, % Tab width
        numbers=none
    },
    % Title for the code cell (e.g., In[1]:)
    title={\color{inpromptcolor}\textbf{In[#1]:}},
    fonttitle=\bfseries\sffamily\small,
    before skip=0.7\baselineskip,
    after skip=0.7\baselineskip,
}

% --- Define Jupyter Output Block Style ---
\newtcolorbox{jupyteroutput}[1]{%
    % #1 will be the cell number (e.g., 1, 2, ...)
    breakable, % Allows the box to break across pages
    colback=jupyteroutbg, % Background color for the output
    colframe=bordercolor, % Frame color
    frame style={line width=0.5pt},
    arc=0mm,
    boxsep=0pt,
    left=3mm, right=3mm, top=2mm, bottom=2mm,
    % Title for the output cell (e.g., Out[1]:)
    title={\color{outpromptcolor}\textbf{Out[#1]:}},
    fonttitle=\bfseries\sffamily\small,
    before upper={\ttfamily\small},
    before skip=0.3\baselineskip, % Slightly less space above output than code
    after skip=0.7\baselineskip,
}

\newcommand{\totalKaggleNotebooks}{585\xspace}
\newcommand{\totalKaggleNotebookAssertions}{535\xspace}
\newcommand{\avgtotalKaggleAssertions}{36.82\xspace}
\newcommand{\datasetAssertionsKaggle}{18239\xspace}
\newcommand{\modelPerfAssertionsKaggle}{2442\xspace}
\newcommand{\modelArchAssertionsKaggle}{861\xspace}
\newcommand{\avgdatasetAssertionsKaggle}{31.18\xspace}
\newcommand{\avgmodelPerfAssertionsKaggle}{4.17\xspace}
\newcommand{\avgmodelArchAssertionsKaggle}{1.47\xspace}
\newcommand{\ksklearnExe}{529\xspace}
\newcommand{\ktensorflowkerasExe}{199\xspace}
\newcommand{\ktorchExe}{30\xspace}
\newcommand{\numKaggleVersionNoExeErrorRev}{531\xspace}
\newcommand{\numKaggleNotebooksVersionNoExeErrorRev}{90\xspace}
\newcommand{\allversions}{2172\xspace}
\newcommand{\totalKaggleNotebooksWithVersionRev}{90\xspace}
\newcommand{\avgInjectedAssertionsKaggleRev}{39.12\xspace}
\newcommand{\totalRatioInjectedAssertionsKaggleRev}{96.42\%\xspace}
\newcommand{\totalVersionKilledKaggleRev}{370\xspace}
\newcommand{\totalPercVersionKilledKaggleRev}{69.68\%\xspace}
\newcommand{\numFNRateVersionRev}{0.30\xspace}
\newcommand{\numFNVersionRev}{161\xspace}
\newcommand{\numVersionKilledDatasetKaggleRev}{277\xspace}
\newcommand{\percVersionKilledDatasetKaggleRev}{52.17\%\xspace}
\newcommand{\numVersionKilledModelArchKaggleRev}{116\xspace}
\newcommand{\percVersionKilledModelArchKaggleRev}{21.85\%\xspace}
\newcommand{\numVersionKilledModelPerfKaggleRev}{119\xspace}
\newcommand{\percVersionKilledModelPerfKaggleRev}{22.41\%\xspace}
\newcommand{\numFPRateVersionRev}{0.08\xspace}
\newcommand{\totalNbvalAssertionsRev}{14332\xspace}
\newcommand{\avgtotalNbvalAssertionsRev}{24.50\xspace}
\newcommand{\totalNbvalPASSRateRev}{86.07\%\xspace}
\newcommand{\mutantScoreNbvalRev}{0.96\xspace}
\newcommand{\assertNbtestOverNbval}{7210\xspace}
\newcommand{\assertNbtestOverNbvalPerc}{50.31\%\xspace}
\newcommand{\avgAssertNbtestOverNbval}{12.33\xspace}
\newcommand{\avgAssertNbtestOverNbvalPerc}{50.31\%\xspace}
\newcommand{\passrateNbtestOverNbval}{13.93\%\xspace}
\newcommand{\RatioNbvalPassZero}{11.55\%\xspace}
\newcommand{\totalNbvalPASSZeroRev}{1655\xspace}
\newcommand{\totalNbvalPASSZeroFiftyRev}{400\xspace}
\newcommand{\totalNbvalPASSFiftyHundredRev}{113\xspace}
\newcommand{\totalNbvalPASSHundredRev}{12164\xspace}
\newcommand{\avgRuntimeAssertionsKaggle}{1.57\xspace}
\newcommand{\ktotal}{1753\xspace}
\newcommand{\kml}{1612\xspace}
\newcommand{\ktrain}{1520\xspace}
\newcommand{\kinference}{1483\xspace}
\newcommand{\ktorch}{97\xspace}
\newcommand{\ktensorflowkeras}{383\xspace}
\newcommand{\ksklearn}{1489\xspace}
\newcommand{\kother}{34\xspace}
\newcommand{\nbvalRepo}{356\xspace}
\newcommand{\deepchecksRepo}{238\xspace}
\newcommand{\deepquRepo}{201\xspace}
\newcommand{\testbookRepo}{181\xspace}
\newcommand{\dynamicRun}{30\xspace}
\newcommand{\confLevel}{0.99\xspace}
\newcommand{\pytestIte}{30\xspace}
\newcommand{\totalKagglePASSRateZeroFiveRev}{96.93\%\xspace}
\newcommand{\mutantScoreKaggleZeroFiveRev}{0.82\xspace}
\newcommand{\FNZeroFiveRev}{0.18\xspace}
\newcommand{\FPZeroFiveRev}{0.1342\xspace}
\newcommand{\totalKagglePASSRateZeroSevenRev}{98.58\%\xspace}
\newcommand{\mutantScoreKaggleZeroSevenRev}{0.78\xspace}
\newcommand{\FNZeroSevenRev}{0.22\xspace}
\newcommand{\FPZeroSevenRev}{0.0712\xspace}
\newcommand{\totalKagglePASSRate}{100.00\%\xspace}
\newcommand{\totalKaggleFAILFiftyOne}{0\xspace}
\newcommand{\totalKaggleFAILFifty}{16\xspace}
\newcommand{\totalKaggleFAILFiftyInspectPass}{10\xspace}
\newcommand{\totalKaggleFAILFiftyInspectNoPass}{6\xspace}
\newcommand{\totalKaggleFAILZero}{21526\xspace}
\newcommand{\totalKaggleFAILOne}{0\xspace}
\newcommand{\totalKaggleAssertions}{21542\xspace}
\newcommand{\failFiftyNBCnt}{8\xspace}
\newcommand{\mutantScoreKaggleZeroNineNineRev}{0.72\xspace}
\newcommand{\mutantScoreKaggleZeroNineNinePercRev}{72.21\%\xspace}
\newcommand{\FNZeroNineNineRev}{0.28\xspace}
\newcommand{\FPZeroNineNineRev}{0.0052\xspace}
\newcommand{\totalKagglePASSRateZeroNineNineNineRev}{100.00\%\xspace}
\newcommand{\mutantScoreKaggleZeroNineNineNineRev}{0.71\xspace}
\newcommand{\FNZeroNineNineNineRev}{0.29\xspace}
\newcommand{\FPZeroNineNineNineRev}{0.0042\xspace}
\newcommand{\outliersModelArchKaggle}{0.00\xspace}
\newcommand{\outliersKilledByModelArch}{0.00\%\xspace}
\newcommand{\outliersDatasetKaggle}{652.47\xspace}
\newcommand{\outliersKilledByDataset}{87.79\%\xspace}
\newcommand{\outliersModelPerfKaggle}{80.67\xspace}
\newcommand{\outliersKilledByModelPerf}{10.85\%\xspace}
\newcommand{\numMutantsOutliersKaggle}{743\xspace}
\newcommand{\numMutantsKilledOutliersKaggle}{655.47\xspace}
\newcommand{\mutationScoreOutliersKaggle}{0.88\xspace}
\newcommand{\repetitionModelArchKaggle}{0.00\xspace}
\newcommand{\repetitionKilledByModelArch}{0.00\%\xspace}
\newcommand{\repetitionDatasetKaggle}{761.03\xspace}
\newcommand{\repetitionKilledByDataset}{81.48\%\xspace}
\newcommand{\repetitionModelPerfKaggle}{39.67\xspace}
\newcommand{\repetitionKilledByModelPerf}{4.25\%\xspace}
\newcommand{\numMutantsRepetitionKaggle}{934\xspace}
\newcommand{\numMutantsKilledRepetitionKaggle}{764.17\xspace}
\newcommand{\mutationScoreRepetitionKaggle}{0.82\xspace}
\newcommand{\addedNullModelArchKaggle}{0.00\xspace}
\newcommand{\addedNullKilledByModelArch}{0.00\%\xspace}
\newcommand{\addedNullDatasetKaggle}{708.37\xspace}
\newcommand{\addedNullKilledByDataset}{86.70\%\xspace}
\newcommand{\addedNullModelPerfKaggle}{38.80\xspace}
\newcommand{\addedNullKilledByModelPerf}{4.75\%\xspace}
\newcommand{\numMutantsAddedNullKaggle}{817\xspace}
\newcommand{\numMutantsKilledAddedNullKaggle}{709.37\xspace}
\newcommand{\mutationScoreAddedNullKaggle}{0.87\xspace}
\newcommand{\labelErrorsModelArchKaggle}{0.00\xspace}
\newcommand{\labelErrorsKilledByModelArch}{0.00\%\xspace}
\newcommand{\labelErrorsDatasetKaggle}{134.80\xspace}
\newcommand{\labelErrorsKilledByDataset}{65.12\%\xspace}
\newcommand{\labelErrorsModelPerfKaggle}{104.17\xspace}
\newcommand{\labelErrorsKilledByModelPerf}{50.32\%\xspace}
\newcommand{\numMutantsLabelErrorsKaggle}{207\xspace}
\newcommand{\numMutantsKilledLabelErrorsKaggle}{181.87\xspace}
\newcommand{\mutationScoreLabelErrorsKaggle}{0.88\xspace}
\newcommand{\removeTorchZeroGradModelArchKaggle}{0.00\xspace}
\newcommand{\removeTorchZeroGradKilledByModelArch}{0.00\%\xspace}
\newcommand{\removeTorchZeroGradDatasetKaggle}{0.00\xspace}
\newcommand{\removeTorchZeroGradKilledByDataset}{0.00\%\xspace}
\newcommand{\removeTorchZeroGradModelPerfKaggle}{8.83\xspace}
\newcommand{\removeTorchZeroGradKilledByModelPerf}{46.49\%\xspace}
\newcommand{\numMutantsRemoveTorchZeroGradKaggle}{19\xspace}
\newcommand{\numMutantsKilledRemoveTorchZeroGradKaggle}{7.90\xspace}
\newcommand{\mutationScoreRemoveTorchZeroGradKaggle}{0.42\xspace}
\newcommand{\modifyHyperparametersModelArchKaggle}{563.97\xspace}
\newcommand{\modifyHyperparametersKilledByModelArch}{46.23\%\xspace}
\newcommand{\modifyHyperparametersDatasetKaggle}{0.00\xspace}
\newcommand{\modifyHyperparametersKilledByDataset}{0.00\%\xspace}
\newcommand{\modifyHyperparametersModelPerfKaggle}{533.57\xspace}
\newcommand{\modifyHyperparametersKilledByModelPerf}{43.73\%\xspace}
\newcommand{\numMutantsModifyHyperparametersKaggle}{1220\xspace}
\newcommand{\numMutantsKilledModifyHyperparametersKaggle}{742.40\xspace}
\newcommand{\mutationScoreModifyHyperparametersKaggle}{0.61\xspace}
\newcommand{\removeHyperparametersModelArchKaggle}{377.23\xspace}
\newcommand{\removeHyperparametersKilledByModelArch}{44.63\%\xspace}
\newcommand{\removeHyperparametersDatasetKaggle}{0.00\xspace}
\newcommand{\removeHyperparametersKilledByDataset}{0.00\%\xspace}
\newcommand{\removeHyperparametersModelPerfKaggle}{263.50\xspace}
\newcommand{\removeHyperparametersKilledByModelPerf}{31.17\%\xspace}
\newcommand{\numMutantsRemoveHyperparametersKaggle}{845\xspace}
\newcommand{\numMutantsKilledRemoveHyperparametersKaggle}{443.17\xspace}
\newcommand{\mutationScoreRemoveHyperparametersKaggle}{0.52\xspace}
\newcommand{\deepLayerRemovalModelArchKaggle}{32.00\xspace}
\newcommand{\deepLayerRemovalKilledByModelArch}{24.24\%\xspace}
\newcommand{\deepLayerRemovalDatasetKaggle}{0.00\xspace}
\newcommand{\deepLayerRemovalKilledByDataset}{0.00\%\xspace}
\newcommand{\deepLayerRemovalModelPerfKaggle}{16.67\xspace}
\newcommand{\deepLayerRemovalKilledByModelPerf}{12.63\%\xspace}
\newcommand{\numMutantsDeepLayerRemovalKaggle}{132\xspace}
\newcommand{\numMutantsKilledDeepLayerRemovalKaggle}{46.33\xspace}
\newcommand{\mutationScoreDeepLayerRemovalKaggle}{0.35\xspace}
\newcommand{\totalKilledModelArchKaggle}{973.20\xspace}
\newcommand{\totalKilledKilledByModelArch}{19.79\%\xspace}
\newcommand{\totalKilledDatasetKaggle}{2256.67\xspace}
\newcommand{\totalKilledKilledByDataset}{45.89\%\xspace}
\newcommand{\totalKilledModelPerfKaggle}{1085.87\xspace}
\newcommand{\totalKilledKilledByModelPerf}{22.08\%\xspace}
\newcommand{\totalDatasetDataMut}{2256.67\xspace}
\newcommand{\DatasetDataMutScore}{0.84\xspace}
\newcommand{\totalPerfDataMut}{263.30\xspace}
\newcommand{\PerfDataMutScore}{0.10\xspace}
\newcommand{\totalArchDataMut}{0.00\xspace}
\newcommand{\ArchDataMutScore}{0.00\xspace}
\newcommand{\totalDataMut}{2701\xspace}
\newcommand{\totalDataMutKilled}{2310.87\xspace}
\newcommand{\DataMutScore}{0.86\xspace}
\newcommand{\totalDatasetCodeMut}{0.00\xspace}
\newcommand{\DatasetCodeMutScore}{0.00\xspace}
\newcommand{\totalPerfCodeMut}{822.57\xspace}
\newcommand{\PerfCodeMutScore}{0.37\xspace}
\newcommand{\totalArchCodeMut}{973.20\xspace}
\newcommand{\ArchCodeMutScore}{0.44\xspace}
\newcommand{\totalCodeMut}{2216\xspace}
\newcommand{\totalCodeMutKilled}{1239.80\xspace}
\newcommand{\CodeMutScore}{0.56\xspace}
\newcommand{\totalMutantsKaggle}{4917\xspace}
\newcommand{\totalMutantsKilledKaggle}{3550.67\xspace}

%\title{\tool: An Automated Assertion Generation Framework for ML Notebooks}
%\title{Towards Testable ML Notebooks}
%\title{Making ML Notebooks Testable with \tool}
%\title{Regression Testing of Machine Learning Notebooks}
\title{Automated Assertion Generation and Regression Testing for Machine Learning Notebooks}
%\title{A Regression Testing Framework for ML Notebooks with Automated Assertion Generation}
%\title{A Regression Testing Framework and Automated Assertion Generation Method for ML Notebooks}

\author{Yingao (Elaine) Yao}
\affiliation{%
  \institution{Cornell University}
  \city{Ithaca}
  \state{NY}
  \country{USA}}
\email{yy2282@cornell.edu}

\author{Vedant Nimje}
\affiliation{%
  \institution{Veermata Jijabai Technological Institute}
  \city{Mumbai}
  \country{India}}
\email{vedantnimjed@gmail.com}

\author{Varun Viswanath}
\affiliation{%
  \institution{Dwarkadas J Sanghvi College of Engineering}
  \city{Mumbai}
  \country{India}}
\email{varunvis2903@gmail.com}

\author{Saikat Dutta}
\affiliation{%
  \institution{Cornell University}
  \city{Ithaca}
  \state{NY}
  \country{USA}}
\email{saikatd@cornell.edu}

% \author{
%   \IEEEauthorblockN{Yingao (Elaine) Yao}
% \IEEEauthorblockA{
% % \textit{dept. name of organization (of Aff.)} \\
% \textit{Cornell University}\\
% Ithaca, NY, USA \\
% yy2282@cornell.edu}
% \and
%   \IEEEauthorblockN{Vedant Nimje}
% \IEEEauthorblockA{
% % \textit{dept. name of organization (of Aff.)} \\
% \textit{Veermata Jijabai Technological Institute}\\
% Mumbai, India \\
% vedantnimjed@gmail.com}
% \and
% \IEEEauthorblockN{Varun Viswanath}
% \IEEEauthorblockA{
% % \textit{dept. name of organization (of Aff.)} \\
% \textit{Dwarkadas J Sanghvi College of Engineering}\\
% Mumbai, India \\
% varunvis2903@gmail.com}
% \linebreakand
% \IEEEauthorblockN{Saikat Dutta}
% \IEEEauthorblockA{
% % \textit{dept. name of organization (of Aff.)} \\
% \textit{Cornell University}\\
% Ithaca, NY, USA \\
% saikatd@cornell.edu}
% }


\begin{abstract}
Jupyter Notebooks have become the de-facto choice for data scientists and machine
learning (ML) engineers for prototyping and experimenting with ML
pipelines. Notebooks provide a rich interactive interface with support for code,
data, and visualization in one place.
% While notebooks provide rich features,they provide very limited support for testing. 
%As a result, during continuous development, many subtle bugs that do not lead to crashes often go
%unnoticed and cause silent errors that manifest as performance regressions.
However, notebooks provide limited support for testing. As a result, during
continuous development, many silent (non-crashing) regressions often go
unnoticed and make notebooks unreliable and results hard to reproduce.

% In this work, we first perform an extensive mutation analysis on a large corpus of \Fix{XX} ML notebooks and study how such bugs manifest. Our analysis reveals that \Fix{...}.
% \deleted{To enable more systematic testing of ML notebooks, we
% introduce \textbf{\tool} -- the first regression testing framework that allows
% developers to write \emph{cell-level} assertions in notebooks and also enables
% testing of such notebooks with pytest as well as in continuous integration
% pipelines.} 
%
To enable more systematic testing of ML notebooks, we introduce
\textbf{\toolgen} -- the \emph{first} automated assertion generation approach
for ML notebooks. \toolgen generates regression-based assertions that check
properties related to data processing, model building, and model evaluation
steps in a typical ML notebook. 
%Such assertions can potentially detect silent regressions and maintain
%notebook reliability without increasing developer burden. 
%
To support systematic integration of such assertions in notebooks, we introduce the
\emph{first} regression testing framework (called \textbf{\tool}) that can be
used as a Jupyter plugin and allows developers to write \emph{\cellscoped}
assertions in notebooks. Three key features of such assertions are that they are
1) \cellscoped: they are linked to specific
notebook cells and execute only after those cells are executed,
2) non-intrusive: they do not block notebook execution (in a Jupyter session),
so that development can continue when they fail, and 3) they
integrate with \texttt{pytest} and CI pipelines, allowing developers to easily
do regression testing of their notebooks.

\Comment{ To enable more systematic testing of ML notebooks,
we introduce \textbf{\tool} -- the \emph{first} regression testing framework
that allows developers to write cell-scoped assertions in Jupyter notebooks. Two
key features of such assertions are that they are 1) non-intrusive: they do not
block notebook execution (in a Jupyter session), so that development can
continue even if they fail, and 2) they integrate with \texttt{pytest} and CI
pipelines, allowing developers to easily do regression testing of their notebooks.
%intermediate states (e.g.,
%integrity checks after data processing) in
their notebooks.
%
%\tool comes with a library of assertion APIs, as well
%as a Jupyter Lab plugin that enables executing assertions within a session.
%
We also develop the \emph{first} automated approach for generating
regression-based assertions that target properties related to data processing,
model building, and model evaluation steps in a typical ML pipeline.
% We also develop the first automated approach for generating regression-based
% assertions for testing different components of ML notebooks, such as
% data processing, model building, and model evaluation. 
%With the \tool framework,
%we aim to improve the reliability and maintainability of ML notebooks without
%increasing developer burden.
%Importantly, \tool makes ML notebooks more reliable and maintainable without
%increasing developer burden.
}

% make ML notebooks more \emph{testable}, we introduce \textbf{\tool}, the first
% framework that allows developers to easily author ML-specific assertions in
% notebooks and also integrate such notebooks (with assertions) with CI. In
% \tool, we also develop the first automated approach for generating
% domain-specific assertions for different components of ML pipelines such as
% data processing, model building, and training. Importantly, \tool makes ML
% notebooks more reliable and maintainable without increasing developer burden.

% In this work, we propose \tool, an approach to generate high-quality
% assertions for jupyter notebooks. \tool generates assertions for different
% stages in a typical data science or ML pipeline -- including data processing,
% model building, and training. Such assertions can serve as regression tests
% for the notebook and allow future users to easily check if they can reproduce
% consistent results as before. 

% Prior studies show that most notebook users find it extremely challenging to install the correct dependencies and have no means to check if they reproduced the same results even if they can run the notebook. Hence, there is an urgent need to develop robust quality assurance tools for jupyter notebooks.

We evaluate \toolgen on a corpus of \totalKaggleNotebooks notebooks from the
popular Kaggle platform. \toolgen generates a total of \totalKaggleAssertions
assertions (\avgtotalKaggleAssertions on average per notebook). The generated
assertions kill \mutantScoreKaggleZeroNineNinePercRev of ML-specific mutations, while
maintaining a high passrate of \totalKagglePASSRate.
% and a low false
%positive rate of \FPZeroNineNineRev\%.
%  obtain a mutation score of \mutantScoreKaggleZeroNineNineRev
% in killing ML-specific mutations while maintaining a high passrate of
% \Fix{\totalKagglePASSRate} and a low false positive rate of \FPZeroNineNineRev.
%
We also show that \toolgen can detect \totalPercVersionKilledKaggleRev of
historical regressions in \numKaggleVersionNoExeErrorRev older versions of Kaggle notebooks.
% We also show that \toolgen could have caught regression bugs in
% \totalPercVersionKilledKaggleRev of buggy versions of the Kaggle
% notebooks using assertions generated for the latest versions. 
%\added{Compared with existing tools like nbval, we show that \toolgen generates higher quality assertions.} 
% \deleted{Because ML pipelines involve non-deterministic computations, the
% assertions can be flaky. Hence, we also show how \toolgen leverages statistical
% methods to minimize flakiness while retaining high
% fault-detection effectiveness.}
A popular ML library, SHAP, integrated \tool into their CI.
Further, we perform a user study with \userstudy ML developers that
shows that such users find \tool highly intuitive 
% (rated $4.3/5$) 
and useful.
% (rated $4.24/5$). 

\end{abstract}

  


\maketitle

\section{Introduction}
\label{sec:intro}
%% notebooks for ml general intro, what are notebooks
Computational Notebooks have become the most popular medium for data scientists
and machine learning engineers to author machine learning
programs~\cite{Perkel2018,siddik2025systematic}. Notebooks are popular because
they are an interactive medium for users to write code and visualize results,
such as training results, plots, and images -- all in the same file. Notebooks are
sequences of \emph{cells}. Each cell contains a code snippet or a system 
command (e.g., to install a dependency) that can be executed independently of
other cells and in arbitrary order. 
Such flexibility makes experimentation and exploration easier, especially for ML code.
% Due to such flexibility, developers prefer
% using notebooks to write and share experimental code, especially when designing
% machine learning pipelines. 
Some popular platforms that support notebooks are
Jupyter~\cite{kluyver2016jupyter}, Kaggle~\cite{kaggle},
Pluto~\cite{pluto_notebooks}, and Google Colab~\cite{googlecolab2024}.

%% talk about problems, no testing, messy, not reproducible, hard to debug
\looseness=-1 Despite their popularity, notebooks are notorious for their poor
reliability and reproducibility~\cite{chattopadhyay2020s,huang2025scientists}.
Due to their out-of-order execution model and absence of tests, users often end
up with buggy notebooks that become hard to reuse and reproduce results
with~\cite{pimentel2019large_scale,head2019managing}. A recent study analyzed
1.4 million notebooks and reported that fewer than 25\% of notebooks could be
executed without execution errors, while only 4\% reproduced the expected
results~\cite{pimentel2019large_scale}. Hence, a \emph{working} notebook may
become unusable due to regressions caused by internal changes (such as code
changes) or external changes (e.g., changes in dependencies or datasets) across
versions. Further, regressions might manifest \emph{within a session}, leading
to silent (non-crashing) bugs (e.g., due to out-of-order execution or stale
states~\cite{head2019managing}).
%without the user noticing them.
%Further, because there are no well-established testing practices, the notebook may suffer from silent regressions over time without the user noticing them.
%the user may not realize if/when the notebook suffers from silent regressions after subsequent changes.
Hence, like regular programs, notebooks also need to follow regression testing practices.
%, akin to tools like \texttt{Pytest}~\cite{pytest} and JUnit~\cite{junit5}.
%provide little to no tools for quality assurance, which limits their usability and hurts their users' productivity. Developers find \cite{chattopadhyay2020s}
%
Currently, for testing and debugging, notebook users follow ad hoc practices,
such as using print statements, writing test cases within the notebook, using a
separate test notebook for validation, or manually inspecting
results~\cite{chattopadhyay2020s,huang2025scientists}. 

\Comment{ Indeed, Netflix recognised this critical need for quality assurance in
notebooks and developed sophisticated internal tools~\cite{netflix2019nbblog} to
enforce reproducibility, manage dependencies, and enable the scheduled,
production-ready execution of notebooks, demonstrating a clear industry push
towards treating notebooks not just as exploratory tools but as deployable
artifacts requiring rigorous validation throughout their life cycle. This
growing demand for robust notebook development is also evident in the increasing
adoption of nbval. This presents a widespread developer desire to enhance the
reliability and maintainability of their notebook-based workflows.}

\Comment{
\looseness=-1 
The reliability problem is especially critical in industry settings. For
example, Netflix uses Jupyter Notebooks in production for data
exploration\Comment{, collaboration,} and workflow  
scheduling~\cite{netflix2019nbblog}. PayPal has developed an enterprise-grade
Jupyter platform -- PayPal Notebooks -- which empowers over 1,300 internal users
to perform data analysis, machine learning, and workflow automation\Comment{ across
diverse data sources}~\cite{paypal2018nbblog}. These examples highlight the
growing need for rigorous quality assurance for
notebooks.%to prevent regressions from impacting production workflows.
%This highlights the importance of quality assurance for notebooks.
}

\Comment{
\mypara{Existing testing tools and their limitations}
Developers have built tools such as \emph{nbval}~\cite{fangohr2020testing},
which is a regression testing tool for Jupyter notebooks. It performs a textual
comparison of a notebook's current output cells against saved values from prior
sessions.  However, its checks are not robust to non-deterministic results
(common in ML pipelines) or fluctuations in the output string format (especially
when considering numerical outputs). Further, it can only check outputs printed
in cells. Deepchecks~\cite{chorev2022deepchecks} is a tool that provides a
library of tests for checking model and data integrity properties in ML
pipelines. However, the developers still need to manually incorporate such
checks, and the framework does not allow them to easily write custom tests.
Deequ~\cite{schelter2018automating} is another tool that is designed for data
validation in production pipelines. Testbook~\cite{testbook} allows developer to
write unit tests in a separate file and potentially execute a subset of cells.
However, none of these tools can 1) automatically generate tests, 2) account for
non-determinism, or 3) allow non-intrusive testing within a Jupyter session. }
% \Fix{recheck if this comparison is accurate}

%\Fix{talk about other tools, deepchecks~\cite{chorev2022deepchecks} , nbval~\cite{fangohr2020testing}, deequ~\cite{schelter2018automating}, fastdup (images)}

\Comment{\NA{Several frameworks exist that are used to validate ML models and
the dataset that is used to train and evaluate them.
\emph{Deepchecks}~\cite{chorev2022deepchecks} validates machine learning models
and data by adding checks related to various issues, such as model predictive
performance, data integrity, data distribution mismatches, and more.
\emph{deequ}~\cite{schelter2018automating} is used for automating the
verification of data quality at scale, which meets the requirements of
production use cases and provides a declarative API, which when combined with
common quality constraints with user-defined validation code, and thereby
enables 'unit tests' for data. For Jupyter notebooks,
\emph{nbval}~\cite{fangohr2020testing} was introduced, that allows developers to
test the notebook's current execution against it's output cells. It integrates
with \texttt{pytest} and thus, enables integration with CI pipelines. But the
very nature of comparing output cells means it is not robust against silent bugs
in the pipeline due to non-determinism, and in a way relies on developers to
specify the property they want to test, i.e., the entities/objects/variables
they want to display in the output.} }

\begin{figure*}[h]
  \footnotesize
  \begin{subfigure}[t]{.79\linewidth}
    \centering
    \includegraphics[width=\textwidth]{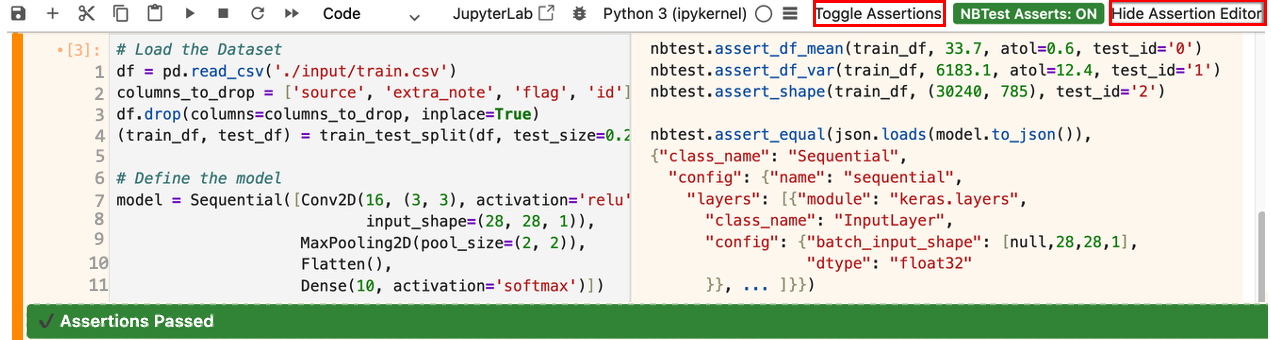}
    \caption{\tool JupyterLab plugin (\tooljupyterplugin)} 
    \label{fig:overview_a}
  \end{subfigure}
  \begin{subfigure}[t]{.2\linewidth}
    \centering
    \includegraphics[width=\textwidth]{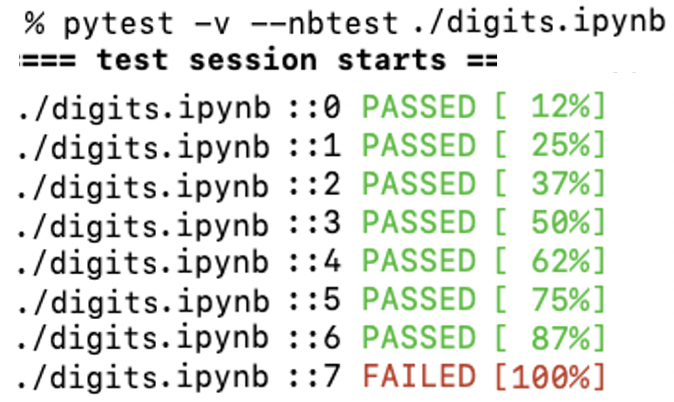}
    \caption{Executing assertions with pytest}
    \label{fig:overview_b}
  \end{subfigure}%
  % \begin{subfigure}{.33\textwidth}
  %   \centering
  %   \includegraphics[width=\linewidth]{figures/assertion_fail.png}
  %   \caption{Failed assertions in JupyterLab}
  %   \label{fig:sfig3}
  % \end{subfigure}
  \vspace{-0.1in}
  \caption{Example usage of \tool.  
  (a) shows \toolgen assertions in a JupyterLab side panel, which can be hidden by \emph{Hide Assertion Editor} button,
   or enabled/disabled with the \emph{Toggle Assertions} button.
  The green message bar (\emph{Assertions Passed}) indicates assertions passed.
  (b) shows pytest output for \tool assertions.
  \vspace{-0.2in}
  }
  \label{fig:nbtest-workflow}
\end{figure*}

% our work: idea: make notebooks testable via cell-level assertions. used regression-based oracles
% integrate with pytest and thereby with CI
% support during development, without breaking developer experience
%\looseness=-1 
\subsection{Our Work}
\mypara{Automated Assertion Generation} 
% A recent study reported that notebook users often follow \emph{cell-based risk
% management}, where users test changes in a new cell before merging with old
% cells to isolate any errors~\cite{huang2025scientists}. Motivated by this
% finding, we 
% Hence,
% \cellscoped assertions can potentially  
% improve the reliability of notebooks by mitigating regressions and nudging
% developers towards better testing practices in notebooks. 
To address the reliability problem in ML notebooks and to assist developers in
writing assertions, we propose an automated technique, \textbf{\toolgen}, to
generate assertions for machine learning (ML) notebooks. A recent study showed that
developers of ML notebooks often implement various integrity
checks for data and ML model-related properties, such as data
shape, model performance, and network architecture, using assertions and print
statements~\cite{shome2024understanding}. Motivated by this finding, we design
an assertion generation technique that first identifies model/data-related APIs
of popular ML/Data libraries (e.g., PyTorch~\cite{paszke2019pytorch},
TensorFlow~\cite{tensorflow2015-whitepaper}, Scikit-Learn~\cite{scikit-learn},
and Pandas~\cite{reback2020pandas}) used in a notebook\Comment{that are related
to machine learning or data science libraries (such as pytorch, sklearn, and
pandas)}, instruments the API calls, executes the notebook to collect expected
results, and generates appropriate assertions in the notebook. A key challenge of testing
ML notebooks is that their execution can often be
non-deterministic due to randomness during training or batching data
% or due to the use of GPUs 
-- causing some assertions to be potentially \emph{flaky}
\cite{dutta2020detecting}. To mitigate this challenge, we execute the notebooks
multiple times to collect execution information and
estimate the variance in the results with statistical
tests. We use the collected data to generate approximate assertions (e.g.,
\texttt{assert\_allclose}) to account for this variance via tolerance bounds
where needed~\cite{dutta2021flex}.

\looseness=-1 
\mypara{Regression Testing Framework} To systematically integrate such
assertions in a notebook environment, we propose a novel regression testing
framework for notebooks: \textbf{\tool}. The design of \tool is guided by a
recent study that found that notebook users often follow \emph{cell-based risk
management}, where users test changes in a new cell before merging with old
cells to isolate any errors~\cite{huang2025scientists}. With \tool, we introduce
the idea that developers can write \emph{\cellscoped} assertions to check the
expected behavior of intermediate computations in a notebook, specifically
within each cell. \Cellscoped assertions are linked to a specific cell and are
executed only after the cell completes execution. For instance, a developer may
want to check that the dataset was pre-processed correctly (e.g., no null
values), if the model training was set up correctly (e.g., model architectures
are defined as expected), or if the training accuracy was close to expectations.
Hence, \cellscoped assertions can potentially  
improve the reliability of notebooks by mitigating regressions and nudging
developers towards better testing practices in notebooks.  
The assertions in \tool are designed to be \emph{non-intrusive} -- they neither
pollute the notebook state nor block notebook execution when they fail.
Non-intrusiveness allows developers to continue development without being
interrupted by assertion failures. 

% With \tool, we introduce the idea that developers can
% write \deleted{\emph{cell-level assertions}}\cellscoped assertions to
% check the expected behavior of intermediate computations in a notebook,
% specifically within each cell. \Cellscoped assertions are linked to a
% specific cell and are executed only after the cell completes execution.
%
% For instance, a developer may want to check that the dataset was pre-processed
% correctly (e.g., no null values), if the model training was set up correctly
% (e.g., model weights are initialized randomly), or if the training accuracy was
% close to expectations.
%  To support such checks, \tool provides 9
% APIs, including common ones like} \texttt{nbtest.assert\_true(exp)} to
% check if \texttt{exp} is true and \texttt{nbtest.assert\_allclose(X, Y, atol,
% rtol) to check if the values X and Y are equal up to the desired tolerance\Fix{, as
% well as specialized APIs that check data and model integrity properties}.
%
% \Fix{see if these categories and examples are consistent?}
\looseness=-1 \emph{\toolpylib} is a Python library in \tool that provides $11$
assertion APIs, covering two categories of assertions: (1) generic assertions
(e.g., \texttt{assert\_true(exp)} to check if \texttt{exp} is true and
\texttt{assert\_allclose(X, Y, atol, rtol)} to check if the values X and Y are
equal up to the desired tolerance), (2) data-related assertions (e.g.,
\texttt{assert\_df\_mean(df, col, expected\_mean)} to check if the mean of a
column in a data frame is close to the expected mean). These APIs allow
developers to easily write assertions for regression testing of data, models,
and other computations in ML notebooks. \tool integrates with \texttt{pytest}
via the \toolpylib plugin, which allows all developer-written tests to be
directly run with \texttt{pytest}. This feature enables notebooks to be
integrated with continuous integration (CI) pipelines.
% in a notebook.
% check for various properties of data,
% models, and other computations in a notebook, which can help them catch
% regressions early during development.
% The APIs cover three categories of assertions: 1) generic assertions (e
% such as common assert APIs
% like \texttt{assert\_equal} and \texttt{assert\_all\_close}, as well as
% specialized assertion APIs for checking properties of data frames and models,
% such as \texttt{assert\_df\_mean} and \texttt{assert\_shape. 

% \deleted{Figure~\ref{fig:nbtest-workflow} illustrates
% the workflow of \tool.}
\looseness=-1 
\mypara{Example} Figure~\ref{fig:nbtest-workflow} presents a simplified example
Jupyter Notebook in JupyterLab with \tooljupyterplugin's panel
(Figure~\ref{fig:overview_a}) and the output of running \tool's assertions using
\texttt{pytest} (Figure~\ref{fig:overview_b}) via \toolpylib plugin. The
notebook is training a convolutional neural network (CNN) model for a digit
recognition task (identify the digit in an image). The left panel of
Figure~\ref{fig:overview_a} shows one simplified cell of the notebook. The cell
performs the following steps: 1) loads the digit recognition dataset
(\texttt{train.csv}) into a pandas data frame called \texttt{df} (Line 1), 2)
drops some unnecessary columns (e.g., \texttt{source}) from the data frame (Lines
2-3), 3) splits the data frame into training and test sets using
\texttt{train\_test\_split} in 80\%-20\% ratio (Line 4), and 4) defines a CNN
model for training using TensorFlow APIs for neural network layers
(\texttt{Sequential}, \texttt{Conv2D}, \texttt{Dense}, etc.) (Lines 7-11). The
rest of the notebook (not shown here) performs training of the model and
computes accuracy of the model on the test set using TensorFlow APIs.  
Figure~\ref{fig:overview_a} right panel shows the \tooljupyterplugin's cell
containing the assertions automatically generated by \toolgen written using
\tool's APIs for this cell. \toolgen automatically generates assertions for
checking (1) data integrity properties (here: mean, variance, and shape of
training data frame (\texttt{train\_df})), (2) model integrity properties (here:
names and shapes of each intermediate layer in the CNN), and (3) model
performance properties (e.g., training accuracy) -- not shown here.
Figure~\ref{fig:overview_b} shows the \texttt{pytest} output of running \tool
assertions, each assertion reported as either \texttt{PASSED} or
\texttt{FAILED}. In this case, the output shows that all but the last assertion passed. Each assertion is labeled by a unique ``test\_id''.
 %The developers can reference the unique ``test\_id'' for each assertion
%to map them to pytest's output (shown as \texttt{FILENAME}::\texttt{TESTID}).

\Comment{
In addition, we developed a JupyterLab plugin (\textit{\tooljupyterplugin}) that
provides two key features: (1) toggling the visibility of assertions, and (2)
enabling or disabling the assertions execution. Within JupyterLab, \tool
assertions are displayed in a separate panel on the right
(Figure~\ref{fig:overview_a}). When developers wish to focus on notebook
development without being distracted by assertions, they can hide this panel
using the \texttt{Hide Assertion Editor} button (Figure~\ref{fig:overview_a}),
thereby avoiding clutter and improving readability. Furthermore, the plugin
allows users to control assertion execution via the \texttt{\tool Asserts: ON}
\space toggle (Figure~\ref{fig:overview_a}). When enabled, the assertions
are executed \emph{with the cell}, allowing developers to quickly validate their
local cell changes without exiting the notebook. A bottom panel will
show results of assertion execution (the green panel in
Figure~\ref{fig:overview_a} bottom indicates that all assertions passed).
% Assertions passed are indicated with green checkmarks
% (Figure~\ref{fig:nbtest-workflow}(b)).
% , while failed assertions display error
% messages in red boxes (Figure~\ref{fig:nbtest-workflow}(c)).
This way, developers can make edits without accidentally triggering the test
failure, but validate the changes by executing \emph{only} the related
assertions as needed.
}

\Comment{
\mypara{\tool vs Inline Tests} More recently, Liu et
al.~\cite{liu2022inline,liu2023extracting} introduced a new testing paradigm:
\emph{inline testing} that executes tests at a lower level of granularity
(statement level) than unit tests. Inline tests only target one statement at a
time and test general PL features (e.g., regexes, string manipulation). In
contrast, \tool's assertions check for the correctness of logic \emph{within a
cell}, they lie in between inline tests and unit tests in granularity and target
ML-related properties. Also, inline tests require developer
inputs/oracles (\tool's automated assertions do not) and are enforced at
test-time (\tool's assertions are enforced at both production-time and
test-time).
}
% Cell-level assertions represent a new \emph{testing paradigm} that
% encourages developers to perform checks within a cell, a natural computation
% unit in Jupyter notebooks.
We envision that \tool's \cellscoped assertions
(automated and manual) can enable a lot of use cases for ML notebooks, such
as documenting developer assumptions of expected behaviors, catching regressions
during interactive sessions, improving the reproducibility of notebooks,
quantifying variance in training performance or intermediate results, and
identifying contracts that can be used for monitoring of ML pipelines. 
% A key philosophy of \tool is to encourage developers to write assertions
% \emph{within each cell} that check the expected behavior of intermediate
% computations in a notebook even during development to detect silent regressions
% early. This is in contrast to writing test cases in a separate file (e.g., in
% TestBook~\cite{testbook}), which is a post-development activity and may not be
% as effective in catching regressions during development. 

% use cases for such assertions

% results: asserts generated, mutation score, etc.
\mypara{Results}
% We collect a corpus of \ktotal competition notebooks from the Kaggle platform
% based on the popularity of Kaggle competitions and the execution status, out of
% which we obtain \totalKaggleNotebooks executable notebooks. 
We collect a corpus of \totalKaggleNotebooks executable notebooks from popular
competitions in Kaggle. \toolgen generates a total of \totalKaggleAssertions
assertions (\avgtotalKaggleAssertions assertions per notebook on average),
including assertions for dataset properties (\datasetAssertionsKaggle), model
architecture properties (\modelArchAssertionsKaggle), and model performance
properties (\modelPerfAssertionsKaggle). We perform mutation analysis of the
notebooks with assertions using ML-specific mutation operations, such as adding
outliers in data and modifying layers of neural networks. Overall, we obtain a
mutation score of \mutantScoreKaggleZeroNineNinePercRev, showing that \toolgen's
generated assertions are robust to a variety of ML-specific mutations.

We also collect \numKaggleVersionNoExeErrorRev older versions of Kaggle
notebooks with potential regressions and evaluate whether our generated assertions could have caught these
real-world regressions. Our generated assertions can successfully detect regressions in 
\totalVersionKilledKaggleRev (\totalPercVersionKilledKaggleRev) buggy versions, showing the effectiveness of 
\toolgen's assertions.
% \Fix{is 2369 the final set of notebooks we can run?}

\mypara{User Study and Adoption} We conducted a user study with $17$ ML
developers. Overall, participants found \tool very intuitive (Rating $4.3/5$),
and useful in checking ML properties of interest (Rating $4.24/5$), and reported
that the show/hide toggle feature in JupyterLab significantly improved
readability (Rating $4.7/5$). Finally, \tool has been adopted by
SHAP~\cite{shap,shap_paper}, a popular ML library (24.4k stars, 3.4k
forks on GitHub) in their CI~\cite{shap_pr}.

\mypara{Contributions}
Our key contributions in this work are as follows:
\begin{itemize}[leftmargin=*,topsep=0pt,label=$\star$]
    \item We propose \toolgen, the first automated approach for generating
    assertions for machine learning notebooks; it accounts for non-determinism
    in ML pipelines using statistical methods.
    \item We develop \tool, the first regression testing framework for machine
    learning notebooks, integrated with \texttt{pytest} and Jupyter. \tool
    allows developers to write non-intrusive \cellscoped assertions to check
    correctness of computations within each cell.   
    \item We evaluate \toolgen on a corpus of \totalKaggleNotebooks Kaggle
    notebooks and show that \toolgen generates \totalKaggleAssertions assertions
    that obtain a mutation score of \mutantScoreKaggleZeroNineNinePercRev and could have caught \totalPercVersionKilledKaggleRev
    real-world regressions.
    \item We perform a user study with \userstudy participants that shows that
    notebook users find \tool intuitive and useful in writing assertions and
    testing notebooks.  Our tool and results are available via
    GitHub~\cite{aNBTest_github} and Zenodo~\cite{nbzenodo}.
\end{itemize}
%\Fix{developer interaction if we get something}

% future: open source, focus on property-based tests, integrate with production

%\input{example}
\section{Methodology}
\subsection{Notebook Dataset Curation}
\label{sec:datacuration}
%To evaluate the effectiveness of our technique in generating assertions for computational notebooks, 
We collect a corpus of runnable real-world notebooks from Kaggle. Kaggle is the
world's leading platform for ML competitions and contains many high-quality,
end-to-end ML pipelines in the form of notebooks\Comment{that are
publicly accessible}. Hence, like prior studies on
notebooks~\cite{yang2022data}, we also collect ML-related Python notebooks from
Kaggle competitions. \Comment{We describe our collection process next.}
We select five popular Kaggle competitions~\cite{kaggle_competition_sorted} that
are related to ML tasks: Titanic~\cite{titanic}, Housing
Prices~\cite{housing_prices}, Spaceship Titanic~\cite{spaceship_titanic}, Jane
Street~\cite{jane_street}, and Digit recognizer~\cite{digit_recognizer}. 
%These competitions include tasks from domains such as forecasting, classification, image recognition, time series analysis, and more.
% \elaine{We rank Kaggle competitions by the number of participating teams, then select the top ones that represent a variety of domains—such as forecasting, classification, image recognition, time series analysis, and more.}
We use the Kaggle API~\cite{kaggle_api} to download all notebooks related to
these competitions. 
%At this stage, we obtain a total of \Fix{XX} notebooks.
%
To avoid dependency issues, we filter out notebooks older than two
years.\Comment{last run more than two years ago -- based on the date in the
\texttt{lastRunTime} field in the Kaggle API response.} We filter out notebooks
with fewer than 10 votes to remove low-quality samples. After these steps, we
obtain \ktotal notebooks. 
%To avoid choosing incomplete or poor quality notebooks, we further filter out notebooks that have fewer than 10 votes on Kaggle. 
%
%Because our work focuses on ML,
We select notebooks that use at least one of the mainstream ML libraries:
PyTorch, TensorFlow, Keras, or Scikit-learn, by checking their import
statements. We identify \kml such notebooks.\Comment{that contain at least one
of the mainstream ML libraries.} 
%The distribution of ML frameworks among them
%\Comment{used by these notebooks }is \ksklearn (Scikit-learn), \ktensorflowkeras
%(TensorFlow/Keras), and \ktorch (PyTorch) (non-exclusive).
% \ktrain
% notebooks involved training, while \kinference involved inference.

\Comment{
\begin{table}[h!]
\centering
\caption{Distribution of Kaggle notebooks in different tasks.}
\label{tab:categorize_kaggle}
\begin{tabular}{cccc|c}
\toprule
           & Training & Inference & others & Total \\ \hline
Kaggle  &   \ktrain         &   \kinference      &   \kother    &   \kml    \\
% Github &    \gtrain         &   \ginference      &   \gother      &  \gml  \\
\bottomrule
\end{tabular}
\end{table}

\begin{table}[h!]
\centering
\caption{Distribution of Kaggle notebooks using different ML libraries.}
\label{tab:api_kaggle}
\begin{tabular}{cccc|c}
\toprule
           & Sklearn & Tensorflow/Keras & Pytorch & Total \\ \hline
Kaggle  &   \ksklearn         &   \ktensorflowkeras      &   \ktorch    &   \kml    \\
% Github &    \gsklearn        &   \gtensorflowkeras      &   \gtorch      &  \gml  \\
\bottomrule
\end{tabular}

\end{table}

\begin{table}[h!]
    \centering
    \caption{Details of Kaggle Notebooks based on ML library used}
    \Fix{can we just merge everything into the current table 3? Delete other two tables}
    \label{tab:ml_api}
    \begin{tabular}{lllrr}
    \toprule
    Libraries & Tasks & APIs & Num &Total \\ \hline
    \multirow{ 2}{*}{Sklearn} & Training & model.fit() & \ksklearnT \\
                              & Inference & model.predict() & \ksklearnI & \ksklearn\\ \hline
    
    \multirow{ 2}{*}{Tensorflow/Keras} & Training & model.fit() & \ktfT &  \\
                              & Inference & model.predict() & \ktfI & \ktensorflowkeras\\ \hline
    \multirow{ 2}{*}{Pytorch} & Training & loss.backward() & \ktorchT  \\
                              & Inference & model.eval() & \ktorchI & \ktorch \\ \hline
    \textbf{Total} & -- & -- & \kml & \kml \\
    \bottomrule
    \end{tabular}
\end{table}
}

\Comment{ Table~\ref{tab:categorize_kaggle} presents the number of Kaggle
notebooks that use each ML library. Table~\ref{tab:ml_api} further classifies
the notebooks based on the tasks that they perform (such as training or
inference) -- inferred from the APIs they use. We observe that \Fix{...} }
\Comment{ For SC2, we added ML-related keywords to our API
searches. For Kaggle, we selected five popular
competitions\cite{kaggle_competition_sorted} focusing on various ML tasks as
prior work\cite{yang2022data} did. These competitions are,
Titanic\cite{titanic}, Housing Prices\cite{housing_prices}, Spaceship
Titanic\cite{spaceship_titanic}, Jane Street\cite{jane_street}, and Digit
recognizer\cite{digit_recognizer}. These competitions cover various tasks such
prediction, classification, and forecasting. The dataset also cover different
modalities including numerical numbers, text, and  images }

% For SC3, we focused on projects from the last two years to ensure compatibility with recent ML libraries. In Kaggle, similarly, we filtered the projects whose value in the "lastRunTime" field in within two years from November 2024. 

% For SC4, we measured project popularity using the "totalVotes" metric on Kaggle. To ensure the project have some popularities, on kaggle, we only chose notebooks whose "totalVotes" is at least 10. 

% One practical consideration when fetching these dataset using API is, Kaggle limit the number of items returned in a query. Based on our empirical observation, the max number of return item for Kaggle is around 2200. This means that given the same filter condition, we can only get at most certain number of items in the response. Kaggle APIs, however, didn't provide such a filter for date range, therefore, the best we can do is iterate over all possible conditions supported by `--sort-by`. 

% We fetch notebooks based on SC1 ~ SC4 until no notebook meet all the criteria is found. After these processes, we collected \ktotal notebooks from Kaggle.
% and \gtotal notebooks across \gtotalrepo repositories on Github.
% \Fix{
% To understand the prevalence of assertions in the notebooks, we also count the number of original assertions of these collected notebooks. \kaggleAssertNbs out of \ktotal Kaggle notebooks contains \kaggleAssert assertions in total, resulting in \kaggleAssertAvg assertion per notebook on average.}
% \Fix{what does this mean? do we need this?}

\Comment{
For SC5, we focus on notebooks that use at least one of the mainstream ML libraries, which are , Pytorch, Tensorflow, Keras and Sklearn. Specifically, we first download the notebooks locally, then search for patterns that import ML libraries mentioned above. We use regular expressions to search for the import patterns such as \inlinecode{import torch}, \inlinecode{import Tensorflow} and \inlinecode{from torch import ...}. Based on SC5, we identify \kml notebooks on Kaggle that contain at least one of the mainstream ML libraries.
}

\Comment{
For SC6, since \tool's assertion generation involves running these notebooks to generate assertions from execution results, we need to ensure these notebooks can be executed without errors. 
}
% Before executing these notebooks, we first performed a sanity check on those notebooks which contain environment setup information. Usually, notebooks can use the three type of files to record environment setup information\cite{pimentel2019large_scale}, which are, (1) requirements.txt, (2) Pipfile, and (3) setup.py. Other than these three, we performed manual inspection and found that the environment setup information can be provided implicitly in other forms, such as in readme and in self-defined files, we marked this type of config information as Others in Table~\ref{tab:dist_nbenv_dep_github} Therefore, we first calculate the number of notebooks that contain any of the environment setup files.  
\Comment{
We find that since kaggle didn't provide the user a place to save the environment information, none of the kaggle notebook contain any environment setup file. 
}

% \textbf{}Since only \LATER{XX} of the notebooks contain environment dependency information, we need to restore notebooks' execution environments. Only after that, can we verify the notebook execution. 
\Comment{
For kaggle notebooks, we use `pipreqsnb` to generate requirements.txt for each notebook. We find this method was able to generate requirements.txt for \kgenerated out \kml notebooks. 

We will discuss this in detail in Section \ref{sec:notebook_execution}.
}
% \begin{table*}[h!]
% \centering
% \begin{tabular}{cccccc|c}
% \toprule
%            & requirements.txt & environments.yml & Pipfile & Other & Fixed by us & Total \\ \hline
% Notebooks  &   \greqrepo         &   \genvrepo      &   \gpiprepo & \gOtherReqRepo   &      &  \gmlrepo  \\
% Installable &     \greqrepoinstall         &     \genvrepoinstall  &  \gpiprepoinstall  & \gOtherReqRepoInstall          &    &  \ginstallrepo  \\
% Executable &                &                &         &   &     &    \\
% \bottomrule
% \end{tabular}
% \caption{Distribution of Github notebooks being provided with environment dependency information.}
% \label{tab:dist_nbenv_dep_github}
% \end{table*}

\mypara{Environment Setup}
\label{sec:notebook_execution}
%Notebooks are well-known to have reproducibility problems.
%It is well-known that notebooks do not come with any guarantee that they will be runnable off-the-shelf, especially because kaggle does not provide the list of dependencies for each notebook they host. 
%To mitigate this issue, 
We use the pipreqs~\cite{pipreqs} tool to generate the
list of dependencies (a \texttt{requirements.txt} file) for each notebook. 
% Using
% this method, we were able to generate dependencies for \kgenerated out of \kml
% notebooks. 
We then create a conda environment for each notebook, install the
identified dependencies, and execute the notebook. Finally, we retain notebooks
that execute without any errors. After this process, we obtain
\totalKaggleNotebooks Kaggle notebooks, out of which \ksklearnExe use
scikit-learn, \ktensorflowkerasExe use TensorFlow/Keras, and  \ktorchExe use
PyTorch (non-exclusive).

\Comment{
\mypara{Filtering Buggy Notebooks} While the notebooks we collected from
Kaggle competitions are generally high quality, they might still contain bugs
that cause them to produce incorrect results and bias our evaluation. To
mitigate this issue, we explored previously known categories of common bugs
found in ML notebooks~\cite{yang2022data,drobnjakovic2024abstract,chorev2022deepchecks,liblit2023shifting}, and
inspected all the notebooks to filter out those that contain such bugs.
Specifically, we looked for the following types of bugs: 1) data leakage, 2) conflicting labels in the dataset, 
3) mixed data type in the dataset,
4) missing zero\_grad calls, 5) missing eval calls. 
To check bug types 1, 4, and 5, we manually inspect all notebooks.
To check bug types 2 and 3, we use the Deepcheck library~\cite{chorev2022deepchecks} to check them.
After this step, we filtered out $6$ notebooks, resulting in a final set of \totalKaggleNotebooks notebooks.
\sd{doesnt make sense, the final notebooks did not change after filtering?}
\sd{also what does it mean to use deepcheck to check the bugs? was it autoamted/manual?}
}
\Comment{To see the diversity of ML libraries covered in these notebooks, we
again count the distribution of ML libraries in these executable notebooks,
shown in Table~\ref{tab:api_kaggle_exe}.} \Comment{
\begin{table}[h!]
\centering
\Fix{merge these numbers with table 3}\elaine{Some notebooks contain multiple libraries.}
\begin{tabular}{cccc|c}
\toprule
           & Sklearn & Tensorflow/Keras & Pytorch & Total \\ \hline
Kaggle  &   \ksklearnExe         &   \ktensorflowkerasExe      &   \ktorchExe    &   \totalKaggleNotebooks    \\
\bottomrule
\end{tabular}
\caption{Distribution of executable Kaggle and Github notebooks using different ML libraries.}
\label{tab:api_kaggle_exe}
\end{table}
}

% \tool's assertion generation involves running these notebooks to generate assertions from execution results, we need to ensure these notebooks can be executed without errors. 
% %
% We find that since kaggle didn't provide the user a place to save the environment information, none of the kaggle notebook contain any environment setup file. 
% %
% For kaggle notebooks, we use `pipreqsnb` to generate requirements.txt for each notebook. We find this method was able to generate requirements.txt for \kgenerated out \kml notebooks. 
%
% We will discuss this in detail in Section \ref{sec:notebook_execution}.

% After the above process, we were able to execute \knbExe kaggle notebooks.

% After identifying a suitable number of notebooks performing specific ML or Deep Learning tasks, we implemented a rigorous process to ensure their executability and compatibility. Because each kaggle notebook relies on a dataset, we download the dataset required for each notebook locally and fix the dataset paths in the notebook to point to the local path. For each notebook, we collect its dependencies,  initialize a conda environment, and install the dependencies. We then execute the notebook using the conda environment to check if it runs without any errors. Because kaggle notebooks often do not specify the precise dependency versions, many notebooks fail to execute at this stage. If we are able to run the notebook, we extract the exact versions of the installed dependencies for posterity. After this step, we obtain \Fix{XYZ} runnable notebooks from Kaggle.

\Comment{
\begin{itemize}
\item Created separate Conda environments for each notebook.
\item Generated an \texttt{environment.yml} file for each notebook to manage dependencies.
\item Tested notebook execution across multiple operating systems:
\begin{itemize}
\item Windows
\item Linux
\item macOS
\end{itemize}
\item Collated notebooks that passed all compatibility checks, resulting in a final set of 30 notebooks for initial testing and result analysis.
\end{itemize}
}

\begin{table}[!ht]
\centering
\scriptsize

\caption{ML-related Mutation Operators}\label{tab:mutationoperators}
% \Fix{remove unused mutations}
\vspace{-0.1in}
\setlength{\tabcolsep}{0.2em}
\begin{tabularx}{\columnwidth}{H p{2cm}| X}
\toprule
\textbf{Category} & \textbf{Mutation Operator} & \textbf{Description} \\
\midrule
\multicolumn{3}{l}{\textbf{Data-based Mutations}}\\ \midrule
\multirow{5}{*}{\textbf{Data-based}} 
& \outliers \cite{zimmermann2023common} & Introduces extreme values in numeric columns by multiplying selected rows with large random values. \\
\cmidrule{2-3}
& \repetition \cite{ma2018deepmutation} & Randomly replaces a portion of rows with other existing rows in the dataset, simulating duplication noise. \\
\cmidrule{2-3}
& \addednull \cite{zimmermann2023common} & Adds NaN/None entries in a subset of rows and columns to simulate missing values. \\
\cmidrule{2-3}
& \labelerrors\cite{ma2018deepmutation}  & Introduces incorrect labels by randomly changing labels for selected samples. \\
% \cmidrule{2-3}
% & \datashift \cite{wiles2021finegrainedanalysisdistributionshift} & Alters the correlation between features and labels across training and test sets, simulating covariate or concept shift. \\
\midrule
\multicolumn{3}{l}{\textbf{Source Code-based Mutations}}\\\midrule
\multirow{10}{*}{\textbf{Source Code-based}} 
% & \preprocdataleakage \cite{yang2022data} & Moves data preprocessing (e.g., scaling or feature selection) before train-test split, causing pre-processing test leakage. \\
% \cmidrule{2-3}
%& \removetorcheval \cite{liblit2023shifting} & Removes \texttt{model.eval()} in PyTorch models, leading to dropout and batchnorm staying active during evaluation. \\
%\cmidrule{2-3}
& \removezerograd \cite{liblit2023shifting} & Removes \texttt{optimizer.zero\_grad()}, allowing unwanted gradient accumulation across training steps. \\
\cmidrule{2-3}
& \modifyhyperparams \cite{zimmermann2023common} & Changes critical hyperparameters like learning rate or dropout rate to harmful but syntactically valid values. \\
\cmidrule{2-3}
& \removehyperparams \cite{zimmermann2023common} & Removes explicit hyperparameter values to fallback on library defaults, potentially reducing model quality. \\
\cmidrule{2-3}
% & \metricswap \cite{ma2018deepmutation} & Replaces ML pipeline components (e.g. Scaling, Encoding, etc..) with functionally similar but different alternatives. \\
% \cmidrule{2-3}
% & \layerinsert \cite{ma2018deepmutation} & Adds redundant, shape-preserving layers (e.g., ReLU, Dropout) into the network to subtly affect behavior. \\
\cmidrule{2-3}
& \layerremove \cite{ma2018deepmutation} & Removes non-critical layers like ReLU or Dropout, potentially weakening feature learning or regularization. \\
%\cmidrule{2-3}
% & \replacelayer \cite{ma2018deepmutation} & Replaces a layer with a different, compatible type (e.g., ReLU → Tanh), affecting model behavior without breaking code. \\
\bottomrule
\end{tabularx}
\vspace{-0.2in}
\end{table}

\subsection{Mutation Testing}
\label{sec:mutant_generation}
To evaluate the fault-detection ability of our generated assertions, we develop
eight mutation operators inspired by prior studies on common bugs found in ML and data science
projects~\cite{zimmermann2023common,ma2018deepmutation,wiles2021finegrainedanalysisdistributionshift,liblit2023shifting}.
We reviewed mutation operators proposed for ML by Ma et al.~\cite{ma2018deepmutation} and
identified two main categories: data-related and model-definition-related
operators. From these categories, we selected a representative subset
that captures key mutation types, including data distribution anomalies,
corrupted training data, inappropriate hyperparameters, and flawed model
definitions. Our mutators subsume those from
~\cite{ma2018deepmutation}, while also extending them based on common ML
development errors identified in~\cite{zimmermann2023common,wiles2021finegrainedanalysisdistributionshift,liblit2023shifting}.

We exclude certain mutation operators and error patterns found in~\cite{zimmermann2023common,ma2018deepmutation,wiles2021finegrainedanalysisdistributionshift,liblit2023shifting} that typically lead to crashes, such as variable redefinition and missing data.
Our goal is to instead explore mutations that lead to silent errors, which enables us
to evaluate whether \toolgen assertions can identify hard-to-detect issues.

\looseness=-1 
Table~\ref{tab:mutationoperators} presents all mutation operators
that we support. We implement two categories of mutation operators: (1)
\emph{Data-based mutations} that mutate the dataset in the notebook,
for example, by adding outliers or introducing spurious data correlations, and
(2) \emph{Source code-based mutations} that mutate the Python code in the notebook,
for example, by modifying training hyper-parameters or removing neural network
layers. Overall, we implement $4$ data-based mutations and
$4$ source code-based  mutations. 
%We provide detailed descriptions and examples of each mutation operator in Appendices~\ref{sec:outlier-mutation}-\ref{sec:deep-layer-removal-mutation}.
% We provide detailed descriptions and examples of each mutation operator in Appendices \S B-K. 
% \Fix{do we have anything beyond whats in table 1?}
\Comment{
Figure~\ref{fig:dataleakage} presents an example of a mutation that introduces
data leakage in the data pre-processing step in a pipeline~\cite{yang2022data}.
We simulate a data leakage issue by applying pre-processing steps, like scaling
or feature selection, to the whole dataset instead of just the training set.
This mimics the error of allowing test data to affect training data
transformations. 

\begin{figure}[h] % Use figure* to span both columns if needed for the caption
\centering
\begin{minipage}{\linewidth}
\centering
\captionsetup{labelformat=simple, labelsep=colon}
\begin{lstlisting}[language=Python, basicstyle=\ttfamily\footnotesize]
from sklearn.model_selection import train_test_split
from sklearn.preprocessing import StandardScaler
x_train, x_test, y_train, y_test = train_test_split(x, y, test_size=0.2)
sc=StandardScaler()
X_train = |\colorbox{yellow}{sc.fit\_transform(x\_train)}| # transform after split
|\colorbox{yellow}{X\_test = sc.transform(x\_test)}|
\end{lstlisting}
\caption*{(a) Before mutation}
\end{minipage} 
~
\begin{minipage}{\linewidth}
\centering
\captionsetup{labelformat=simple, labelsep=colon}
\begin{lstlisting}[language=Python, basicstyle=\ttfamily\footnotesize]
from sklearn.model_selection import train_test_split
from sklearn.preprocessing import StandardScaler
sc = StandardScaler()
|\colorbox{yellow}{x = sc.fit\_transform(x)}| # transform before split
x_train, x_test, y_train, y_test = train_test_split(x, y, test_size=0.2)
\end{lstlisting}
\caption*{(b) After mutation}
\end{minipage}    
\caption{An example mutation for adding data leakage}
\label{fig:dataleakage}
\end{figure}

Figure \ref{fig:modify_hyperparams} demonstrates the \emph{\modifyhyperparams}
mutation. We simulate ML training related bugs by randomly removing or altering
hyper-parameters for different training or data processing steps. We target model
initialization, method calls, and optimizer configurations across popular
frameworks. We always preserve required parameters (like `units' for Dense
layers). Instead, we apply legal changes such as increasing epochs, shifting learning
rates, or swapping activation functions, causing models to overfit, underfit, or
fail to converge properly. Because modifying a single hyperparameter may not
significantly affect the training accuracy, this mutation modifies up to 4
hyper-parameters in a single notebook.

\begin{figure}[h]
\centering

\begin{minipage}{\linewidth}
\centering
\captionsetup{labelformat=simple, labelsep=colon}
\begin{lstlisting}[language=Python, basicstyle=\ttfamily\footnotesize]
model = Sequential([
Dense(128, activation=|\colorbox{yellow}{'relu'}|, input_shape=(784,)),
Dropout(rate=|\colorbox{yellow}{0.2}|),
Dense(64, activation='relu'),
Dense(10, activation='softmax')
])

model.compile(
optimizer=Adam(learning_rate=|\colorbox{yellow}{0.001}|),
loss=|\colorbox{yellow}{'categorical\_crossentropy'}|,
metrics=['accuracy']
)
\end{lstlisting}
\caption*{(a) Before mutation}
\end{minipage}
        
\begin{minipage}{\linewidth}
\centering
\captionsetup{labelformat=simple, labelsep=colon}
\begin{lstlisting}[language=Python, basicstyle=\ttfamily\footnotesize]
model = Sequential([
Dense(128, activation=|\colorbox{yellow}{'sigmoid'}|, input_shape=(784,)),
Dropout(rate=|\colorbox{yellow}{0.45}|),
Dense(64, activation='relu'),
Dense(10, activation='softmax')
])

model.compile(
optimizer=Adam(learning_rate=|\colorbox{yellow}{0.01}|),
loss=|\colorbox{yellow}{'sparse\_categorical\_crossentropy'}|,
metrics=['accuracy']
)
\end{lstlisting}
\caption*{(b) After mutation}
\end{minipage}    

\caption{Modifying hyperparameters mutation}
\label{fig:modify_hyperparams}
\end{figure}
}
% For evaluating \tool's generated assertions to effectively capture bugs in Jupyter Notebooks, we perform Mutation Testing, using mutation operators that simulate real world bugs found in ML projects in general.

\Comment{
\subsubsection{Mutation Testing Operators}
\begin{enumerate}
    \item Data-based Mutation Operators:
        \begin{itemize}
            \item Introducing Outliers: We inject outliers by randomly selecting rows and multiplying their numerical values by a large, random factor.
            
            \item Data Repetition: To simulate data duplication, we remove a specified number of rows and replace them with existing rows, effectively repeating data points while maintaining the dataset's size.
            
            \item Adding NULL values: We introduce missing data by replacing values in a random subset of data points and attributes with NULL (\verb|np.nan| or \verb|None|). 

            \item Spurious Data Correlation \cite{wiles2021finegrainedanalysisdistributionshift}: We emulate spurious correlations by decoupling attributes that are highly correlated in the training set within the test set, eliminating the original correlation.  

            \item Label errors: We identify the target column in the dataset and actual labels 
        \end{itemize}
        
    \item Code Mutation Operators:
        \begin{itemize}
            \item Removing \verb|zero_grad()|: PyTorch accumulates gradients during backpropagation. Developers typically use \verb|zero_grad()| to reset these gradients before each update. Omitting \verb|zero_grad()| results in indefinite gradient accumulation, preventing batch updates. To simulate this common error, we intentionally remove \verb|zero_grad()| calls in this operator.

            \begin{figure}[h] % Use figure* to span both columns if needed for the caption
            \centering
        
            \begin{minipage}{\linewidth}
                \centering
                \captionsetup{labelformat=simple, labelsep=colon}
                \caption*{(a) Before mutation}
                \begin{lstlisting}[language=Python, basicstyle=\ttfamily\footnotesize]
for train, test in loader:
    loss = metric(model(train), test)
    optimizer.zero_grad()
    loss.backward()
    optimizer.step()
                \end{lstlisting}
            \end{minipage}
        
            \begin{minipage}{\linewidth}
                \centering
                \captionsetup{labelformat=simple, labelsep=colon}
                \caption*{(b) After mutation}
                \begin{lstlisting}[language=Python, basicstyle=\ttfamily\footnotesize]
for train, test in loader:
    loss = metric(model(train), test)
    # Missing zero_grad() call
    loss.backward()
    optimizer.step()
                \end{lstlisting}
            \end{minipage}    
            
            \caption{Removing zero\_grad() calls}
            \label{fig:remove_eval}
        \end{figure}
            
            \item Removing \verb|eval()|: PyTorch models often require \verb|eval()| to ensure layers behave correctly during evaluation. Omitting this call can lead to inaccurate performance assessment. To simulate this potential bug, we remove all instances of \verb|eval()|.

            \begin{figure}[h] % Use figure* to span both columns if needed for the caption
            \centering
        
            \begin{minipage}{\linewidth}
                \centering
                \captionsetup{labelformat=simple, labelsep=colon}
                \caption*{(a) Before mutation}
                \begin{lstlisting}[language=Python, basicstyle=\ttfamily\footnotesize]
for batch_num in enumerate(dataloader):
    model.train()
    # forward, backwards and optimization steps
    if batch_num % 50 == 0:
        model.eval()
        precision, recall, f1 = model.evaluate_on(data, batch_size)
                \end{lstlisting}
            \end{minipage}
        
            \begin{minipage}{\linewidth}
                \centering
                \captionsetup{labelformat=simple, labelsep=colon}
                \caption*{(b) After mutation}
                \begin{lstlisting}[language=Python, basicstyle=\ttfamily\footnotesize]
for batch_num in enumerate(dataloader):
    model.train()
    # forward, backwards and optimization steps
    if batch_num % 50 == 0:
        # Missing eval() call
        precision, recall, f1 = model.evaluate_on(data, batch_size)
                \end{lstlisting}
            \end{minipage}    
            
            \caption{Removing eval() calls}
            \label{fig:remove_eval}
        \end{figure}

            \item Removing/Modifying hyper-parameters: 
            We simulate ML optimization errors by randomly removing or altering hyperparameters. Our implementation targets model initialization, method calls, and optimizer configurations across popular frameworks. When removing, we preserve essential parameters (like 'units' for Dense layers). When modifying, we apply context-aware changes such as increasing epochs, shifting learning rates, or swapping activation functions, causing models to overfit, underfit, or fail to converge properly.

            \begin{figure}[h]
            \centering
        
            \begin{minipage}{\linewidth}
                \centering
                \captionsetup{labelformat=simple, labelsep=colon}
                \caption*{(a) Before mutation}
                \begin{lstlisting}[language=Python, basicstyle=\ttfamily\footnotesize]
        model = Sequential([
            Dense(128, activation='relu', input_shape=(784,)),
            Dropout(rate=0.2),
            Dense(64, activation='relu'),
            Dense(10, activation='softmax')
        ])
        
        model.compile(
            optimizer=Adam(learning_rate=0.001),
            loss='categorical_crossentropy',
            metrics=['accuracy']
        )
                \end{lstlisting}
            \end{minipage}
        
            \begin{minipage}{\linewidth}
                \centering
                \captionsetup{labelformat=simple, labelsep=colon}
                \caption*{(b) After mutation}
                \begin{lstlisting}[language=Python, basicstyle=\ttfamily\footnotesize]
        model = Sequential([
            Dense(128, activation='sigmoid', input_shape=(784,)),
            Dropout(rate=0.45),
            Dense(64, activation='relu'),
            Dense(10, activation='softmax')
        ])
        
        model.compile(
            optimizer=Adam(learning_rate=0.01),
            loss='sparse_categorical_crossentropy',
            metrics=['accuracy']
        )
                \end{lstlisting}
            \end{minipage}    
            
            \caption{Modifying hyperparameters}
            \label{fig:modify_hyperparams}
        \end{figure}
            
            \item Pre-processing leakage \cite{yang2022data}: We simulate a data leakage issue by applying pre-processing steps, like scaling or feature selection, to the whole dataset instead of just the training set. This mimics the error of allowing test data to affect training data transformations. 

            \begin{figure}[h] % Use figure* to span both columns if needed for the caption
            \centering
        
            \begin{minipage}{\linewidth}
                \centering
                \captionsetup{labelformat=simple, labelsep=colon}
                \caption*{(a) Before mutation}
                \begin{lstlisting}[language=Python, basicstyle=\ttfamily\footnotesize]
from sklearn.model_selection import train_test_split
from sklearn.preprocessing import StandardScaler
x_train, x_test, y_train, y_test = train_test_split(x, y, test_size=0.2)
sc=StandardScaler()
X_train = sc.fit_transform(x_train)
X_test = sc.transform(x_test)
                \end{lstlisting}
            \end{minipage}
        
            \begin{minipage}{\linewidth}
                \centering
                \captionsetup{labelformat=simple, labelsep=colon}
                \caption*{(b) After mutation}
                \begin{lstlisting}[language=Python, basicstyle=\ttfamily\footnotesize]
from sklearn.model_selection import train_test_split
from sklearn.preprocessing import StandardScaler
sc = StandardScaler()
x = sc.fit_transform(x)
(x_train, x_test, y_train, y_test) = train_test_split(x, y, test_size=0.2, random_state=2)
                \end{lstlisting}
            \end{minipage}    
            
            \caption{Pre-processing data leakage}
            \label{fig:remove_eval}
        \end{figure}
            
        \end{itemize}

\end{enumerate}

}

\subsection{Collecting Kaggle Notebook Versions}
\label{sec:kaggle_version_collection}
To evaluate the effectiveness of our generated assertions in detecting
real-world regressions, we need a dataset of Jupyter Notebooks that contain
regression bugs. Unfortunately, such a dataset is not publicly available.
Besides, unlike regular programs, developers do not frequently commit changes in
notebooks on GitHub, and even if they do, the absence of tests often makes the
regressions go unnoticed.
% Even if they do commit changes, due to the absence of tests, they may be unaware that the notebooks contain regression bugs. 
So, instead we leverage Kaggle, where users often submit multiple public
versions of notebooks to improve competition performance. Since the latest
version is likely to be the most stable and well-tested, we hypothesize that the
older versions may contain real-world silent regression bugs. Based
on this hypothesis, we collect older notebook versions, then apply assertions
generated from the latest version to each older version, to evaluate if these
assertions can detect regressions.

\mypara{Collection and Filtering} We develop a web scraping tool to extract all
available versions of notebooks we consider in \S~\ref{sec:datacuration}. Out of
the \allversions available versions, we do not consider versions with execution errors
or large diffs (more than 500 diff lines) or duplicate versions. This leaves us with
\numKaggleVersionNoExeErrorRev notebook versions.
% and thus only download \numKaggleVersionRev versions. 
% We also remove the duplicate versions, leaving us
% with \versionNoDuplicate unique notebook versions.

\mypara{Assertion transfer} 
We consider an assertion transferable between versions if either (1) the
statements it checks remain unchanged, or (2) the statements checked differ
syntactically but are semantically equivalent, such as variable renaming or minor
keyword change. For the first case, since \tool inserts the assertion
immediately after the statement, we can simply check whether the preceding
statement is identical between versions.
To decide about assertions in the second case, we require careful reasoning
about program semantics to determine whether the assertion is still valid in the
older version, despite syntactic differences. 
% which requires nuanced reasoning about program semantics.
Given the strong capabilities of LLMs in code understanding, we leverage an LLM
(Gemini 3.0 Flash) for this task.
%to determine whether an assertion is transferable.

Specifically, we provide the LLM with (1) the latest version with generated
assertions, and (2) the older version with already transferred assertions in the
first case, and ask the LLM to decide which remaining assertions are
transferable (prompt in~\cite{nbzenodo}). When an assertion is
deemed transferable, the LLM inserts a lightweight macro with the assertion ID
(later replaced with the full assertion) at the appropriate location, and
outputs the modified old version. Then, we perform two validation checks. First,
we execute the notebook to verify that it does not have formatting issues. We
found only six notebooks with missing closing brackets, which we corrected
manually. Second, to detect hallucinations, we verify that all transferred assertions are a subset of assertions to be transferred, then remove any
unexpected assertions. This produces the final notebook with transferred
assertions. While this process is not perfect, it allows us to easily transfer
many assertions for evaluation. In a realistic setting, developers would
directly modify the notebook with assertions, so the transfer process would not
be needed.

\subsection{Assertion-Statement Dependency Mapping}
\label{sec:assertion_statement_mapping}
In the presence of mutation or regression in a notebook, a failing assertion may
be caused either by the mutation itself, i.e., true positive (TP), or by
unrelated factors such as randomness, i.e., false positive (FP). To distinguish
FP from TP, we construct a mapping between each failed assertion and the set of
preceding statements that can reach it via control or data flow. Then we check
if the mutation or regression is within these statements. If so, we can classify
the assertion failure as a TP, otherwise as an FP. To build such a mapping, we
develop a custom Python tool that extracts notebook AST, and performs standard
data-flow analysis~\cite{allen1976program}. Similarly, we can classify a passing
assertion as a true negative (TN) or false negative (FN).

\subsection{Metrics}
\label{sec:metrics}
% \Fix{metrics can be streamlined a bit more i think to reuse one metric in another. }
%We define the metrics that we use to evaluate our approach.

\mypara{Passrate $p$} 
Given an assertion, let $N$ be the total number of executions, and let $P$ be the number of executions in which the assertion passes.  
The \emph{passrate} $p$ of the assertion is defined as: $p = \frac{P}{N}$.

\mypara{Mutation Score $m$} 
Given an assertion, let $M$ be the total number of generated mutants.  
For each mutant $M_i$ ($i \in [1, M]$), let $p_i$ be the passrate of the assertion on $M_i$.  
The \emph{mutation score} $m$ of the assertion is defined as: $m = \frac{1}{M}
\sum_{i=1}^{M} (1 - p_i)$. %For a type of assertion, we define its mutation score
%as the average mutation score of all assertions of this type.
% Let $N$ be the total number of executions, 
% and let $A$ be the number of assertions in a notebook.
% For each iteration $i \in \{1, \dots, N\}$, $P_{i} \in \{1, \dots, A\}$ is the number of assertions that pass.
% % and let $P$ be the number of iterations in which a given assertion passes.  
% We define the \emph{passrate} of an assertion as: 
% $\frac{\sum_{i=1}^{N} P_{i}}{N * A}$.
% \mypara{Mutation Score}  
% Given a mutation type and an assertion, 
% let $M$ be the total number of generated mutants. For each mutant $i \in \{1, \dots, M\}$, 
% let $K_i$ be the number of executions killed out of $N$ executions.  
% The score of mutant $i$ is defined as: $s_i = \frac{K_i}{N_i}$, where $0 \leq
% s_i \leq 1$. 
% \begin{equation}
% s_i = \frac{K_i}{N_i}, \qquad 0 \leq s_i \leq 1.
% \label{eq:mutant_score_individual}
% \end{equation}  
% The \emph{mutation score} of an assertion for this mutation type is:  
% \begin{equation}
% $S = \frac{1}{M} \sum_{i=1}^{M} (\frac{K_i}{N})$.
% \label{eq:mutation_score}
% \end{equation}  
% \Fix{is this for each assertion? or overall?}

\mypara{Number of Versions Killed $K_V$} 
Let $\mathcal{V}$ denote the set of versions of a notebook.  
For each version $v \in \mathcal{V}$, let $A^{v}$ be the number of assertions,
and let $p_i$ be the passrate of assertion $i \in [1, A^{v}]$ on $v$.  
The \emph{number of versions killed} $K_V$ is defined as:\\ $K_{V} = \sum_{v \in
\mathcal{V}} \mathds{1}_{(\exists i \in [1, A^{v}],\, p_i < 1)}$,
where $\mathds{1}$ is the indicator function.

\mypara{False Positive and Negative Ratios}
Let \textit{\#FP} denote the number of failed assertions not caused by mutations or regressions, and \textit{\#Failed} the total number of failed assertions.
Let \textit{\#FN} denote the number of passed assertions in the presence of mutations or regressions, and \textit{\#Passed} the total number of passed assertions.
Both are determined by the assertion-statement dependency mapping in Section~\ref{sec:assertion_statement_mapping}.
The \textit{FPR} and \textit{FNR} are defined as:
$\textit{FPR} = \frac{\textit{\#FP}}{\textit{\#Failed}}$, \quad
$\textit{FNR} = \frac{\textit{\#FN}}{\textit{\#Passed}}$.

\Comment{
\subsubsection{Evaluating extracted notebook versions}

For evaluation, we use \tool to first generate assertions for the latest version
and transfer these assertions to each older version to check if they can detect
the regression. Because the code structures and cell contents (as well as number
of cells) across the two versions are different, we first identify the cells
that are unchanged between the two versions. Then, for the remaining cells, we
match cells if the AST of their contents are \emph{similar}. We do this for each
cell pair by initially preprocessing the code through comment removal and
separating main logic from test assertions (lines starting with
\texttt{nbtest.assert}).

\Comment{It then extracts three key structural features: assignment targets (variables being assigned), function calls (with their argument counts and keyword parameters), and variable references.
The primary matching criteria needs identical assignment target sets - if both cells assign to variables like {df, model, accuracy}, they're considered similar. A secondary validation is then used to compare function call patterns, which means that the cells must have the same number of function calls, and each call must match in name and argument structure (e.g., \texttt{model.fit(X, y)} matches \texttt{model.fit(X\_train, y\_train, validation\_data=val)} as long as the core function and argument count align).}
Once we match cells across a given pair of versions, we transfer the assertions
from the latest to the older version by matching the preceding statement of the
assertions.
\Comment{This is done by creating a hybrid cell that combines the old cell's original logic with the new cell's test assertions. The algorithm extracts the main code from the old cell (excluding any existing assertions) and the assertion statements from the new cell (lines starting with \texttt{nbtest.assert}). Then it constructs a new code cell by concatenating the old cell's main code with the new cell's assertions, separated by newlines.}
This is a best-effort process, so it may not be possible to transfer some assertions if the cell contents were modified too much or the changes were too large. However, we minimize such situations by limiting the maximum diff size.
% those situations by 
% There could be extreme cases where due to a high degree of modification of the notebook, no assertions are able to be injected into the older version; however such situations are mostly avoided due to the max diff limit considered while downloading the notebooks.

We execute each notebook version using \tool and record which assertions fail in the older version. A notebook version is considered \emph{killed}, i.e.,  regression is detected, if at least one transferred assertion fails.
}
\Comment{
\subsection{Metrics}
% \Fix{describe all metrics we use in eval}
% \mypara{Pass rate of a notebook} Defined as the ratio of number of times all tests in the notebook pass to number of times the notebook was tested
\mypara{Pass Rate of an assertion} It is the ratio of the number of iterations an assertion passes to the total number of the \texttt{pytest} iterations.
\mypara{Mutation Score} It is the proportion of killed mutants to the number of generated mutants. If a mutant is executed multiple times using \texttt{pytest}, its individual score is computed as the fraction of runs in which it is killed (at least one assertion failed). For instance, suppose we generate two mutants and execute each one 10 times. If the first mutant is killed in 8 out of 10 executions and the second in 5 out of 10, the overall mutant score is calculated as $\left(\frac{8}{10} + \frac{5}{10}\right) / 2 = 0.65$.
}

\section{\toolgen: Automated Assertion Generation} 
% We describe the challenges and our solutions for building our automated
% assertion generation approach.
%\Fix{requirements, properties are heterogenous , vary by type of pipeline, choice of framework, task being solved}

%\mypara{Challenges} %There are two key challenges for automating the generation of assertions for ML Notebooks. 
%Second, it is unknown what criteria to use to generate assertions for a given ML notebook.
There are two key challenges in automating the generation of assertions for ML
Notebooks. First, there is a lack of clear criteria for selecting which program
variables or computations to target when generating assertions in ML notebooks. 
%ML pipelines are used for a large variety of tasks and domains, including classification, regression,  and speech recognition, and use a variety of frameworks such as Scikit-Learn, Pandas, Tensorflow, and PyTorch.
In principle, one could add assertions for \emph{every} intermediate
result/variable in the notebook, which would provide comprehensive coverage.
%which makes testing very rigorous but also clutters the notebook with too many assertions that may not be meaningful and limits the usability of the notebook interface. 
However, this brute-force strategy may introduce assertions that are not
meaningful to the developer, cluttering the notebook interface, and reducing its
readability. Conversely, a minimal approach that adds assertions only for the
final outputs -- such as checking the final evaluation metrics of the trained
model -- may preserve usability but sacrifice bug-detection capability. So, bugs
that manifest in intermediate computations might be hard to detect or localize.
% might only be detected at the final
% stage or be missed completely. 
Hence, an assertion generation approach must strike a balance between the fault
detection capability, testing key properties of the ML pipeline, and preserving
the notebook's usability.

Second, it is well-known that ML pipelines often exhibit nondeterminism due to
various operations like randomized batching, random model initialization, and
parallel execution on GPUs. Assertions generated without accounting for this
randomness can lead to \emph{flaky assertions}, where repeated executions of the
same notebook yield inconsistent outcomes. 
%This flakiness undermines the developer’s trust in the testing framework and
%can mask genuine bugs. %Hence, the generated assertions must account for this
%randomness to avoid \emph{flakiness}.
Researchers have previously identified \emph{flaky} behavior in unit tests in ML
libraries~\cite{dutta2020detecting,dutta2021flex,nejadgholi2019study} and ML
training pipelines~\cite{pham2020problems} showing that assertions must
account for underlying nondeterminism.

% the importance of designing robust assertions that can tolerate or account for
% controlled nondeterminism.

% \begin{figure}[htbp]
%     \centering
%     % \vspace{-0.1in}
%     % \includegraphics[width=0.7\textwidth]{figures/workflow.png}
%     \includegraphics[width=\linewidth]{figures/workflow.png}
%     % \vspace{-0.2in}
%     \caption{Assertion Generation Workflow}
%     \label{fig:nbtest-gen_workflow}
%     % \vspace{-0.2in}
% \end{figure}

\begin{algorithm}
    \renewcommand{\algorithmicrequire}{\textbf{Input:}}
\renewcommand{\algorithmicensure}{\textbf{Output:}}
    \caption{\toolgen's Assertion Generation}
    \label{alg:assertgen}
    \begin{algorithmic}[1]
    
    \Require $\mathcal{P}$: Jupyter Notebook, $n$: num. of runs, $APIs$: set of APIs to check for properties,
             $c$: confidence level
    \Ensure $\mathcal{P}'$: Jupyter Notebook with generated assertions
    
    \State $\mathcal{I} \gets \Call{PropertyFinder}{\mathcal{P}, \textit{APIs}}$ \label{alg:pro}
    \State $\mathcal{P}' \gets \Call{AssertionGenerator}{\mathcal{P}, \mathcal{I}, n, c}$ \label{alg:assert}
    
    \Procedure{PropertyFinder}{$\mathcal{P}$, $\textit{APIs}$}
    
        \State $\textit{AST}_\mathcal{P} \gets \Call{ParseAST}{\mathcal{P}}$ \label{alg:line:parse_ast}
        \State $\{(v_{1}, l_{1}, \tau_{1}), \ldots, \textit{up to N} \} \gets \Call{FindAPI}{\textit{APIs}, \textit{AST}_\mathcal{P}}$ \label{alg:line:find_api}
        \State \Return $\{(v_{1}, l_{1}, \tau_{1}), \ldots, \textit{up to N} \}$ \label{alg:line:return_property}

    \EndProcedure

    \Procedure{AssertionGenerator}{$\mathcal{P}, \mathcal{I}, n, c$}
        \State $\mathcal{P}_I \gets \Call{Instrument}{\mathcal{P}, \mathcal{I}}$ \label{alg:line:instrument}
        \State $\mathcal{D} \gets \Call{RunCollect}{\mathcal{P}_I, n}$ \algorithmiccomment{$\mathcal{D}$ is a set of 4-tuples ($v_i$, $l_i$, $\tau_i$, $s_i$)}     \label{alg:line:runcollect}   
        \State $\mathcal{P}' \gets \mathcal{P}$ \label{alg:line:initp}
        \For{$(v_i, l_i, \tau_i, s_i) \in \mathcal{D}$} \label{alg:line:for}
            \State $\epsilon_i \gets \Call{ComputeTolerance}{s_i, c}$ \label{alg:line:compute_tolerance}
            \State $a_i \gets \Call{GenerateAssertion}{v_i, \tau_i, s_i, \epsilon_i}$ \label{alg:line:generate_assertion}
            \State $\mathcal{P}' \gets \Call{InsertAssertion}{\mathcal{P'}, a_i, l_i}$ \label{alg:line:insert_assertion}
        \EndFor \label{alg:line:endfor}
        \State \Return $\mathcal{P}'$
    \EndProcedure
    
    \end{algorithmic}
\end{algorithm}

\label{sec:method_approach}

\mypara{Our Approach} To solve the above two challenges, 
we propose \toolgen's assertion generation algorithm (Algorithm~\ref{alg:assertgen}).
% To solve the first challenge, we leverage statistical
% techniques to estimate the variance in the notebook outputs. To solve the second
% challenge, we identify popular APIs in ML libraries such as
% Scikit-Learn~\cite{scikit-learn}, Pandas~\cite{reback2020pandas},
% TensorFlow~\cite{tensorflow2015-whitepaper}, and
% PyTorch~\cite{paszke2019pytorch} that are used in the notebook to process data
% and models. We then generate assertions checking the integrity properties of
% such components. A prior study~\cite{shome2024understanding} showed that
% developers often write assertions or print statements to perform similar tasks.
Algorithm~\ref{alg:assertgen} takes a notebook ($\mathcal{P}$), the number of
runs ($n$) for tolerance calculation, a set of APIs to check for properties
($\textit{APIs}$), and a confidence level ($c$) as input, and outputs a notebook
with generated assertions ($\mathcal{P}'$). It consists of two stages: (1)
\propfinder that identifies ML-specific \entities in the notebook to target
for assertion generation (Line~\ref{alg:pro}), and (2) \assertgen that
generates assertions for the identified \entities (Line~\ref{alg:assert}) while
using statistical techniques to account for randomness. 

% Our assertion generation approach consists of two stages: 

\empara{(1) \propfinder} \label{sec:propertyfinder} \propfinder extracts each
notebook cell, analyzes its Abstract Syntax Tree (AST)
(Line~\ref{alg:line:parse_ast}), passes the $\textit{AST}_\mathcal{P}$ and the
APIs identifying key ML-related \entities to \texttt{FINDAPI}
(Line~\ref{alg:line:find_api}), and outputs a set of ($v_i$, $l_i$, $\tau_i$),
where $v_i$ is a variable in the notebook related to the \entity, $l_i$ is the
line number of $v_i$ in the notebook, and $\tau_i$ is the type of the \entity.
% (i.e., \datasetAssert, \modelArchAssert, or \modelPerfAssert).

A \entity represents the program computation related to some data or models.
\footnote{The notion of \entity in \tool is different from what is typically
used in property-based testing. Our properties are regression-based (hence
strongly tied to previous execution(s) and the notebook), whereas the ones used
for property-based testing specify more general behavioral properties that may
apply across different executions and notebooks.}
% represents a program component in the form of a variable or expression (e.g., method call). 
Typically, ML notebooks contain various types of
computations such as data pre-processing, model initialization and training, and
model evaluation. So, for each type, we identified specific \entities to
check by understanding common data quality issues in machine learning reported
by Polyzotis et al.~\cite{polyzotis2019data}, as well as common
properties checked by developers via print statements and
assertions~\cite{shome2024understanding}.
We identified the three \emph{types} of \entities based on popular APIs used by
ML Notebooks:
%t goes through each cell of the Jupyter Notebook and methodically analyses it's Abstract Syntax Tree (AST). The classification of properties that are identified by PropertyFinder is as follows:
\begin{itemize}[leftmargin=*,topsep=0ex]
    \item\textbf{\datasetAssert}: We consider APIs that initialize datasets from
    a CSV file (using \verb|pandas.read_csv|), or from a dataset library (like
    \verb|sklearn.datasets|), or by splitting a dataset using
    \texttt{train\_\allowbreak test\_\allowbreak split} from scikit-learn.
    %Datasets play a key role in deciding the performance of ML pipelines and it is therefore important that there are no discrepancies in its composition and size. 
    \item\textbf{Model Architecture (\modelArchAssert)}: 
    %We consider statements
    %that initialize ML models with hyperparameters. 
    We consider APIs that initialize models using
    Scikit-Learn (e.g., \texttt{Logistic\allowbreak Regression}) as well as Deep Neural
    Network models using PyTorch/TensorFlow APIs.
    % We collect ML models with initialization hyperparameters. We consider traditional ML models created using  Scikit-Learn APIs (e.g., \texttt{LogisticRegression}) as well as Deep Neural Network models created using PyTorch and TensorFlow APIs.      
    \item\textbf{Model Performance (\modelPerfAssert)}: 
    %We collect expressions
    We consider APIs that define metrics such as \verb|accuracy_score| and
    \verb|r2_score|, from \verb|sklearn.metrics|. Sometimes, they also define
    custom metrics that build on top of common numerical APIs. So, we implement
    a lightweight static analysis that parses such expressions to identify
    custom metrics used in the notebook.
    % Evaluation metrics are measures of model performance, and hence, crucial properties in a pipeline. Hence, we target APIs like \verb|accuracy_score|, \verb|r2_score|, \verb|recall_score| etc. from \verb|sklearn.metrics|, \space \verb|model.evaluate()| from \verb|tensorflow|, etc.. 
\end{itemize}
% \Fix{use both longer and shorter names for the properties above}

% \Fix{unclear what the following sentence means. do we collect all statements or
% expressions using the APIs or only when they appear in these contexts? also what
% does it mean to extract a property?} 
% \propfinder extracts \entities that appear in
% common contexts where developers inspect intermediate computations, such as
% assignment statements, print statements, and the last line in a
% cell~\cite{shome2024understanding}. 
\looseness=-1 
\propfinder collects statements that either call one of the above
APIs, or appear in contexts where developers typically inspect intermediate
computations, such as print statements or the last line in a
cell~\cite{shome2024understanding}. The latter cases produce outputs in the
cell. To capture variance in computations, we assign a new random seed to each
notebook run. %, replacing all existing seeds.
\toolgen currently supports 100 APIs
across four popular ML/Data libraries: Scikit-Learn, PyTorch, TensorFlow, and
Pandas. \tool can be easily extended to similar APIs by adding them to a
configurable JSON file. \Comment{For example, to support XGBoost’s
\texttt{XGBClassifier} in \tool, we simply add the \texttt{XGBClassifier} API to
the configuration file. Since \texttt{XGBClassifier()} returns a scikit-learn
compatible model, which \tool already supports, it can automatically recognize
the output type and generate model architecture assertions without further
changes.}
% \Fix{add how we can extend the set of APIs}
% Additionally, a random seed is generated and used to set or replace seeds in APIs that have non-deterministic behavior. Examples include \verb|train_test_split|, models imported or instantiated from \verb|sklearn| or \verb|tensorflow|, etc.. 
\Comment{This step ensures that we capture the variance in the instrumented
values and that our assertions remain valid even if the seeds are changed by the
developer(s).} \Comment{
\begin{jupytercode}{1}
import pandas as pd
import numpy as np

# Create a sample DataFrame
df = pd.read_csv('input.txt')
print(df)
\end{jupytercode}
\begin{jupyteroutput}{1}
Hello world
\end{jupyteroutput}
}
%\Fix{@vedant: can we say how many APIs of each kind we support for each library?}

\empara{(2) Assertion Generation}\label{sec:assertgen} For detected \entities in
\propfinder, \assertgen first instruments the notebook
(Line~\ref{alg:line:instrument}), executes the instrumented notebook for $n$
iterations and collects the set of outputs ($v_i$, $l_i$, $\tau_i$, $s_i$), where
$s_i$ is the set of $n$ values collected for variable $v_i$ across the $n$ runs
(Line~\ref{alg:line:runcollect}).
% The notebook generated by \propfinder, comprising of instrumentation APIs, is executed by the \nbrunner multiple times. This leads to collection of a range of values for each property, and is used to determine the type of check and tolerance bounds needed for testing the same. 
For each set of collected values $s_i$, \assertgen computes a tolerance bound
$\epsilon_i$ (Line~\ref{alg:line:compute_tolerance}) based on the confidence
level $c$. Based on the \entity type $\tau_i$, the tolerance bound
$\epsilon_i$, the checked variable $v_i$ and the corresponding collected values
$s_i$, \assertgen generates an assertion
(Line~\ref{alg:line:generate_assertion}) to check the integrity of the \entity.
Finally, it inserts the generated assertion in the original notebook
(Line~\ref{alg:line:insert_assertion}).

Specifically, Line~\ref{alg:line:generate_assertion} generates one or more assertions to check the
\emph{integrity} of each \entity as follows:
%. Listed below are the checks that are added for each class of property:
\begin{itemize}[leftmargin=*,topsep=0ex]
    \item\textbf{\datasetAssert}: For each dataset variable, it inserts
    assertions to check properties such as the dataset's size, column names and
    types, and the mean and variance of numeric columns.     
    \item\textbf{\modelArchAssert}: For model definitions, it inserts assertions to validate the architecture, including the number and types of layers, as well as associated hyperparameters.
    % extract and check initialization parameters.     
    % For neural networks, we extract their configuration and architecture, i.e., type and sequence of their constituent layers. We add assertions checking for all these attributes.
    \item\textbf{\modelPerfAssert}: It inserts assertions to check whether
    evaluation metric values fall within an expected range. Given the inherent
    randomness in ML pipelines, we use approximate assertions (e.g.,
    \texttt{assert\_allclose})  to capture expected variance in the
    results.
    %to allow for small numerical deviations in the     results.
\end{itemize}

\mypara{Assertion Bound Computation} 
Chebyshev's inequality~\cite{chebyshev1867valeurs} is a well-known technique in
probability theory that guarantees that no more than a certain proportion of
values will exceed a certain distance from the mean. Therefore, for any \entity
with variance in outputs (such as \modelPerfAssert), we use Chebyshev's
Concentration Inequality to compute a tolerance bound for the assertion. Given
the confidence level $c$, Equation~\ref{eq:chebyshev} shows how to compute a
value such that the probability of the \entity exceeding it is at most $1-c$, where
$X$ is a scalar random variable with a finite mean $\mu$ and variance
$\sigma^2$.
\begin{equation}
    \vspace{-0.1in}
\label{eq:chebyshev}
P_r(|X - \mu| \geq k\sigma) \leq 1 - c,       (k>0)
\end{equation}

Chebyshev's Inequality is a conservative choice when the underlying distribution
of the data is unknown. If the distribution is known, more precise bounds can be
computed~\cite{dutta2021flex}.
% Since the underlying distribution of the data is
% unknown~\cite{dutta2021flex}, we choose a more conservative bound to reduce the
% chances of flaky failures.

\mypara{Test Oracle} Because property-based oracles are scarce in the ML domain,
in this work, we rely on \emph{regression-based oracles} to check for
correctness. 
%So, we use the data from the runs of the instrumented notebook as
%expected values in the assertions we generate.
%instrument the current version of the notebook to collect the values of program variables and use the collected values as test oracles in the generated assertions.
Naturally, this approach assumes that the current version of the notebook is
correct, which may not be true. However, regression-based oracles can help
document the developer's assumptions, as well as capture the current state of
the notebook, and their violation can indicate regressions or reveal incorrect
assumptions in the future. Regression oracles have also been proposed in prior
test generation
techniques~\cite{xie2006augmenting,fraser2011evosuite,lukasczyk2022pynguin,liu2023extracting}.
Further, developers can also modify the generated assertions to use manual
oracles.

% $$
% P_r(|X - \mu| \geq k\sigma) \leq \frac{1}{k^2}
% $$
% So, given the samples of a \entity, we use Chebyshev’s Inequality to compute a value such that the probability of the \entity exceeding it at most $1-C$ (which is provided by the developer). We use this value to set the absolute or relative tolerance value in the assertion. This approach is most commonly needed for \entities involving evaluation metrics.

% \Fix{@vedant: add an example code with all kinds of assertions}
\Comment{
\begin{lstlisting}[language=Python, caption={Examples of generated assertions}, label={lst:gen-examples}]
import pandas as pd
import matplotlib.pyplot as plt
df = pd.read_csv('./heart_failure.csv')
nbtest.assert_df_var(df, 5655280085.135918, atol=0.0, test_id='4')
nbtest.assert_df_mean(df, 20331.504464625676, atol=0.0, test_id='3')
nbtest.assert_column_types(df, ['int64', 'float64', 'int64', 'int64', 'int64', 'int64', 'int64', 'float64', 'float64', 'int64'], test_id='2')
nbtest.assert_column_names(df, ['DEATH_EVENT', 'age', 'anaemia', 'creatinine_phosphokinase', 'diabetes', 'ejection_fraction', 'high_blood_pressure', 'platelets', 'serum_creatinine', 'serum_sodium'], test_id='1')
nbtest.assert_shape(df, (299, 13), test_id='0')

_____

from sklearn.linear_model import LogisticRegression
logr = LogisticRegression(random_state=0)
nbtest_tmpvar_4 = logr.fit(X_train, y_train)
nbtest.assert_sklearn_model(nbtest_tmpvar_4, '/home/usr/sample/output/sklearn_model_17.pkl', test_id='17')
y_pred = logr.predict(X_test)
cm = confusion_matrix(y_test, y_pred, labels=logr.classes_)
print('accuracy: ', accuracy_score(y_test, y_pred))
nbtest.assert_allclose(
    accuracy_score(y_test, y_pred), 0.85,
    atol=0.16329931618554516, test_id='19'
)

_____

model = keras.Sequential([
    layers.Dense(512, activation='relu', input_shape=[train_shape]), layers.Dense(512, activation='relu'), layers.Dense(512, activation='relu'), layers.Dense(512, activation='relu'), layers.Dense(10, activation='sigmoid')
])

nbtest.assert_nn_model(model, [
    ('Dense', (None, 512), 401920), ('Dense', (None, 512), 262656), 
    ('Dense', (None, 512), 262656), ('Dense', (None, 512), 262656), 
    ('Dense', (None, 10), 5130)], test_id='35'
)

history = model.fit(
    X_train, y_train, validation_data=(X_valid, y_valid), batch_size=512, epochs=100)

nbtest.assert_allclose(
    history.history['val_accuracy'][-1], 0.9743809461593628, atol=0.01265448871819571, test_id='39'
)
nbtest.assert_allclose(
    history.history['val_loss'][-1], 0.11317245960235596, 
    atol=0.04637266056347483, test_id='38'
)
nbtest.assert_allclose(
    history.history['accuracy'][-1], 0.9961130976676941, atol=0.0033877620540605853, test_id='37'
)
nbtest.assert_allclose(
    history.history['loss'][-1], 0.011598876677453518, 
    atol=0.006875411967107568, test_id='36'
)
\end{lstlisting}
\Fix{cut this into a 2-3 very small snippets}
}

%the Chebyshev’s Inequality is applied in \tool's Automated Assertion Generation to provide a conservative estimate for the tolerance bounds of each varying numerical property. 

% If a property varies between executions, as is the case for evaluation metrics, it's tolerance bounds are estimated using Chebyshev's Concentration Inequality, which is described below. 

%\mypara{Chebyshev's Inequality}
% Chebyshev’s inequality is a well-known statement in probability theory that guarantees that, for a broad class of distributions, no more than a certain proportion of its values will exceed a certain distance from the mean. More formally, let $X$ be a scalar random variable with a finite mean $\mu$ and variance $\sigma^2$. Then for any real number $k > 0$: 

% $$
% P_r(|X - \mu| \geq k\sigma) \leq \frac{1}{k^2}
% $$
\section{Regression Testing Framework}
\label{sec:framework}

%  Deepchecks~\cite{chorev2022deepchecks} is a tool that provides a
% library of tests for checking model and data integrity properties in ML
% pipelines. However, the developers still need to manually incorporate such
% checks, and the framework does not allow them to easily write custom tests.
% Deequ~\cite{schelter2018automating} is another tool that is designed for data
% validation in production pipelines. Testbook~\cite{testbook} allows developer to
% write unit tests in a separate file and potentially execute a subset of cells.
% However, none of these tools can 1) automatically generate tests, 2) account for
% non-determinism, or 3) allow non-intrusive testing within a Jupyter session. 

We developed a
regression testing framework, \tool, for Jupyter Notebooks that consists of
three components: (1) \textbf{\toolpylib}: a Python library that provides APIs
for writing \cellscoped assertions in Jupyter Notebooks. It works as a
\texttt{pytest} plugin that collects these assertions like standard unit tests;
(2) \textbf{\toolgen}: a Python library that automatically generates \cellscoped
assertions for a Jupyter Notebook by identifying commonly used ML-related APIs (implementing the technique discussed in
\S~\ref{sec:assertgen}); and (3) \textbf{\tooljupyterplugin}: a JupyterLab plugin
for running \tool's assertions in Jupyter Notebooks. 
Users can install these three components independently via \textit{pip}.
We describe the design of
\toolpylib and \tooljupyterplugin in \S~\ref{sec:framework-lib} and
\S~\ref{sec:framework-jupyter} respectively.

\Comment{
\begin{figure*}
\begin{subfigure}{.4\textwidth}
  \centering
  \includegraphics[width=1\linewidth]{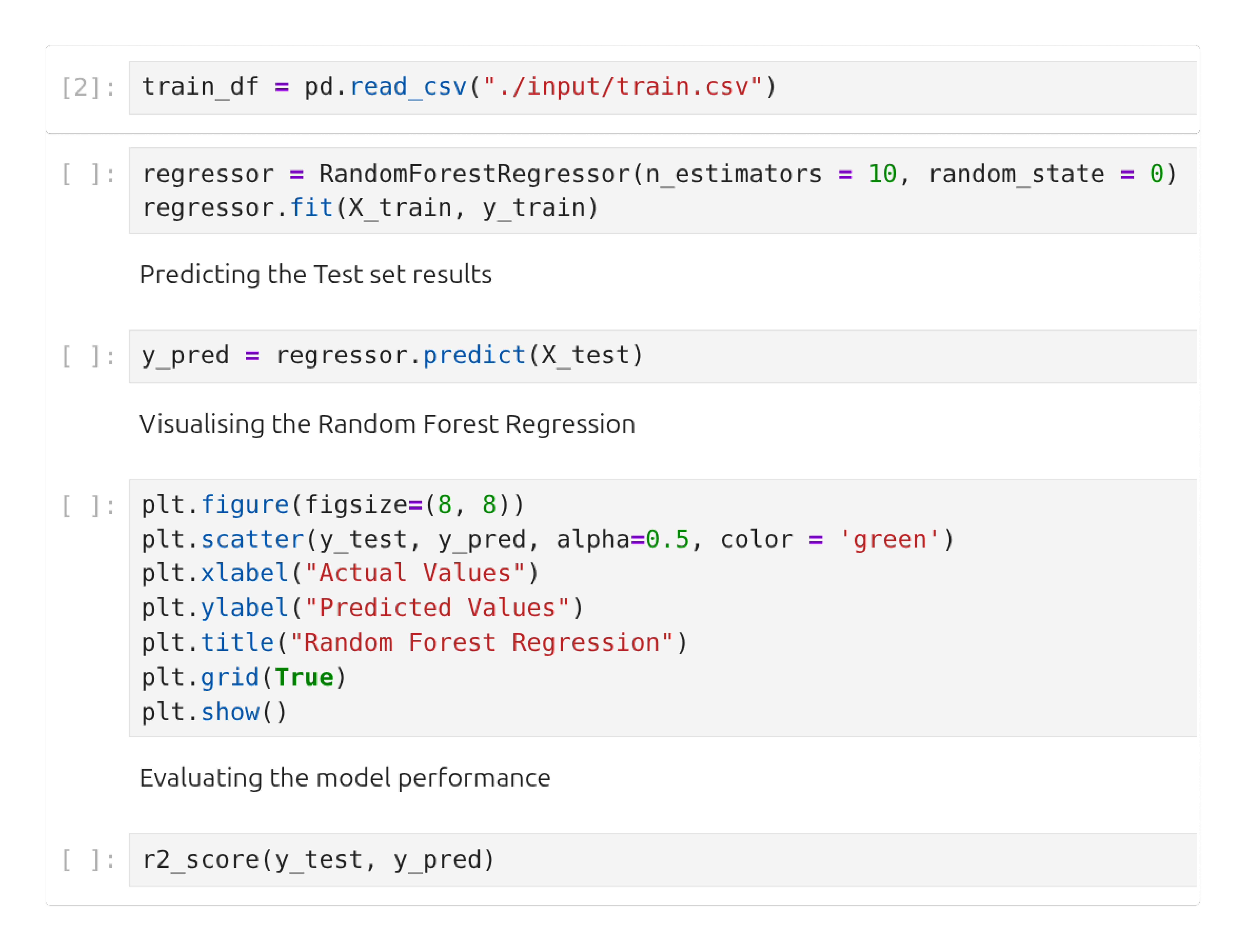}
  \caption{Input Notebook}
  \label{fig:sfig1}
\end{subfigure}%
\begin{subfigure}{.6\textwidth}
  \centering
  \includegraphics[width=1\linewidth]{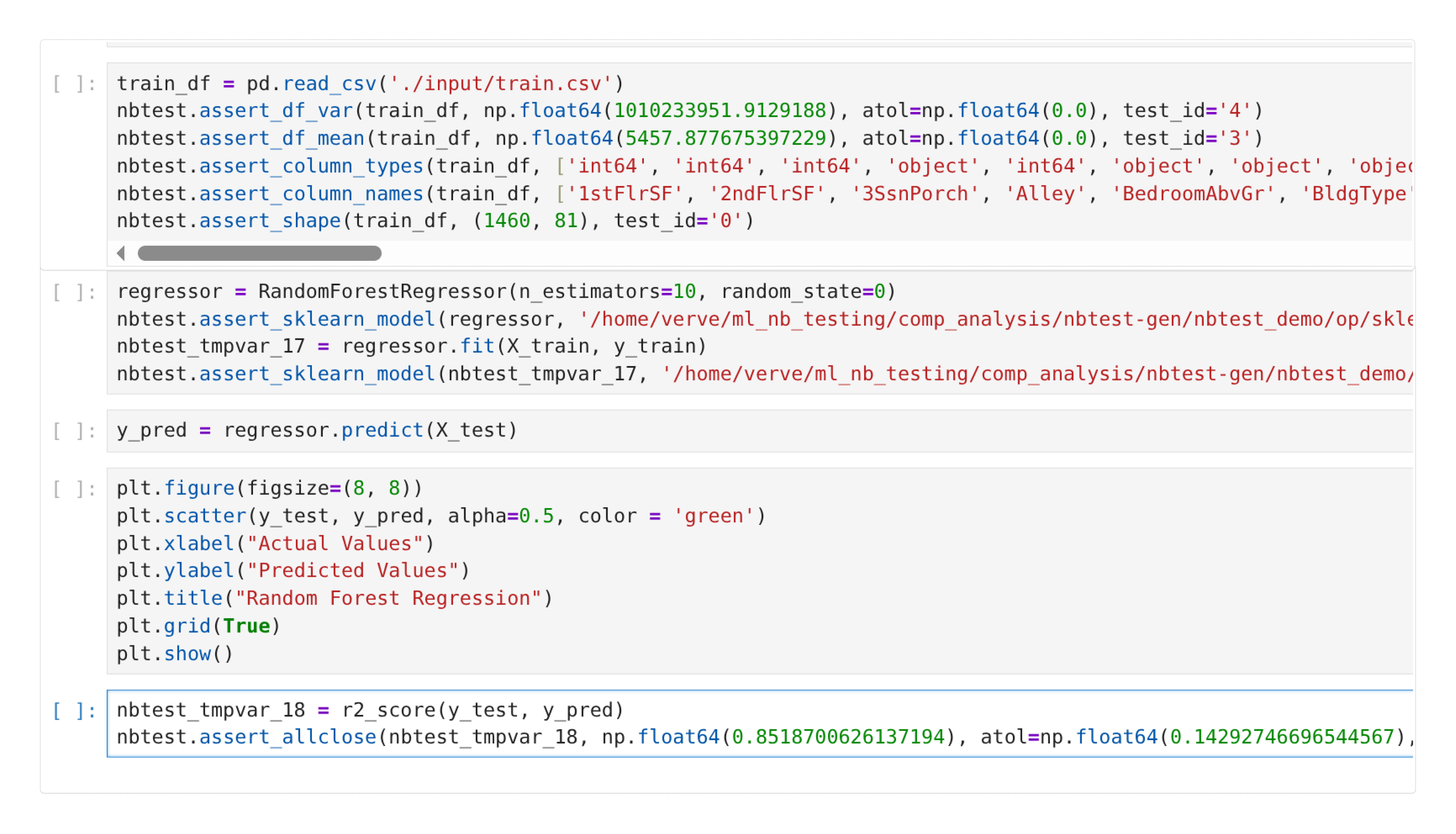}
  \caption{Generated Assertions}
  \label{fig:sfig2}
\end{subfigure}
\begin{subfigure}{.5\textwidth}
  \centering
  \includegraphics[width=1\linewidth]{figures/nbtest_jupyter.png}
  \caption{Executing assertions in Jupyter}
  \label{fig:sfig2}
\end{subfigure}
\begin{subfigure}{.5\textwidth}
  \centering
  \includegraphics[width=1\linewidth]{figures/nbtest_pytest.png}
  \caption{Executing assertions using pytest}
  \label{fig:sfig2}
\end{subfigure}
\caption{Overview of NBTest workflow}
\label{fig:nbtest-workflow}
\end{figure*}
}
\Comment{
\subsection{Design Principles of \tool}
% Refer to Inline tests Section 3.1
We developed \tool based on the following design principles:
%\tool frameworks should meet this minimum set of requirements:
\begin{enumerate}[label=(\arabic*),leftmargin=*,itemsep=0ex,topsep=0ex]
    \item The framework should allow developers to write expressive cellscoped assertions. \textcolor{blue}{\checkmark}
    \item It should be possible to run such
    assertions with \texttt{pytest} and CI. \textcolor{blue}{\checkmark}
    \item The assertions should be non-intrusive; they should not
    disrupt the interactive nature of notebooks or clutter the notebook. \textcolor{blue}{\checkmark} 
    \item It should serve as a plugin in the IDE (e.g., Jupyter Lab). The developers can choose to turn on and off the assertions as they want. \textcolor{blue}{\checkmark}
    \item The framework should automatically generate assertions for a given ML notebook to assist developers. \textcolor{blue}{\checkmark}
    \item When enabled, the overhead of assertions should be low.  \textcolor{blue}{\checkmark}
    %\item When disabled, developers should not be interrupted by the assertions.  \textcolor{blue}{\checkmark}
    \item The \cellscoped assertions should account for randomness in machine learning code to mitigate flakiness. \textcolor{blue}{\checkmark}
    \item The assertion APIs should follow common assertion patterns for the ease of adoption. \textcolor{blue}{\checkmark}
    % The cell-level assertion API should be similar to common assertions ML developers are used to, for easy adoption. \textcolor{blue}{\checkmark}
    %\item fail-safe execution: continue to execute notebook cells even if assertions fail so that all failures can be collected. \textcolor{blue}{\checkmark}
    \item It should identify developers' intentions and only generate a few but relevant assertions.  \textcolor{red}{\xmark}
    % \item It should allow developers to run a subset of tests when needed.
    % \textcolor{red}{\xmark}
    %\item \tool should provide an intuitive UI for users to select a subset of generated assertions. \textcolor{red}{\xmark}
    \item It should allow developers to automatically only re-run cell-level
    assertions affected by a change (Evolution-Aware). \textcolor{red}{\xmark}   
    \item The assertions should be environment or configuration-aware, e.g.,
    different assertions should be enabled for different versions of PyTorch.
    \textcolor{red}{\xmark}   
    %\item The tool should allow checking for and isolating side-effects such as memory usage, disk usage, and networking. \textcolor{red}{\xmark}   
\end{enumerate}
\tool currently supports all features marked as  \textcolor{blue}{\checkmark}
via its suite of three tools, but does not support features marked as
\textcolor{red}{\xmark}. \tool is not evolution-aware, so it always runs all
assertions. \tool is based on regression oracles, or requires developers to
provide a test oracle when manually writing assertions. These principles are
based on our experience and understanding of Machine Learning and the Notebook
environment, and prior studies~\cite{chattopadhyay2020s,huang2025scientists}. In
the future, we want to evolve \tool in collaboration with ML developers. 
% To this end, 
We have started reaching out to ML project developers, have received
some positive engagements, and integrated \tool into the CI of one ML
project.
}

% \Fix{shorten this to one para:}

\Comment{
\sd{move to appendix if needed}
\mypara{Assumption of Linear Execution} \tool assumes linear execution
of notebooks when generating assertions, which we consider both valid and
practical. Mainstream open-source projects already test
notebooks in CI by executing them
sequentially~\cite{linear_order_1}~\cite{linear_order_2}~\cite{linear_order_3},
and Jupyter Notebook itself provides no mechanisms for custom execution orders.
\Comment{Likewise, NBval~\cite{fangohr2020testing}, a popular pytest plugin for testing
notebooks, and Kaggle platform both assume linear execution implicitly.
%  also assumes linear execution implicitly.
% The Kaggle platform follows the same assumption, and thus provides no metadata about the execution order.
In practice, users often linearize notebook cell order before sharing or
rerunning, to improve reproducibility, and tools like ~\cite{head2019managing}
can further help with cleaning and reordering notebooks.} However, once
the assertions are generated, they can be executed in any order in a session
using our Jupyter plugin when developers may edit individual cells.
}
%  We believe this is a valid and practical assumption for the following reasons:}
% \begin{enumerate}
%     \item \elaine{We find that many mainstream open-source projects currently test notebooks in CI by executing them in a linear order ~\cite{linear_order_1}~\cite{linear_order_2}~\cite{linear_order_3}. Further, Jupyter Notebook lacks a built-in mechanism to specify a custom execution order. Similarly, NBval~\cite{fangohr2020testing}, a popular pytest plugin that tests notebooks by checking stored input/output consistency, also assumes linear execution implicitly.}
%     \item \elaine{Kaggle platform assumes linear execution. Hence, Kaggle notebooks include no information about execution order. Therefore, when notebooks fail to run during our dataset curation process, we cannot determine whether the failure is caused by execution order or other factors. In practice, users often linearize notebook cell order before sharing or rerunning, to improve reproducibility and avoid issues from inconsistent execution order. This linearization is considered as good practice and tools like ~\cite{head2019managing} can help with cleaning and reordering notebooks.}
% \end{enumerate}

\Comment{
\subsection{Overview}
Our \tool framework provides three distinct tools: 
\begin{enumerate}[label=(\arabic*),leftmargin=*,itemsep=0ex,topsep=0ex]
    \item \textbf{Python library (\toolpylib):} provides APIs for writing \cellscoped assertions in Jupyter Notebooks 
    and works as a \texttt{pytest} plugin that collect these assertions like standard unit tests.
    \item \textbf{Assertion generator (\toolgen):} a tool that automatically generates \cellscoped assertions for a Jupyter Notebook by identifying commonly used machine learning-related APIs.
    \item \textbf{Jupyter plugin (\tooljupyterplugin):} 
    a JupyterLab plugin for evaluating \tool assertions.
    Assertions can be shown or hidden to improve notebook readability, 
    as well as enabled/disabled during execution.
    When executed, each assertion displays its result inline in the notebook.
    % a plugin for Jupyter Notebooks that allows users to enable/disable \tool's assertions during an active session to allow users to validate their changes without executing the entire notebook. The extension provides real-time feedback on assertion status, highlighting cells with assertions, and immediately shows the assertion results in the notebook.
    % The extension integrates with JupyterLab's cell execution mechanism to provide real-time feedback on assertion status. Developers can execute individual cells containing assertions and immediately see the results, facilitating rapid iteration during development. The extension visually highlights notebook cells containing \tool assertions. These cells are marked with a distinct left border and a subtle amber background, providing a clear visual distinction from regular code cells.
\end{enumerate}
%We next discuss the design and implementation of \toolpylib and \toolgen.
}
% \Comment{
% }
\subsection{\tool Python Library (\toolpylib)}
\label{sec:framework-lib}
We support two categories of assertions, and 11 assertions in total.
\begin{enumerate}[label=(\arabic*),leftmargin=*,topsep=0ex]
    \item\textbf{Generic Assertions:} \toolpylib provides standard assertions
    including \texttt{assert\_equal}, \texttt{assert\_allclose},
    \texttt{assert\_true}, and \texttt{assert\_false}. These functions follow
    the familiar semantics of traditional unit testing frameworks but are
    tailored for execution within notebooks. Users can also use these assertions
    to check model architecture and model performance properties.
    % are specifically designed to work
    % within the notebook execution environment.
    \item\textbf{Data-related Assertions:} Because data processing is common in
    machine learning workflows, \toolpylib includes specialized assertions for
    validating properties of data represented using \texttt{pandas} or \texttt{numpy}. 
    The \texttt{assert\_shape} function checks the expected shape of the dataset.
    The \texttt{assert\_df\_var} and
    \texttt{assert\_df\_mean} functions check the statistical properties of
    dataframes while handling missing values appropriately. The
    \texttt{assert\_column\_types} and \texttt{assert\_column\_names} assertions
    verify the structural integrity of data, ensuring that schema changes or
    data loading errors are detected early in the development process.
    The \texttt{assert\_in} assertion verifies the presence of a column name in the dataset.
    The \texttt{assert\_no\_class\_balance} assertion checks if the label distribution in the dataset is overly skewed towards one class.
    
    % \item\textbf{Model-related Assertions:} Models used in ML pipelines 
    % are important for the performance, therefore, NBTest also provides assertions for checking model architectures.
    % The \texttt{assert\_sklearn\_model} and \texttt{assert\_nn\_model} functions check if the model architectures (e.g., the layer type and size, activation function, etc) defined in \texttt{sklearn} and \texttt{TensorFlow} respectively, are aligned with the expected value.
  
\end{enumerate}

We designed \toolpylib's APIs to mirror common assertions such as
Python's \texttt{assert} and Numpy's \texttt{assert\_allclose}. 
Unlike those assertions, \toolpylib's assertions are inactive during normal execution, 
to not interrupt the development process.
% The complete API reference is in Appendix \S A. 
\toolpylib integrates with \texttt{pytest} through a custom plugin.
% \tool's assertions can be run with Pytest. 
% \tool is designed as a Pytest plugin that with test discovery and execution
% capabilities. 
% \tool with Pytest. The integration with pytest is achieved through a
% custom plugin architecture that extends pytest's test discovery and execution
% capabilities.
When invoked with the \texttt{--nbtest} flag, \texttt{pytest}
parses the Jupyter Notebook, locates cells with
\tool assertions, and executes them as tests. 
% So, once developers finish their edits, they can validate their
% changes by running the \texttt{pytest --nbtest [path to notebook]} command.  
Using \texttt{pytest}'s collection hooks~\cite{pytesthooks}, each assertion is
registered as an individual test with a unique identifier.
%This identifier allows developers to pinpoint exactly which assertions pass or fail.
\tool executes all notebook cells regardless of execution failures,
and then reports results.

\subsection{JupyterLab Plugin (\tooljupyterplugin)}
\label{sec:framework-jupyter}
We developed a JupyterLab plugin (\textit{\tooljupyterplugin}) that
provides two key features: (1) toggling the visibility of assertions via the
\texttt{Hide Assertion Editor} button (Figure~\ref{fig:overview_a}), improving
readability by hiding assertions in a separate panel, and (2) enabling or
disabling assertion execution via the \texttt{\tool Asserts: ON} toggle. When
enabled, assertions are executed \emph{with the cell}, allowing developers to
quickly validate local changes without exiting the notebook while not being
interrupted by failures during normal development.
% ORIGINAL:
% We also developed a JupyterLab plugin (\textit{\tooljupyterplugin}) that
% provides two key features: (1) toggling the visibility of assertions, and (2)
% enabling or disabling assertion execution. Within JupyterLab, \tool
% assertions are displayed in a separate panel on the right
% (Figure~\ref{fig:overview_a}). When developers wish to focus on notebook
% development without being distracted by assertions, they can hide this panel
% using the \texttt{Hide Assertion Editor} button (Figure~\ref{fig:overview_a}),
% thereby avoiding clutter and improving readability. Furthermore, the plugin
% allows users to control assertion execution via the \texttt{\tool Asserts: ON}
% \space toggle (Figure~\ref{fig:overview_a}). When enabled, the assertions
% are executed \emph{with the cell}, allowing developers to quickly validate their
% local cell changes without exiting the notebook. A bottom panel will
% show results of assertion execution (the green panel in
% Figure~\ref{fig:overview_a} bottom indicates that all assertions passed).
% This way, developers can make edits without accidentally triggering the test
% failure, but validate the changes by executing related assertions \emph{only}.

\mypara{Comparison with Existing Tools}
There are several existing tools for testing Jupyter Notebooks or ML pipelines.
For instance, nbval~\cite{fangohr2020testing}
% \elaine{(deployed in \nbvalRepo repos)} 
performs regression testing by doing a textual comparison of a notebook's
current cell outputs against saved values from prior executions. 
% However, its checks are not robust to
% non-deterministic results (common in ML pipelines) or fluctuations in the output
% string format (especially when considering numerical outputs). Further, it can
% only check outputs printed in cells. 
Deepchecks~\cite{chorev2022deepchecks}
% \elaine{(deployed in \deepchecksRepo repos)} 
and Deequ~\cite{schelter2018automating}
% \elaine{(deployed in \deepquRepo repos)} 
provide libraries of model and data
integrity checks for ML pipelines, but they require developers to manually
incorporate such checks. Testbook~\cite{testbook}
% \elaine{(deployed in \testbookRepo repos)} 
allows developers to write unit
tests in a separate file. % and potentially execute a subset of cells. 
However, none of these tools can (1) automatically generate ML-specific tests
like \toolgen (except \nbval which can perform basic output checking), (2)
account for non-determinism like \toolgen, or (3) allow non-intrusive and
\cellscoped testing within a Jupyter session like \tooljupyterplugin. We provide
a comparison with \nbval in \S~\ref{sec:rq_nbval}. In the future, we plan to
extend \toolgen to generate assertions using
Deepchecks' and Deequ's APIs.

% Our plugin extends pytest's collection hooks \cite{pytesthooks} to traverse notebook
% cells and extract assertion statements. Each assertion is registered as an
% individual test item with a unique identifier (e.g., \texttt{nbtest\_id\_0\_5})
% that corresponds to its location within the notebook. This approach
% allows developers to identify precisely which assertions pass or fail during
% test execution.  By default, \tool does not stop executing the notebook after a test failure. Instead, it executes all the notebook cells, collects all passes and failures, and generates a final test report.

%During test execution, the plugin creates an isolated execution
%environment for each notebook cell containing assertions.
% The notebook's execution state is preserved across cells, ensuring that
% variables and imports from earlier cells remain available to later assertions.
% This approach maintains the natural flow of notebook execution while enabling
% systematic testing. \Comment{Listing~\ref{lst:nbtestoutput} shows an example output of
% running \tool pytest plugin.}
%To preserve the semantics of notebook workflows, the plugin maintains the notebook’s execution state across cells. Thus, variables, imports, and intermediate computations from earlier cells remain accessible to subsequent assertions.

%\begin{figure}
\Comment{
\begin{lstlisting}[language=bash, caption=pytest Integration Example,label={lst:nbtestoutput}]
# Execute notebook normally (assertions are not run)
$ jupyter execute example.ipynb --output=run.ipynb
[NbClientApp] Executing example.ipynb
[NbClientApp] Save executed results to run.ipynb

# Run tests using pytest with NBTest plugin
$ pytest -v --nbtest example/nbtest_plugin_example.ipynb
============== test session starts ===============
platform linux -- Python 3.11.11, pytest-8.3.4
plugins: nbtest-0.1.0, anyio-4.6.2
collected 2 items

example/nbtest_plugin_example.ipynb::nbtest_id_0_5 PASSED  [ 50%]
example/nbtest_plugin_example.ipynb::nbtest_id_0_6 FAILED  [100%]

================== FAILURES ===================
____ example/nbtest_plugin_example.ipynb::Cell 0 ____
Assertion failed
Cell 0: Assertion error
AssertionError: False is not true
===============================================
\end{lstlisting}
%\caption{Example output of \tool}
%\label{lst:nbtestoutput}
%\end{figure}
}
\Comment{
\mypara{Error Handling and Reporting}
%NBTest provides comprehensive error reporting that bridges the gap between
%notebook-style development and traditional testing frameworks. 
When assertions fail, the plugin captures the full execution context, including variable states,
stack traces, and cell inputs. This information is formatted according to
pytest's reporting standards, providing developers with familiar error messages
while preserving notebook-specific context.
By default, \tool does not stop executing the notebook after a test failure. Instead, it executes all the notebook cells, collects all passes and failures, and generates a final test report.
}
%The framework handles notebook
%execution errors gracefully, distinguishing between assertion failures and other
%runtime exceptions. %This distinction is crucial in machine learning contexts
%where code may contain intentional randomness or depend on external data sources
%that may occasionally be unavailable.

% The NBTest framework is implemented as a collection of three interconnected Python packages that provide comprehensive testing capabilities for Jupyter Notebooks. This section details the implementation of the core library components and their integration with the pytest testing framework.

% \subsubsection{Architecture Overview}

% NBTest consists of three main components: (1) \texttt{nbtest-plugin}, a pytest plugin that enables test discovery and execution within notebooks, (2) \texttt{nbtest-gen}, a command-line tool for automated assertion generation, and (3) \texttt{nbtest-lab-extension}, a JupyterLab extension for interactive test management. These components work together to provide a seamless testing experience that integrates with existing development workflows.

\Comment{
\subsubsection{Core Assertion Library}
The \texttt{nbtest-plugin} package provides the foundational assertion functions that developers can embed directly within notebook cells. The library implements two categories of assertions to address the specific testing needs of machine learning workflows.
\textbf{Generic Assertions:} The framework provides standard assertion functions including \texttt{assert\_equal}, \texttt{assert\_allclose}, \texttt{assert\_true}, and \texttt{assert\_false}. These functions follow the familiar semantics of traditional unit testing frameworks but are specifically designed to work within the notebook execution environment. Each assertion includes metadata tracking capabilities that enable pytest integration and provide detailed failure reporting.

\begin{lstlisting}[language=Python, caption=Generic Assertions Usage Example]
import nbtest
import math
import numpy as np

# Test exact equality
nbtest.assert_equal(round(math.pi, 2), 3.14)

# Test boolean conditions
nbtest.assert_true(math.pi > 3)
nbtest.assert_false(math.pi == 3)

# Test numerical closeness with tolerance
nbtest.assert_allclose(np.sin(np.pi/2), 1.0, atol=1e-10)
\end{lstlisting}

\textbf{DataFrame-Specific Assertions:} Recognizing that data manipulation is central to machine learning workflows, NBTest includes specialized assertions for pandas DataFrame objects. The \texttt{assert\_nanvar} and \texttt{assert\_nanmean} functions test statistical properties of dataframes while handling missing values appropriately. The \texttt{assert\_column\_types} and \texttt{assert\_column\_names} assertions verify the structural integrity of data, ensuring that schema changes or data loading errors are detected early in the development process.

\begin{lstlisting}[language=Python, caption=DataFrame Assertions Usage Example]
import pandas as pd
import nbtest

# Load data
df = pd.read_csv("./heart_failure.csv")

# Test DataFrame shape
nbtest.assert_equal(df.shape, (299, 13))

# Test column names
nbtest.assert_column_names(df, ['DEATH_EVENT', 'age', 'anaemia', 
    'creatinine_phosphokinase', 'diabetes', 'ejection_fraction', 
    'high_blood_pressure', 'platelets', 'serum_creatinine', 
    'serum_sodium', 'sex', 'smoking', 'time'])

# Test column data types
nbtest.assert_column_types(df, ['int64', 'float64', 'int64', 
    'int64', 'int64', 'int64', 'int64', 'float64', 'float64', 
    'int64', 'int64', 'int64', 'int64'])

# Test statistical properties
nbtest.assert_nanmean(df, 20331.504464625676, atol=0.0)
nbtest.assert_nanvar(df, 5655280085.135918, atol=0.0)
\end{lstlisting}
}

\Comment{
\subsubsection{pytest Integration Mechanism}

The integration with pytest is achieved through a custom plugin architecture
that extends pytest's test discovery and execution capabilities. When pytest is
invoked with the \texttt{--nbtest} flag, the plugin parses Jupyter Notebook
files (\texttt{.ipynb}) and identifies cells containing NBTest assertions.

\begin{lstlisting}[language=bash, caption=pytest Integration Example]
# Execute notebook normally (assertions are no-ops)
$ jupyter execute example.ipynb --output=run.ipynb
[NbClientApp] Executing example.ipynb
[NbClientApp] Save executed results to run.ipynb

# Run tests using pytest with NBTest plugin
$ pytest -v --nbtest example/nbtest_plugin_example.ipynb
============== test session starts ===============
platform linux -- Python 3.11.11, pytest-8.3.4
plugins: nbtest-0.1.0, anyio-4.6.2
collected 2 items

example/nbtest_plugin_example.ipynb::nbtest_id_0_5 PASSED  [ 50%]
example/nbtest_plugin_example.ipynb::nbtest_id_0_6 FAILED  [100%]

================== FAILURES ===================
_______ example/nbtest_plugin_example.ipynb::Cell 0 _______
Assertion failed
Cell 0: Assertion error
AssertionError: False is not true
===============================================
\end{lstlisting}

The plugin implements pytest's collection hooks to traverse notebook cells and
extract assertion statements. Each assertion is registered as an individual test
item with a unique identifier (e.g., \texttt{nbtest\_id\_0\_5}) that corresponds
to its location within the notebook. This granular approach allows developers to
identify precisely which assertions pass or fail during test execution.

During test execution, the plugin creates an isolated execution environment for
each notebook cell containing assertions. The notebook's execution state is
preserved across cells, ensuring that variables and imports from earlier cells
remain available to later assertions. This approach maintains the natural flow
of notebook execution while enabling systematic testing. }

\Comment{
\subsection{JupyterLab Extension (\tooljupyterplugin)}
The \texttt{\tooljupyterplugin} plugin provides support for \tool's assertions within the JupyterLab environment~\cite{jupyterlab}, which is commonly used to run notebooks in the browser. %Built using TypeScript and the JupyterLab
%extension framework,
The plugin allows developers to enable or disable
assertions during development without modifying the notebook code so that they can execute the assertions (within the session) once they are satisfied with their edits, but disable them otherwise.

The extension integrates with JupyterLab's cell execution mechanism to provide real-time feedback on assertion status. Developers can execute individual cells containing assertions and immediately see the results, facilitating rapid iteration during development. The extension visually highlights notebook cells containing \tool assertions. These cells are marked with a distinct left border and a subtle amber background, providing a clear visual distinction from regular code cells.
}
\Comment{
\begin{figure}[H]
    \centering
    \includegraphics[width=\linewidth]{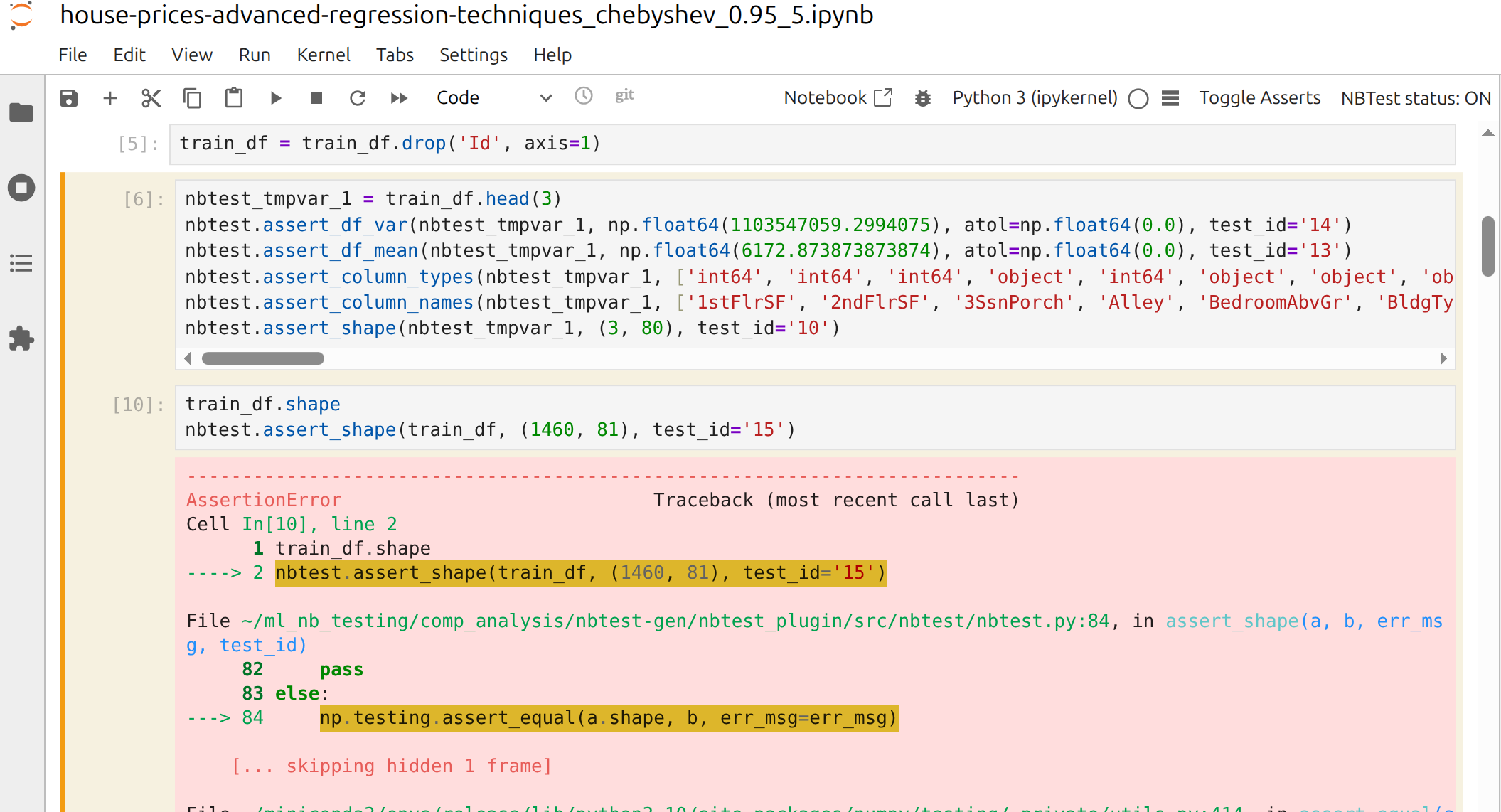}
    \caption{Using \texttt{\tooljupyterplugin} to execute assertions in a cell}
    \label{fig:extension}
\end{figure}
}

\section{Evaluation}
% \Fix{convert all kaggle links to references}

\newcommand{\rqassertgen}{How many assertions does \toolgen generate? How does their passing rate change with different confidence thresholds?\xspace}
\newcommand{\rqassertgennum}{RQ1\xspace}

\newcommand{\rqmutation}{How effective are the generated assertions in detecting domain-specific mutations?\xspace}
\newcommand{\rqmutationnum}{RQ2\xspace}
\newcommand{\rqversions}{How effective are the generated assertions in detecting real-world regressions in ML notebooks?\xspace}
\newcommand{\rqversionsnum}{RQ3\xspace}
\newcommand{\rqnbval}{How does \toolgen compare with \nbval?\xspace}
\newcommand{\rqnbvalnum}{RQ4\xspace}

%\mypara{Research Questions}
We answer the following research questions in this work:
\begin{itemize}[leftmargin=*,topsep=0ex]
\item \textbf{\rqassertgennum}: \rqassertgen
\item \textbf{\rqmutationnum}: \rqmutation
\item \textbf{\rqversionsnum}: \rqversions
\item \textbf{\rqnbvalnum}: \rqnbval
% \item \textbf{\rqconfigsnum}: \rqconfigs
\end{itemize}

\mypara{Experimental Setup}
We run experiments on a large cluster of diverse machines with GPUs and Ubuntu
20.04 OS. To run each notebook, we set up a Conda environment with Python 3.9
and install the dependencies for the notebook. We run \toolgen with \dynamicRun
iterations and use the Chebyshev's statistical method with the confidence
intervals of $0.5$, $0.7$, $\confLevel$, and $0.999$ during assertion generation
for RQs 1-3. For mutation analysis, we need to balance running many notebooks
and reliably estimating the mutation score. Hence, similar to prior
work~\cite{dutta2020detecting}, we run each notebook 30 times to compute the
mutation score.
% We use Python 3.9 and the python dependencies required by each project to run experiments individually. We run \tool with the number of dynamic iterations to be \dynamicRun and use Chebyshev statistic method with the confidence levels of \confLevel.

\mypara{Implementation Details}
\toolgen uses NBFormat~\cite{nbformat} to parse the notebooks and Python's AST library to
parse Python code in each notebook cell.
Specifically, \toolgen uses AST's~\cite{ast_python} visitor design pattern API to
detect the variables related to the ML-specific \entities, and the transformer
design pattern API to insert the generated assertions back into the original
notebook. 

% \tooljupyterplugin is implemented as a JupyterLab extension in
% \texttt{TypeScript}. It registers two commands and integrates them into the
% notebook toolbar. One command activates the kernel and updates the UI status
% indicator (i.e., \texttt{NBTest Asserts: ON/OFF}). The other toggles assertion
% visibility by storing or removing assertions in the cell metadata (i.e.,
% \texttt{Hide Assertion Editor}). The total lines of code in \tool is $2877$.
% \Fix{what does activate the kernel mean? }
% \Fix{how does it actually execute the assertions?}

\looseness=-1 
\tooljupyterplugin is implemented as a JupyterLab extension in
\texttt{TypeScript}. It registers two commands and integrates them into the
notebook toolbar. The \texttt{Toggle Assertions} command controls assertion
execution. It sends a \texttt{Python} snippet to the notebook's running kernel
that flips an environment variable and updates the \texttt{\tool Asserts:
ON/OFF} toolbar indicator accordingly. When the variable is enabled, \tool
assertions are executed after the cells containing them are executed. The
\texttt{Hide Assertion Editor} command toggles assertion visibility by moving
\tool assertions between a cell's source and its metadata. Assertions stored in
the metadata are not executed. The total lines of code in \tool are $2877$.

\subsection{\rqassertgennum: Passrates of \toolgen's Assertions}

% ------ Table ------

% \begin{wraptable}[8]{r}{0.45\textwidth}
%   \scriptsize
% \begin{wraptable}[6]{r}{0.5\textwidth}

% \end{wraptable}

% \begin{wraptable}[8]{r}{0.5\textwidth}
    \begin{table}[t!]
    \centering
    \tablesize
    %\vspace{-0.1in}
    \caption{\#Assertions generated per notebook by \toolgen}
    \vspace{-0.1in}
    \label{tab:rq1_assert_gen}
    \begin{tabular}{l|r|r}
    \toprule
    \textbf{Type} & \textbf{Total \#Assertions} & \textbf{\#Assertions/Notebook} \\ \midrule
    \datasetAssert &  \datasetAssertionsKaggle & \avgdatasetAssertionsKaggle \\\midrule
    \modelPerfAssert & \modelPerfAssertionsKaggle & \avgmodelPerfAssertionsKaggle \\ \midrule
    \modelArchAssert & \modelArchAssertionsKaggle & \avgmodelArchAssertionsKaggle \\  
    \midrule
    \textbf{Total} & \totalKaggleAssertions & \avgtotalKaggleAssertions \\
    \bottomrule
    \end{tabular}
    \vspace{-0.1in}  
    \end{table}
    % \end{wraptable}

\begin{table}[t!]
    \centering
    \tablesize
    %\vspace{-0.1in}
    \caption{Distribution of Passrates}
    \label{tab:rq1_pass_rate}
    \vspace{-0.1in}
    \begin{tabular}{l|r|r|r|r|r}
    \toprule
            \textbf{Tool} & \textbf{$0$\%} & \textbf{$(0-50]$\%} & \textbf{$(50-100)$\%} & \textbf{$100$\%} & \textbf{Total (Overall)} \\ \midrule
            \tool & \totalKaggleFAILOne & \totalKaggleFAILFiftyOne      &      \totalKaggleFAILFifty  &  \totalKaggleFAILZero  & \totalKaggleAssertions (\totalKagglePASSRate) \\ \midrule
            \nbval & \totalNbvalPASSZeroRev & \totalNbvalPASSZeroFiftyRev    & \totalNbvalPASSFiftyHundredRev  & \totalNbvalPASSHundredRev  & \totalNbvalAssertionsRev (\totalNbvalPASSRateRev) \\
    \bottomrule
    \end{tabular}
    \vspace{-0.2in}
    \end{table}

Table~\ref{tab:rq1_assert_gen} presents the statistics of the assertions
generated by \toolgen for \totalKaggleNotebooks Kaggle
notebooks. Out of \totalKaggleNotebooks, \toolgen generates at
least one assertion (\totalKaggleAssertions in total) for
\totalKaggleNotebookAssertions notebooks. The remaining 50
notebooks did not use any of the ML APIs that \toolgen tracks (as described
in \S~\ref{sec:method_approach}). As a result, \toolgen was unable to generate any
assertions for them. On average, \toolgen generates \avgtotalKaggleAssertions
assertions per notebook, showing that \toolgen's approach of leveraging ML APIs to
track relevant \entities allows it to generate many assertions for a majority of
ML Notebooks.

\toolgen generates the highest number of assertions for datasets
(\datasetAssertionsKaggle). This is because, for each dataframe (a dataset
object loaded using \texttt{pandas.read\_csv}) used in a notebook, \toolgen will
generate five assertions that check the column names, column types, mean and
variance of numeric columns, as well as the shape of the dataframe. \toolgen
generates \modelPerfAssertionsKaggle \modelPerfAssert{} assertions
(\avgmodelPerfAssertionsKaggle on Avg.). Most notebooks consist of (1) one or
more metrics of model performance, such as model accuracy or precision, and (2)
a prediction dataframe that contains predictions of the trained model on test
sets. In the latter case, similar to dataset dataframes, \toolgen generates
multiple assertions for the prediction dataframe, which are considered as
\modelPerfAssert.

The average time \toolgen takes for generating the assertions per
notebook is \avgRuntimeAssertionsKaggle hours when
using \dynamicRun runs per notebook. \toolgen runtime is proportional to the
notebook execution time
% , which may be higher for code using larger models and datasets
and the number of runs. However, like other test generation tools, we expect
\toolgen to be used offline. The generated assertions perform simple property
checks\Comment{without complex computations}, hence they do not add any
significant overhead to notebook execution time. Table~\ref{tab:rq1_pass_rate}
shows the distribution of passrates of assertions generated by \toolgen and \nbval (we compare them in
\S~\ref{sec:rq_nbval}).

% \looseness=-1
% We run the notebooks $30$ times with the \toolgen generated assertions using
% \texttt{pytest}, and record how often each 
% % of \totalKaggleAssertions 
% assertion in the notebooks passes. We categorize the assertions based on their
% passrate: $[0-50]\%$, $(50-100)\%$, and $100\%$. The passrate is computed as
% the ratio of times the assertions pass to the total runs.
% Table~\ref{tab:rq1_pass_rate} shows the results for \toolgen and \nbval (we
% compare them in \S~\ref{sec:rq_nbval}).

We observe that the passrate of all the assertions is \totalKagglePASSRate,
showing that most assertions generated by \toolgen have high reliability. Out of
these, \totalKaggleFAILZero assertions always pass.
\totalKaggleFAILFifty assertions have a passrate between 50\% and 100\%.
%  with an average passrate of \avgPASSRate.
% while \totalKaggleFAILFifty assertions have a passrate of between 50 and 100\%. The average passrate of this \totalKaggleFAILFifty assertions is $61.1\%$.  
These \totalKaggleFAILFifty assertions, spread across \failFiftyNBCnt notebooks, are of
type \texttt{assert\_allclose} that check model performance. 
% Even though \tool
% accounts for randomness during assertion generation, 
Due to the small number of runs (30), in some cases, \tool under-approximates
the variance in the results and produces overly strict bounds. We inspect these
\totalKaggleFAILFifty assertions and find that in
\totalKaggleFAILFiftyInspectPass of them, the assertion bounds were overly
strict. When rerun with $50$ iterations at a $0.999$ confidence level and
validated with $30$ \texttt{pytest} runs, all assertions passed. 
The remaining \totalKaggleFAILFiftyInspectNoPass assertions failed due to
non-deterministic assertions on training history values, where even very loose
bounds could not ensure consistency. The checked variables exhibited highly
irregular distributions, so even increasing the confidence level to $0.999$ with
$50$ iterations did not help.

\mypara{Assertion Quality vs Confidence Thresholds}
Table~\ref{tab:rq1_assert_quality} shows how the quality of the generated
assertions, specifically, the passrate, mutation score, estimated false positive
rate (FPR), estimated false negative rate (FNR), changes with different
confidence thresholds used during assertion generation. Because the static
analysis for Python is imprecise, the reported FPR only estimates the true FPR.
Our manual inspection of a sample of the reported FPs revealed that many are in
fact true positives, indicating that the reported FPR overestimates the true
one. Overall, when the confidence threshold increases from $0.5$ to $0.999$, the
passrate increases from \totalKagglePASSRateZeroFiveRev to
\totalKagglePASSRateZeroNineNineNineRev, as expected. However, the mutation
score drops significantly from \mutantScoreKaggleZeroFiveRev to
\mutantScoreKaggleZeroNineNineNineRev, along with the FPR. Because Chebyshev's
method is very conservative, the passrates are already very high even at
$0.5$.%, similarly for the mutation score. 
An ideal confidence threshold should result in high passrate, high mutation
score, low FPR and low FNR. Both thresholds $0.99$ and $0.999$ have
very high passrates. However, the mutation score at $0.99$
is slightly higher than at $0.999$.
% However, in terms for FPR, confidence threshold $0.7$ is much higher than confidence threshold $0.99$. 
% Since higher FPR can cause inconvenience to developers,  
Hence, we recommend using a confidence threshold of $0.99$ for generating
assertions.

% \added{We evaluate how the quality of generated assertions varies with different
% confidence thresholds used during assertion generation.
% Table~\ref{tab:rq1_assert_quality} shows how the passrate, mutation score,
% false positive rate (FPR), and false negative rate (FNR) (defined in
% Section~\ref{sec:metrics}) of the generated assertions change with different
% confidence thresholds used during assertion generation. Overall, we observe that
% as the confidence threshold increases, the passrate of assertions also
% increases, as expected. However, the mutation score drops significantly from confidence threshold 0.5
% to 0.7, while the drop is not significant from confidendence threshold 0.7 to 0.99. 
% While the passrate and mutation score do not change significantly from confidence threshold 0.7 to 0.99, 
% the FPR drops significantly from \FPZeroSevenRev to \FPZeroNineNineRev. But the FNR does not increase
% significantly (only 5\% relative difference). This shows that using a higher
% confidence threshold can significantly reduce the FPR without significantly
% increasing the FNR, which is desirable in practice.  Because Chebyshev's method
% is known to provide a conservative bound, hence the passrates are already very
% high even with 0.5 threshold, and the mutation score is also high. However, the
% looser threshold also leads to higher FPR, which can cause inconvenience to
% developers. Hence, we recommend using a confidence threshold of 0.99 for
% generating assertions.}

%\begin{wraptable}[6]{r}{0.6\textwidth}
\begin{table}[t!]
    \tablesize
    \centering
    % \vspace{-0.2in}
    \caption{Quality of assertions generated by \toolgen}
    \vspace{-0.1in}
    \label{tab:rq1_assert_quality}
    % \begingroup
% \color{blue}
% \arrayrulecolor{blue}
    \begin{tabular}{lrrrr}        
        \toprule
        \makecell[l]{\textbf{Threshold}}&  \makecell[r]{\textbf{passrate}} & \makecell[r]{\textbf{mutation score}} & \makecell[r]{\textbf{Est. FPR}} & \makecell[r]{\textbf{Est. FNR}} \\ \hline
        0.5        &  \totalKagglePASSRateZeroFiveRev & \mutantScoreKaggleZeroFiveRev  &  \FPZeroFiveRev & \FNZeroFiveRev\\
        0.7       &  \totalKagglePASSRateZeroSevenRev & \mutantScoreKaggleZeroSevenRev  &  \FPZeroSevenRev & \FNZeroSevenRev\\
        0.99     &    \totalKagglePASSRate   & \mutantScoreKaggleZeroNineNineRev &  \FPZeroNineNineRev & \FNZeroNineNineRev\\
        0.999    &    \totalKagglePASSRateZeroNineNineNineRev & \mutantScoreKaggleZeroNineNineNineRev &  \FPZeroNineNineNineRev & \FNZeroNineNineNineRev\\
        \bottomrule
    \end{tabular}
 %   \endgroup
    \vspace{-0.2in}
\end{table}
%\end{wraptable}

\subsection{\rqmutationnum: Mutation Analysis of \toolgen's Assertions}
\label{sec:rq_mutation}
% \Fix{double check this RQ to remove old mutation data}
We apply the mutations described in \S~\ref{sec:mutant_generation} to all the
notebooks. This process generates many \emph{mutant notebooks}, each with a
single mutation, which we simply refer to as mutants hereafter.  Some mutations are
randomized (e.g., \outliers) and/or generate multiple mutants for the same
notebook (e.g., \modifyhyperparams). Hence, we apply such mutations multiple
times and generate a maximum of 4 mutants per mutation per notebook where
applicable. We run each mutant \pytestIte times and compute the ratio of
runs in which the mutant is killed by an assertion.

Table~\ref{tab:mut_score} presents the results. Each row presents the mutation
score for one mutation. A mutant is \emph{killed} if at least one assertion
fails. The first four rows represent Data-based mutations, while the next four
represent Source Code-based mutations.
% ORIGINAL:
% Table~\ref{tab:mut_score} presents the results. Each row presents the mutation score for one mutation. 
% The second, third, and fourth columns present the number of mutants killed by each type of assertion.
% One mutant may be killed by one or more assertions of a given type. So, we
% consider a mutant as \emph{killed} if at least one assertion of that type fails.
% The \textbf{Mut. Gen.} and \textbf{Mut. Killed} columns present the total
% mutants generated by a mutation and mutants killed by one or more assertions. 
% The final column \textbf{Mut. Score} presents the mutation score for each
% mutation. The final row \textbf{Total} presents the mutants killed by each type
% of assertion, the mutants generated/killed mutants, and the overall mutation
% score. The first four rows represent Data-based mutations, the next four
% represent Source Code-based mutations.

\begin{table}[t!]
    \centering
    \tablesize
    % \vspace{-0.2in}
    \caption{Distribution of Mutation Scores}
    \label{tab:mut_score}
    \vspace{-0.1in}
    % \begingroup
    % \color{blue}
    \setlength{\tabcolsep}{0.3em}
    \begin{tabular}{l|rrr|r|r|r}
    \toprule
    \textbf{Mut.\textbackslash Assert.} 
    & \textbf{\datasetAssert} 
    & \textbf{\makecell[r]{Model\\Perf.}} 
    & \textbf{\makecell[r]{Model\\Arch.}}  
    & \textbf{\makecell[r]{Mut.\\Gen.}} 
    & \textbf{\makecell[r]{Mut.\\Killed}} 
    & \textbf{\makecell[r]{Mut.\\Score}} \\ \midrule
    \outliers   &   \outliersDatasetKaggle    &      \outliersModelPerfKaggle          &   \outliersModelArchKaggle        & \numMutantsOutliersKaggle & \numMutantsKilledOutliersKaggle & \mutationScoreOutliersKaggle \\
    \repetition  &    \repetitionDatasetKaggle            &   \repetitionModelPerfKaggle             &   \repetitionModelArchKaggle         & \numMutantsRepetitionKaggle  & \numMutantsKilledRepetitionKaggle & \mutationScoreRepetitionKaggle \\
    \addednull &      \addedNullDatasetKaggle          &    \addedNullModelPerfKaggle            &     \addedNullModelArchKaggle        & \numMutantsAddedNullKaggle & \numMutantsKilledAddedNullKaggle & \mutationScoreAddedNullKaggle\\
    \labelerrors &     \labelErrorsDatasetKaggle           &   \labelErrorsModelPerfKaggle             &   \labelErrorsModelArchKaggle         &  \numMutantsLabelErrorsKaggle & \numMutantsKilledLabelErrorsKaggle & \mutationScoreLabelErrorsKaggle\\ \midrule
    \textbf{\makecell[l]{Data-based\\Total/Mutation score}} &  \makecell[r]{\totalDatasetDataMut\\(\DatasetDataMutScore)}     &  \makecell[r]{\totalPerfDataMut\\(\PerfDataMutScore)}             &    \totalArchDataMut (\ArchDataMutScore)      & \totalDataMut  & \totalDataMutKilled & \DataMutScore \\ \midrule
    \removezerograd &    \removeTorchZeroGradDatasetKaggle            &    \removeTorchZeroGradModelPerfKaggle              &  \removeTorchZeroGradModelArchKaggle   & \numMutantsRemoveTorchZeroGradKaggle & \numMutantsKilledRemoveTorchZeroGradKaggle & \mutationScoreRemoveTorchZeroGradKaggle\\
    \modifyhyperparams &       \modifyHyperparametersDatasetKaggle         &     \modifyHyperparametersModelPerfKaggle            &  \modifyHyperparametersModelArchKaggle   & \numMutantsModifyHyperparametersKaggle  & \numMutantsKilledModifyHyperparametersKaggle & \mutationScoreModifyHyperparametersKaggle \\
    \removehyperparams &     \removeHyperparametersDatasetKaggle           &     \removeHyperparametersModelPerfKaggle              &  \removeHyperparametersModelArchKaggle   &  \numMutantsRemoveHyperparametersKaggle & \numMutantsKilledRemoveHyperparametersKaggle & \mutationScoreRemoveHyperparametersKaggle\\
    \layerremove &     \deepLayerRemovalDatasetKaggle           &     \deepLayerRemovalModelPerfKaggle           &      \deepLayerRemovalModelArchKaggle      & \numMutantsDeepLayerRemovalKaggle & \numMutantsKilledDeepLayerRemovalKaggle & \mutationScoreDeepLayerRemovalKaggle\\
    \midrule
    \textbf{\makecell[l]{Code-based\\Total/Mutation score}} & \makecell[r]{\totalDatasetCodeMut\\(\DatasetCodeMutScore)}              &   \makecell[r]{\totalPerfCodeMut\\(\PerfCodeMutScore)}    &  \makecell[r]{\totalArchCodeMut\\(\ArchCodeMutScore)}      & \totalCodeMut  & \totalCodeMutKilled & \CodeMutScore \\ \midrule
    \midrule
    \textbf{Total (All Mutations)}  &       \totalKilledDatasetKaggle         &       \totalKilledModelPerfKaggle         &    \totalKilledModelArchKaggle        & \totalMutantsKaggle & \totalMutantsKilledKaggle & \mutantScoreKaggleZeroNineNineRev\\ 
    \bottomrule
    \end{tabular}
   % \endgroup
    \vspace{-0.2in}
    \end{table}

\mypara{General Trends} Overall, the assertions generated by \toolgen obtain a
mutation score of \mutantScoreKaggleZeroNineNineRev (or
\mutantScoreKaggleZeroNineNinePercRev) across all mutations. We generate a total
of \totalDataMut dataset-related mutants. The mutants are
either killed by \datasetAssert or \modelPerfAssert assertions.
% all of which are either killed by \datasetAssert or \modelPerfAssert assertions.
%
Because ML pipelines are often resilient to small data errors, we
observe that corrupting the dataset does not always cause an observable
degradation of the model performance. So, \modelPerfAssert assertions only have
a mutation score of \PerfDataMutScore for data-based mutations. In contrast, the
\datasetAssert assertions obtain a mutation score of \DatasetDataMutScore. They
are generally stricter, e.g., they check if the mean and variance of data
columns are close to expected values. Hence, they can detect data-based
mutations more accurately and earlier in the ML pipeline. \modelArchAssert
assertions are unable to detect such mutations because they only check for the
integrity of the ML model architecture and its parameters.

We generate \totalCodeMut code-related mutants that either change the training hyperparameters, or the model
architecture (e.g., \layerremove). The overall mutation score is \CodeMutScore.
Expectedly, dataset assertions fail to detect most of these because
these mutations typically occur later in the pipeline. These mutations are
mostly detected by \modelArchAssert assertions (mutation score
\ArchCodeMutScore) or \modelPerfAssert assertions (mutation score
\PerfCodeMutScore). The \modelArchAssert assertions are stricter in general, so
they can detect bugs that affect the model architecture easily.
% The \modelArchAssert assertions are stricter in general, so they can almost detect bugs that affect the model architecture easily. 

\mypara{Trends across Data-based mutants} We observe that almost \outliersKilledByModelPerf of
\outliers, \addedNullKilledByModelPerf of \addednull mutants, and \repetitionKilledByModelPerf of \repetition are killed by
\modelPerfAssert assertions. In such cases, the model performance changes
significantly. For example, for one notebook~\cite{outliersKilled_model_perf}
the accuracy dropped to $0.197$ from $0.63$ after applying the \outliers
mutation. In another example~\cite{outliersKilled_model_perf_2}, the \outliers
caused the accuracy to increase to $0.982$. 
% \deleted{For \datashift mutations,
% they only reorder the dataset rows, so they are also hard to detect. In the
% future, one can generate assertions that check the relationships between data
% columns instead of single-column properties (e.g., mean and variance) to detect
% such subtle regressions. An interesting future work could be to generate
% assertions that check the correlation between data columns.}

% \looseness=-1 
\mypara{Trends across Code-based mutants} Some mutations like
\removezerograd only impact the training process by affecting how model
gradients are updated across training iterations. Hence, they often cause
non-deterministic assertion failures, where the model performance changes
drastically but only in a fraction of the cases. For \modifyhyperparams mutants,
around \modifyHyperparametersKilledByModelArch of them are killed by \modelArchAssert, and around \modifyHyperparametersKilledByModelPerf of them
are killed by \modelPerfAssert. Since most of the \modifyhyperparams or
\removehyperparams mutants modify the hyperparameters in the model definition
(such as choice of activation layer), the regression is detected by the
\modelArchAssert easily. In some other cases, a change in hyperparameters also
causes drastic changes in model performance, which is detected by
\modelPerfAssert assertions. Among the surviving mutants, the major reason is
typically when the hyperparameter change does not alter the training results
beyond the expected range in \modelPerfAssert assertions. 
% \deleted{We find similar reasoning behind the lower mutation score of \metricswap mutation.} 
Because our mutations are applied statically, there is no guarantee that these will alter
the pipeline behavior. An interesting future work would be to design mutations
that leverage dynamic information to generate behavior-modifying mutants.

\mypara{Surviving mutants} 
We manually inspect $70$ surviving mutants. We find that $27$
mutants survived because the involved APIs are not tracked by \toolgen -- these
can potentially be killed by improving API coverage. The remaining
$43$ are equivalent mutants ($8$
data-based, $35$ code-based) that either do not semantically modify the pipeline
or do not create an observable difference during the execution. Among the data-based
equivalent mutants, three mutants targeted dataframes with fewer than ten
columns, so 10\% selection yielded no columns to mutate. Three mutants were
applied to empty dataframes and two mutants introduced anomalies which were later
removed by preprocessing. Among the code-based equivalent mutants, $28$
\modifyhyperparams/\removehyperparams mutants only changed the random seed.
Three \removehyperparams removed explicit hyperparameters already set to their
API defaults (e.g., \texttt{learning\_rate=0.1}), and four \modifyhyperparams
mutants merely changed the verbosity or formatting. The presence of equivalent
mutants can underestimate \toolgen's true mutation score, but computing the true
score is generally undecidable~\cite{equivalent_mutants}. % and more challenging in
%stochastic ML pipelines.
Hence, we leave this for future work.

\subsection{\rqversionsnum: Detecting Real-World Regressions with \toolgen's Assertions}

% \end{wraptable}

\looseness=-1
Using the methodology described in \S~\ref{sec:kaggle_version_collection}, we
collect \numKaggleVersionNoExeErrorRev versions of Kaggle notebooks from \numKaggleNotebooksVersionNoExeErrorRev notebooks (out of \totalKaggleNotebooks).
% containing at least one older version. We execute each older version to check if
% it can run without errors. \Fix{Out of \versionNoDuplicate versions}, we find
% \numKaggleVersionNoExeErrorRev can run successfully, and
% \numKaggleNotebooksVersionNoExeErrorRev notebooks have at least one version that
% can execute without error. 
For these remaining notebook versions, we generate
assertions in the latest versions using \toolgen and transfer the assertions to the
older versions using the methodology described in
\S~\ref{sec:kaggle_version_collection}. 
We successfully injected \totalRatioInjectedAssertionsKaggleRev of the original
assertions into \numKaggleVersionNoExeErrorRev versions, with
\avgInjectedAssertionsKaggleRev assertions per version on average. 
% The injected assertions consisted of
% \ratioInjectedDatasetRev \datasetAssert assertions, \ratioInjectedModelArchRev \modelArchAssert assertions, and
% \ratioInjectedModelPerfRev \modelPerfAssert assertions.
% , closely matching the original assertion distribution ($77.53\%$, $18.55\%$, and $3.92\%$, respectively). 
\label{sec:rq_kaggle_version}

% ------ Table ------

% \begin{wraptable}[9]{r}{0.25\textwidth}
\begin{table}[t!]
    \centering
    \tablesize
    % \vspace{-0.2in}
    \caption{Regression Testing of Kaggle versions}
    \label{tab:version}
    \vspace{-0.1in}
    \setlength{\tabcolsep}{0.1cm}
    %\begingroup
%\color{blue}
%     \begin{tabular}{l|rHH}
%     \toprule
%     \textbf{Assert Type} & \textbf{Versions killed}(Out of \numKaggleVersionNoExeErrorRev) & \textbf{FP}  & \textbf{FN} \\
%    % & (Out of \numKaggleVersionNoExeErrorRev ) &  & \\
%     \midrule   
%     \datasetAssert & \numVersionKilledDatasetKaggleRev (\percVersionKilledDatasetKaggleRev) & &  \\
%     \modelArchAssert & \numVersionKilledModelArchKaggleRev(\percVersionKilledModelArchKaggleRev) &  &   \\
%     \modelPerfAssert  & \numVersionKilledModelPerfKaggleRev(\percVersionKilledModelPerfKaggleRev)  & &  \\
%     \midrule
%     \textbf{Total} & \totalVersionKilledKaggleRev~(\totalPercVersionKilledKaggleRev) & \Fix{\numFPVersionRev~(\numFPRateVersionRev)} &  \numFNVersionRev~(\numFNRateVersionRev)    \\
%     \bottomrule
%     \end{tabular}
\begin{tabular}{l|rrr|r}
\toprule
\textbf{Assert Type} 
& \textbf{\datasetAssert }
& \textbf{\modelArchAssert }
& \textbf{\modelPerfAssert }
& \textbf{Total} \\ \midrule
\textbf{Versions killed} 
& \numVersionKilledDatasetKaggleRev~(\percVersionKilledDatasetKaggleRev)
& \numVersionKilledModelArchKaggleRev~(\percVersionKilledModelArchKaggleRev)
& \numVersionKilledModelPerfKaggleRev~(\percVersionKilledModelPerfKaggleRev)
& \totalVersionKilledKaggleRev~(\totalPercVersionKilledKaggleRev) \\

\bottomrule
\end{tabular}
%    \endgroup
\vspace{-0.2in}
\end{table}

Table~\ref{tab:version} presents how many of these versions (which denote
regressions) can be \emph{killed}/detected by \toolgen's assertions. Each cell
denotes the number of versions killed by each type of assertion. 
% The third column shows the number of original notebooks for which at least one
% version was killed. The final column shows the number of original notebooks for
% which all versions were killed.
Overall, \totalVersionKilledKaggleRev versions (out of
\numKaggleVersionNoExeErrorRev,
\totalPercVersionKilledKaggleRev) are successfully
killed by our assertions, showing that our automatically generated assertions
are effective at detecting many real-world regressions. 
% For \totalNotebookVersionKilledKaggle notebooks,
% the assertions kill at least one of their versions, while for
% \totalNotebookVersionAllKilledKaggle notebooks, the assertions can kill all
% their versions.
%
Overall, we observe a relatively low false positive rate of
\numFPRateVersionRev. However, our assertions also missed many regressions, with
a false negative rate of \numFNRateVersionRev~(i.e., \numFNVersionRev
versions).

\looseness=-1
%We analyze all results and categorize them into success and failure modes. 
Among versions that were killed by our assertions, the changes typically
involved dataset pre-processing modifications (e.g., dropping different dataset
columns~\cite{versiondataset}) and model architecture changes (e.g., changing
the number of estimators in scikit-learn's Random Forest
Classifier~\cite{version_modelarch}). These are typically detected via our
\datasetAssert and \modelArchAssert assertions, respectively. Often, such
changes also significantly affect the performance of the ML pipeline, which is
then detected by our \modelPerfAssert assertions. For example, a notebook
version~\cite{version_modelperf} was killed by all three kinds of assertions.
Most missed regressions fall into one of the two categories: (1) the changes made
did not modify the program's semantics, and (2) significant changes
to data pre-processing or model architecture hindered the automatic
transfer of original assertions. 
%  \Fix{do we have any numbers for these?}
% Among the missed regressions, they were typically either due to no semantics modifying changes or caused due to significant changes in data pre-processing or model architecture that did not allow us to easily or automatically transfer some original intermediate assertions to the notebook. 
% In the latter cases, these changes did not significantly modify the model performance. 
% Manually transferring the original assertions may improve the detection of such regressions.

% \Fix{i deleted 5.4.1. we can add maybe a couple of sentences describing the conclusion. move the rest to an appendix.}
\Comment{
We also manually inspected a random sample of $20$ versions to evaluate the correctness of our assertion transfer mechanism. 
Overall, the transfer mechanism is highly effective, with only very rare cases of incorrect transfers.
These incorrect transfers mainly occur when preceding statements are textually identical 
across versions, but the surrounding context differs semantically.
For assertions that are not transferred, the main reason is that 
the corresponding statements in the latest version are missing in the older version, making the assertions non-transferrable.
In a small of number cases, the corresponding statements are only partially similar 
but involve substantial refactoring across versions, 
making it difficult to determine whether a transfer is valid, even under manual inspection.
}
\Comment{
\subsubsection{Evaluating the transfer mechanism}
Our transfer mechanism may not be as effective as desired. So, to
evaluate the correctness of our assertion transfer mechanism, we randomly
selected 20 notebook versions and manually investigated two questions: (1) among
the transferred assertions, what percentage are correctly transferred? and (2)
among the untransferred assertions, what are the main reasons for non-transfer,
and what percentage of them are actually transferable?

Considering the first question, across the 20 versions, 687 assertions
were automatically transferred. Among them, we observe that only two were incorrectly
transferred. In the first case, the assertion was transferred because the
preceding statements were textually identical between the original and the target
versions, however, the surrounding context in the target version was
semantically different, which made the assertion invalid in the target version.
In the second case, the original version contained an assignment statement,
whereas the target version contained the same right-hand side expression, but it
was not assigned to a variable. Although the context was otherwise identical,
creating an assertion would require modifying the target version with a variable
assignment, which we did not allow in our tool.

Considering the second question, across the 20 versions, 370 assertions
were not transferred. Among them, we identified two assertions that were in fact
transferable. In these two cases, the codes differed by a single keyword in an
API call. So, while our simple transfer mechanism failed, the LLM-based approach
should have worked here. However, we find that the LLM-based approach simply
missed to transfer two out of 12 assertions associated with this statement.
Out of the remaining 368 assertions, 356 assertions were not transferable because
the statements they checked in the original version were completely missing in
the target, and the remaining 12 assertions were not transferable because the
corresponding statements in the target version were only partially similar but
had substantially refactored logic, which made it difficult to determine whether a
transfer was valid. In such cases, even manual determination is difficult.
Overall, we observe that the quality of assertion transfer is generally
very high, with only minor cases of incorrect transfer or missed transfer.
}

\subsection{\rqnbvalnum: \toolgen vs. \NBval}
\label{sec:rq_nbval}
To compare \nbval with \toolgen, we use the same experimental steps and collect
the same metrics as in RQ1 for \nbval. Table~\ref{tab:rq2_cell_cov} presents the
results. 
Overall, we observe that \toolgen generates \assertNbtestOverNbval more
assertions than \nbval (\assertNbtestOverNbvalPerc greater). On average,
\toolgen generates \avgAssertNbtestOverNbval more  (or
\avgAssertNbtestOverNbvalPerc more) assertions than \nbval per notebook. More
importantly, the passrate of \nbval is \passrateNbtestOverNbval lower than that
of \toolgen, indicating that \toolgen produces more reliable assertions. While
\nbval seems to achieve a higher mutation score, on a closer inspection, we find
that most of the failures are unrelated to the mutants, meaning these assertion
failures are false positives. Because \nbval performs strict output comparisons,
it often makes brittle checks that break due to differences in plot metadata and low-level logging data, which are not meaningful to developers. This can be
further confirmed by the distribution of passrate of \nbval's assertions in
Table~\ref{tab:rq1_pass_rate}. We find \RatioNbvalPassZero of assertions fail in
all runs, even though the code remains unchanged. 

While we computed FPR and FNR for \toolgen, we cannot compute them for
\nbval because \nbval asserts on cell outputs instead of specific variables and
in general it is hard to determine which statements are producing outputs.
% Consequently, we cannot perform data flow analysis in this scenario to map
% assertions to statements.
%So, we perform manual inspection of \nbval's results.
%So instead, we manually inspect the \nbval's results for qualitative analysis.
So, we manually inspect $20$ notebooks to understand why \nbval's assertions
fail consistently. We find three main sources of failures: (1) notebook cells
often contain model outputs, such as prediction results or training loss, which
are typically non-deterministic and exact matches almost always fail, (2)
comparing plots almost always fails because of differences in plot metadata, and
(3) comparing log data (e.g., timestamps, training logs) that also cannot be
compared across runs. Beyond these cases, \nbval often checks outputs that are
not meaningful and more importantly, not useful for debugging. In contrast,
\toolgen targets important ML-related properties and also accounts for
non-determinism, thereby avoiding such problematic scenarios and producing
higher quality assertions.

\begin{table}[t!]
    \tablesize
    \centering
    \setlength{\tabcolsep}{0.05cm}
    % \vspace{-0.1in}
    \caption{\toolgen vs. \nbval: Assertion Statistics and Quality}
    \label{tab:rq2_cell_cov}
    \vspace{-0.1in}
    % \begingroup
% \color{blue}
    \begin{tabular}{lrrrr}
    \toprule
    \makecell[l]{\textbf{Tool}}  & \makecell[r]{\textbf{\# Assertions}} &  \makecell[r]{\textbf{\#Assertions/Notebook}} &  \makecell[r]{\textbf{passrate}}  &  \makecell[r]{\textbf{mutation score}}\\ \midrule
    nbval    &     \totalNbvalAssertionsRev &  \avgtotalNbvalAssertionsRev &    \totalNbvalPASSRateRev &    \mutantScoreNbvalRev \\
    \tool (0.99)    &   \totalKaggleAssertions & \avgtotalKaggleAssertions &    \totalKagglePASSRate  & \mutantScoreKaggleZeroNineNineRev  \\
    \bottomrule
    \end{tabular}
%    \endgroup
    \vspace{-0.2in}   
\end{table}

\section{User Study and Real-World Adoption}\label{sec:userstudy} The goal of our user
study was to understand: 1) if developers of ML notebooks find \tool's Python
library intuitive and convenient to use in their daily workflows, and 2) if
\toolgen's generated assertions are useful compared to manually written assertions.
To fulfill this goal, we recruited $17$ developers who have prior experience
with writing ML code in Jupyter notebooks. 

\mypara{Recruitment} 
% We contacted students across multiple universities via email. 
% Finally, we recruited two Masters and five PhD students for our study. 
% Overall, five students have $1$ to $5$ years of experience with using Jupyter Notebook for ML or Data science-related tasks, while the remaining two have less than one year of experience. 
We recruited 17 participants (15 graduate students, 2 software engineers)
through online advertisements on social media and word-of-mouth referrals.
Overall, three participants have over five years of experience in machine
learning, and four have more than five years of experience using Jupyter
notebooks for ML-related tasks. None of the participants is an author of this
paper or contributed to this paper in any other way.

\mypara{Study Design} For each participant, we sent them a document containing
the setup instructions, a tutorial of \tool's different components and examples
of API usage, the assertion generator, and the JupyterLab plugin. We designed four
tasks. In the first task, we provided participants with a notebook and asked them
to manually write assertions using \tool APIs for three cases to check for data
integrity, model integrity, and model performance.\Comment{They need to use
\tool's assertion APIs, then run the notebook using \texttt{pytest}.} In the
second task, we asked participants to evaluate \toolgen's generated assertions and
compare them with their manual assertions. In the third task, we injected a bug in
the notebook and asked participants to detect this bug using \toolgen's generated
assertions. The third task demonstrates that a bug can fail silently and remain
undetected during typical execution, while \toolgen's assertions can detect it
effectively. In the fourth task, we asked participants to use \tooljupyterplugin's
hide/show feature.

After completing the tasks, we asked participants to fill out a survey with 7
questions: (1) time taken to write assertions (in minutes with a timer); (2)
ease of writing assertions using \tool's APIs (1 = very difficult, 5 = very
easy); (3) whether \tool's \cellscoped assertions better capture developer
assumptions compared to notebooks without them (Yes/No); (4) usefulness of
automatically generated assertions (1 = not useful, 5 = very useful); (5)
willingness to integrate \tool into their Jupyter Notebook workflow (Yes/No);
(6) intuitiveness of using \tool with Pytest (1 = not intuitive, 5 = very
intuitive); and (7) readability of the Jupyter Notebook with hide/show assertions
feature (1 = poor, 5 = very good). 
%The survey also included a text box for general feedback.

\mypara{Study Results} 
On average, participants spent $7.9$ minutes manually writing the assertions. 
% Six participants finished in under $10$ minutes, whereas one participant took $20$ minutes.
The average rating for manually writing assertions using \tool's APIs was
$3.94$. This suggests that while \tool's APIs are useful, coming up with
assertions manually might be challenging for users. All participants agreed that
the notebook with \cellscoped assertions provided a better sense of developer
assumptions in the notebook than without them. Participants found \toolgen's
generated assertions very useful (avg. rating $4.24$, minimum $3$). 16
participants said that they would like to integrate \tool into their workflow.
Finally, the participants found \tool's interface very intuitive (avg. score
$4.3$). The average readability score was $4.7$, suggesting that the hide/show
feature is highly effective in improving the readability of \tool's assertions.
% Specifically, one of the developers rated the tool highly, with an average score
% of $4.8$ across all questions. 
Overall, we conclude that notebook users found \tool's interface intuitive and
its assertions useful.
%, and also appreciated the usefulness of \tool's generated assertions.

\Comment{ One participant commented,\emph{``When I used Jupyter Notebooks to
write ML-related code, I often ran into inexplicable errors or couldn’t tell
which cell execution had overwritten a value. Having this tool makes it easy to
reliably validate the correctness of my data''}. Another participant appreciated
the generated assertions, saying: \emph{``... I also really like this idea,
because sometimes I want to check whether my code is correct, and I'm too lazy
to manually write tests ...''}. \deleted{The participants also suggested many
improvements/features such as, interactive view of test results based on
code-diffs, generating assertions for a subset of dataset columns instead of all
(which is the default behavior of \tool), and integrating the assertion
generator in JupyterLab. Two participants asked whether \tool could be made
available as a VSCode plugin. Two participants suggested extending \tool to
support more types of assertions. Another two participants recommended adding
inter-procedural assertions and providing explanations for the generated
assertions.} \added{The participants also suggested many improvements, such as
interactive UI, VSCode plugin, explanations for assertion failures, and more.}
\deleted{Another participant pointed out that they would need to check both the
generated assertions and their code for correctness, and they wondered if it is
possible to generate assertions that are based on the correct expected output.}
\added{Another participant wondered if it is possible to generate assertions
that are based on the correct expected output.} Currently, \tool's generated
assertions are regression-based. \deleted{However, in the future, we plan to
consider these suggestions to improve \tool, such as incorporating
property-based oracles.}\added{However, in the future, we plan to work on these
suggestions.} }

\mypara{Real-world Adoption} We reached out to developers of open-source machine
learning (ML) projects, and the initial responses were encouraging. 
% For example, we contacted the maintainers of a widely popular repository focused on machine learning explainability, which has over 23.9k stars on GitHub. 
% The lead maintainer responded positively, stating: \emph{``We are happy to use tools that simplify our workflow here. ... Would be interesting to have a call about this for sure. Can we schedule something for the week ...?''} 
% We also heard back from a developer from a project focused on computational graph construction: \emph{``We will take a look at this evaluation component then. It will also likely be of interest in other projects.''} Lastly, the developers from two other ML projects replied that they will \emph{``happily check it.''} These positive responses show that our work has the potential for deeper collaboration with the ML open-source community.
We submitted a pull request (PR)~\cite{shap_pr} to integrate \tool into the
continuous integration (CI) pipeline of SHAP~\cite{shap,shap_paper}, a popular
open-source ML project (24.4k stars). The developers have merged our PR and
integrated \tool into SHAP's CI pipeline. We also have an ongoing
PR~\cite{ydata_pr} for fg-data-profiling~\cite{fg-data-profiling}, a popular
open-source ML project (13k stars). The developer labeled our issue with the
\textit{code quality} tag and commented that they ``tag it as part of the roadmap
so the community is aware that this is something we are keen to
integrate''~\cite{ydata_issue}. Developers of the NVIDIA-BioNeMo
library~\cite{bionemo} (820 stars) have also shown interest in
integrating \tool into their CI~\cite{bionemo_issue}. These positive responses
demonstrate the promise of \tool in the real world.

\Comment{
\begin{enumerate}
    \item How long does it take to write three assertions?

        Seven out of nine participants took less than ten minutes to write all three assertions. One of the eight participants took twenty minutes. The other participants didn't respond.
    \item How convenient was it to use the \tool API for writing assertions manually? (Rate on a scale from 1 to 5, where 1 = Not convenient at all and 5 = Very convenient.)

        The average rating is $3.33$. Two out of nine participants gave ratings below 3, and the the other seven out of nine participants gave ratings at least $3$.

    \item Do you think the notebook with cell-level assertions can better encode the developer's assumptions compared to the ones without assertions?

        All participants answered yes.

    \item How useful were the assertions automatically generated by \tool?(Rate on a scale from 1 to 5, where 1 = Not useful at all and 5 = Very useful.)

        The average rating is $4.44$. All participants gave at least $4$.

    \item Would you consider integrating \tool into your Jupyter Notebook workflow?

        Eight out of nine participants answered yes. The rest one answered no.

    \item How intuitive was the \tool tool with pytest? (Rate on a scale from 1 to 5, where 1 = Not intuitive at all and 5 = Very intuitive.)

        The average rating is $4.67$. All participants gave at least $4$.
    
\end{enumerate}
}
\section{Discussion} \label{sec:discussion} 
\mypara{Limitations}
% While our assertion APIs can be used by any Jupyter Notebook user, our assertion
% generation technique focuses on ML-related components such as datasets, models,
% and performance metrics. Hence, it may not be useful for other domains yet. 
%\toolgen may not be useful for non-ML domains yet. 
Our assertion generation relies on regression oracles. Hence, it is unaware
of the expected correct outputs. In the future, we plan to integrate
property-based oracles for data and training validations.   
Also, while the assertion generation only runs once offline, which is realistic
for many workflows, the overhead of the bound computation is still high for
long-running notebooks. To reduce the overhead of long-running notebooks and
updating the assertions after the notebook evolves, we plan to extend \tool to
support: \emph{(1) cost amortization via CI}: collecting multiple samples across
CI runs of the notebook and using such samples to estimate bounds, \emph{(2)
evolution-awareness}: obtaining the backward and forward slice of changed cells
and only re-running those cells to update assertions, \emph{(3) API-specific
variance models}: learning expected output variance from historical notebook
runs for each API.

\looseness=-1
\mypara{Threats to Validity} \textit{(1) Internal:} The internal threat comes
from the implementation of \tool. To address this threat, multiple co-authors
carefully performed testing and code reviews to check that it was correctly
implemented.
% There may be bugs in our code and scripts. To mitigate this threat, multiple co-authors of the paper reviewed each others' code over the course of six months and resolved many bugs.
\textit{(2) External:} 
%The main external threat comes from our studied
%notebooks. 
Since our findings are based on ML-focused notebooks, they may not
generalize to non-ML workflows; however, this is beyond the scope of this work.
% However, generalization beyond ML workflows is
%beyond the scope of this work.
Also, the studied notebooks might be limited to the covered ML domains and
frameworks. To mitigate this problem, we selected four diverse and
representative domains, i.e., tabular classification, time series forecasting,
computer vision, and deep learning, which is broader than in prior
work~\cite{yang2022data,drobnjakovic2024abstract}.
% Our results (generated assertions and mutation analysis) may not generalize beyond the selected notebooks. Because we focus on machine learning only, our results might be biased due to the selected notebooks. To mitigate this threat partially, we collected as many executable notebooks as possible. However, we do not expect our results to generalize beyond ML notebooks, which is beyond the scope of this work.
\textit{(3) Construct:} Mutants may not be representative of real bugs. To
mitigate this, we choose historical versions of Kaggle notebooks, which also
showed high mutation scores. To mitigate risk due to non-determinism, we run the
notebooks multiple times and use statistical methods to compute metrics. 

\Comment{
\mypara{Usability}
Too many assertions may clutter the notebook\Comment{, as reflected in our user
study response}. To mitigate this problem, the current NBTest JupyterLab plugin
allows users to toggle the visibility of assertions using a side-by-side layout
(Figure~\ref{fig:overview_a})\Comment{, following the design of prior
work~\cite{readability_2}}. This functionality allows users to hide or show
assertions as needed, improving notebook readability and reducing visual
clutter. In our user study, this hide/show feature received an average score of
$4.7$, highlighting its effectiveness. Moving forward, we plan to further refine
the NBTest user interface based on ongoing feedback from the community.
}
% However,our experimental hide/show feature in NBTest
% JupyterLab plugin as well as organizating the assertions in the side-by-side panel is highly effective in improving the readability of notebooks
% with NBTest assertions, as suggested by an average score of $4.7$ in the user
% study. We plan to continually keep improving the user interface of NBtest with
% additional interactions from the community. For example, we could consider
% seperating the assertions from the notebooks core logic~\cite{testbook} or
% organize the assertions using side-by-side layouts~\cite{readability_2}.
% \Fix{revisit this after the UI change}

\Comment{
\mypara{Future work} 
\Comment{\textit{(1) Generalizing to unseen ML libraries:}  
Currently, \tool supports the four most popular ML/Data libraries and can be
easily extended to similar APIs as mentioned in
Section~\ref{sec:propertyfinder}. However, \tool cannot directly support APIs
whose semantics differ substantially. To bridge this gap, we plan to formulate
assertion generation as conditional synthesis from API-usage context and train a
model that maps context to assertion templates.}
% we propose using machine learning techniques to automatically generate
% appropriate assertions. One possible approach is to train a neural network on
% the currently supported APIs and the assertions generated by \tool, enabling
% it to generalize to new, structurally different APIs.
\textit{(1) Evolution of the generated assertions:}  
% As the first tool to introduce cell-level assertions for Jupyter notebooks, 
\tool does not yet support  evolution-aware assertion updates. 
Currently, users need to re-run NBTest after notebook changes to re-generate assertions. 
% We acknowledge the importance of evolution-awareness and plan to address this in future work. 
We plan to add evolution-awareness by 1) detecting cells affected by code
changes and only re-running those cells to update assertions, and 2) caching and
restoring cell states from previous runs to reduce re-running cost. \textit{(2)
Reducing assertion generation cost:} \tool current requires $\sim$96 minutes per
notebook to generate assertions. To reduce this cost, we plan to learn library-
and training algorithm-specific variance models from experimental data to
predict the expected variance of outputs for given inputs and training
configurations. This method will allow us to generate assertions with expected
bounds without needing to run the notebook multiple times.
}
\section{Related Work}
\Comment{
\mypara{Static analysis of Jupyter Notebooks} 
Yang et al.~\cite{yang2022data} proposed a static data-flow analysis technique
to detect data leakage bugs in Jupyter Notebooks. Drobnjakovi{\'c} et
al.~\cite{drobnjakovic2024abstract} developed NBLYZER, an abstract
interpretation-based approach to prove the absence of data leakages in
notebooks. Liblit et al.~\cite{liblit2023shifting} identified six static
analysis rules to detect various defects in notebooks such as invalid execution
order of cells, and misuses of common deep learning library APIs. Unlike these
works, \tool focuses on dynamic detection of regression bugs in Notebooks.
}

\mypara{Dynamic analysis of Jupyter Notebooks}
Many prior techniques improve the executability and reproducibility of Jupyter
Notebooks. Snifferdog~\cite{wang2021restoring} automatically infers
% is a tool for automatically identifying 
correct library dependencies for Jupyter Notebooks.
Osiris~\cite{wang2020restoring} identifies possibly correct execution orders 
of Jupyter Notebook cells that reproduce the correct results.
NBSAFETY~\cite{macke2020fine} combines dynamic and static analyses to identify
unsafe cell executions, such as out-of-order executions.
% , often caused by
% interactive nature of notebooks. 
\tool plays a complementary role to these tools
by automatically generating assertions that can detect regression bugs.

\mypara{Empirical analysis of notebooks}
Robinson et al.~\cite{robinson2022error} studied how student users debug various
issues in notebooks\Comment{, and found seven debugging strategies with varying
success rates}. 
% Similarly, other works have analyzed the common bugs and pain
% points in notebooks~\cite{chattopadhyay2020s,de2022bug}.
Chattopadhyay et al.~\cite{chattopadhyay2020s} interviewed
\Comment{data scientists and software engineers who use notebooks} various
notebook users and identified several pain points, which included difficulty in
debugging and reproducibility of results. De Santana et al.~\cite{de2022bug}
analyzed numerous GitHub commits/StackOverflow posts to identify common sources
of bugs in Jupyter Notebooks.
 \toolgen's generated assertions can
potentially improve the quality and debuggability of notebooks.

% Chattopadhyay et al.~\cite{chattopadhyay2020s} interviewed
% \Comment{data scientists and software engineers who use notebooks} various
% notebook users and identified several pain points, which included difficulty in
% debugging and reproducibility of results. De Santana et al.~\cite{de2022bug}
% analyzed numerous GitHub commits/StackOverflow posts to identify common sources
% of bugs in Jupyter Notebooks. \deleted{The assertions generated by \tool can
% potentially improve the quality and debuggability of notebooks by pinpointing
% faulty code cells.}

\mypara{Bug Detection for Notebooks} Prior work has targeted specific types of
bugs in Jupyter Notebooks,
% developed techniques for detecting special classes of bugs in Jupyter Notebooks, 
such as name-value
inconsistencies~\cite{patra2022nalin}, data
leakage~\cite{yang2022data,subotic2022static,drobnjakovic2024abstract}, and syntax or dependency
errors~\cite{grotov2024debug,grotov2024untangling}. In contrast, \toolgen uses a
regression-based oracle, but has the potential to detect a broader range of
bugs, as shown by our mutation analysis and experiments with Kaggle versions.
% \NA{name value inconsistencies: Nalin~\cite{patra2022nalin}
% data leakage~\cite{yang2022data}\cite{subotic2022static}, 
% llm-based fixin~\cite{grotov2024debug,grotov2024untangling}
% }

\looseness=-1
\mypara{Assertion/Test Generation}
Many ML-based techniques have been proposed for generating assertions for unit
tests~\cite{dinella2022toga,watson2020learning,tufano2020unit,ibrahimzada2022perfect,white2020reassert}.
However, these techniques focus on simpler assertions for Java. \Comment{In
contrast, in this work we focus on ML-specific assertions that test data
pre-processing, model training, and model evaluation steps in a typical ML or
data science pipeline -- which prior tools do not support.} Tools such as
Pynguin~\cite{lukasczyk2022pynguin} and DynaPyt~\cite{test_gen_dynapyt} are
general test generation tools for Python, but they are not specialized for ML
programs or notebooks. In contrast, \toolgen automatically generates
assertions to detect regression bugs in notebooks and accounts for ML
randomness.

\Comment{
\mypara{Python Test Generation/Analysis Tools}
Pynguin~\cite{lukasczyk2022pynguin} is a unit test generation tool for Python.
However, it is a general tool and not specialized for ML programs
or Jupyter notebooks. Pynguin is unable to generate any meaningful tests even
when the notebook is converted to Python program. DyLin~\cite{test_gen_dylin}
detects general Python anti-patterns, and DynaPyt~\cite{test_gen_dynapyt} offers
a flexible framework for building dynamic analyses for Python. In contrast,
\tool automatically generates assertions to detect regression bugs in Jupyter
Notebook and accounts for randomness.
}
% \elaine{DyLin~\cite{test_gen_dylin} applies dynamic analysis to detect
% programming anti-patterns, but it primarily targets general Python constructs,
% with only three patterns related to machine learning. In contrast, \tool checks
% regressions across notebook versions and accounts for randomness in machine
% learning. DynaPyt~\cite{test_gen_dynapyt} provides a general framework for
% building dynamic analyses but offers no built-in analyses and is not specialized
% for machine learning or Jupyter notebooks.}
% \Fix{maybe combine dylin and dynapyt? better to say those are complementary to our work?}

% \added{We compared \tool with Inline
% Tests~\cite{liu2022inline,liu2023extracting} in Section~\ref{sec:intro}. While
% there are fundamental differences between the two approaches, our motivation is
% similar to inline testing in \emph{spirit} -- \tool allows developers to write
% tests at a lower level of granularity instead of unit or end-to-end tests.}
\mypara{Inline Tests} 
More recently, Liu et
al.~\cite{liu2022inline,liu2023extracting} introduced a new testing paradigm:
\emph{inline testing} that executes tests at a lower level of granularity
(statement level) than unit tests. Inline tests only target one statement at a
time and test general PL features (e.g., regexes, string manipulation). In
contrast, \toolgen's assertions check for the correctness of logic \emph{within a
cell}, and they lie in between inline tests and unit tests in granularity and target
ML-related properties. Also, inline tests require developer
inputs/oracles (\toolgen's automated assertions do not) and are enforced at
test-time (\toolgen's assertions are enforced at both production-time and
test-time).

\mypara{Testing Non-deterministic Systems} Prior work has
tested ML frameworks~\cite{pham2019cradle,wei2022freelunch,liu2023nnsmith,mltestingsurvey} and
probabilistic programming systems~\cite{dutta2018testing,dutta2019storm,llerena2018verifying}.
However, \tool focuses on detecting regressions.
Flaky test techniques detect and repair tests or assertions that fail non-deterministically~\cite{dutta2020detecting,9438576,eck2019understanding,dutta2021flex}.
However, these focus on tests and assertions developers have already written.
In contrast, \toolgen automatically identifies important properties of an ML
pipeline and generates regression-based assertions that account for non-determinism.

% \mypara{Testing Non-deterministic Systems} 
% \elaine{Prior work has tested ML frameworks~\cite{pham2019cradle,wei2022freelunch,liu2023nnsmith,mltestingsurvey} and probabilistic programming systems~\cite{dutta2018testing,dutta2019storm,llerena2018verifying}.
% and flaky-test techniques detect~\cite{dutta2020detecting,9438576,eck2019understanding,spade} or repair~\cite{dutta2021flex} non-deterministically failing assertions by widening their bounds. 
% However, these focus on tests and assertions developers have already written. 
% In contrast, \toolgen automatically identifies important properties of an ML pipeline and generates regression-based assertions that account for non-determinism.}
%\input{threats}
\section{Conclusion}
We presented \toolgen, an automated assertion generation approach (and \tool
framework) for generating \cellscoped and non-intrusive assertions for Jupyter
Notebooks. We showed that \toolgen is effective at detecting mutations and
real-world regressions. \tool has also been adopted by SHAP, a popular ML library, and
integrated into its CI pipeline, showing its promise for real-world adoption. We
envision that \tool will encourage developers to adopt test-driven development
of ML Notebooks and researchers to expand \tool to detect other classes of bugs.

\Comment{
We presented \tool, the first regression testing framework for Jupyter Notebooks
that allows developers to write cell-scoped and
non-intrusive assertions. \tool is specialized for ML-related
notebooks, especially for assertion generation. Through mutation analysis and
evaluating Kaggle notebook versions, we showed that \tool can effectively detect
many regressions. We envision that \tool will encourage developers to adopt test-driven development of ML Notebooks.
\tool has also been adopted by SHAP, a popular ML library, and integrated
into its CI pipeline. \tool has the potential to
enhance the reliability, reproducibility, and quality
of ML notebooks.}
% \tool has the potential to make ML Notebooks more reliable and reproducible, and improve their quality. 

\mypara{Data Availability} We share our artifacts and data
here~\cite{nbzenodo,aNBTest_github}. 
\section*{Acknowledgements}
We thank Owolabi Legunsen, Kevin Guan, Pengyue Jiang, Shinhae Kim and the anonymous reviewers for their feedback. 
This work is partially supported by Google Research Award, Meta LLM Evaluation Research Grant, and Google Gemini Cloud Credits.
\balance
\newpage
\bibliographystyle{ACM-Reference-Format}
\bibliography{references}
% \bibliographystyle{IEEEtran}
%\balance
% \clearpage
% \newpage
% \appendix
% \input{appendix}
\end{document}